\documentclass[twocolumn]{aastex631}

\usepackage{gensymb}
\usepackage{subfig}
\usepackage{multirow}
\usepackage{amsmath}
\usepackage{color}

\newcommand{\e}{$e^{-}$}

\newcommand{\nw}{nW m$^{-2}$ sr$^{-1}$}

\newcommand{\nh}{New Horizons}
\newcommand{\gaia}{Gaia}
\newcommand{\simmod}{\raise.17ex\hbox{$\scriptstyle\sim$}} 

\shorttitle{COB and DGL from New Horizons-LORRI}
\shortauthors{Symons et al.}
\accepted{for publication in \apj}

\begin{document}

\title{A Measurement of the Cosmic Optical Background and Diffuse Galactic Light Scaling from the $\mathbf{R < 50}$ AU New Horizons-LORRI Data}

\correspondingauthor{Teresa Symons}
\email{tas4514@rit.edu}

\author[0000-0002-9554-1082]{Teresa Symons}
\affiliation{Center for Detectors, School of Physics and Astronomy, Rochester Institute of Technology, 1 Lomb Memorial Drive, Rochester, NY 14623, USA}

\author[0000-0001-8253-1451]{Michael Zemcov}
\affiliation{Center for Detectors, School of Physics and Astronomy, Rochester Institute of Technology, 1 Lomb Memorial Drive, Rochester, NY 14623, USA}
\affiliation{Jet Propulsion Laboratory, California Institute of Technology, 4800 Oak Grove Drive, Pasadena, CA 91109, USA}

\author[0000-0002-3892-0190]{Asantha Cooray}
\affiliation{Department of Physics and Astronomy, University of California, Irvine, CA 92697, USA}

\author[0000-0002-9548-1526]{Carey Lisse}
\affiliation{Space Exploration Sector, Johns Hopkins University Applied Physics Laboratory, 11100 Johns Hopkins Road, Laurel, MD 20723, USA}

\author[0000-0001-8137-8176]{Andrew R. Poppe}
\affiliation{Space Sciences Laboratory, University of California at Berkeley, 7 Gauss Way, Berkeley, CA 94720, USA}



\begin{abstract}
Direct photometric measurements of the cosmic optical background (COB) provide an important point of comparison to both other measurement methodologies and models of cosmic structure formation, and permit a cosmic consistency test with the potential to reveal additional diffuse sources of emission. The COB has been challenging to measure from Earth due to the difficulty of isolating it from the diffuse light scattered from interplanetary dust in our solar system. We present a measurement of the COB using data taken by the Long-Range Reconnaissance Imager (LORRI) on NASA’s \nh\ mission, considering all data acquired to $47$ AU. We employ a blind methodology where our analysis choices are developed against a subset of the full data set, which is then unblinded. Dark current and other instrumental systematics are accounted for, including a number of sources of scattered light. We fully characterize and remove structured and diffuse astrophysical foregrounds including bright stars, the integrated starlight from faint unresolved sources, and diffuse galactic light. 
For the full data set, we find the surface brightness of the COB to be $\lambda I_{\lambda}^{\rm{COB}} = 21.98 \pm 1.23\ ({\rm stat.}) \pm 1.36\ ({\rm cal.})$ \nw. 
This result supports recent determinations that find a factor of $2 {-} 3 \times$ more light than expected from the integrated light from galaxies and motivate new diffuse intensity measurements with more capable instruments that can support spectral measurements over the optical and near-IR.
\end{abstract}

\section{Introduction}
\label{S:intro}
 

The extragalactic background light (EBL) is the sum of all light emitted by sources beyond the Milky Way integrated over the history of the universe. EBL sources include faint residual radiation from the universe's early evolution, such as the cosmic microwave background \citep{cmb}, as well as later emission from stellar and galactic evolution through cosmic time \citep{hauser_dwek,cooray_review}, and as a result is a powerful probe of cosmic structure formation. 
The EBL measured at optical wavelengths, called the cosmic optical background (COB), is thought to be largely sourced by stellar nucleosynthesis from stars in galaxies throughout cosmic history, but also includes emission from active galactic nuclei (AGN) and all other forms of blackhole activity, such as mini-quasars \citep{Tyson, cooray_yoshida}. Previously unaccounted sources such as diffuse populations of stars \citep{conselice, Roman2021} or the products of particle astrophysics \citep{Boddy2022} may contribute a non-negligible amount to the COB intensity. The COB therefore provides an important point of comparison to the summed emission from known populations of galaxies \citep{Driver2016} that can reveal additional diffuse sources of emission. 

Direct photometry of the COB has been difficult to accomplish from Earth due to complications arising from local bright foregrounds, including Earth’s atmosphere and the Zodiacal Light (ZL; diffuse light scattered from dust in our solar system, see \citealt{leinert}), which are generally $> 100 \times$ brighter than the expected level of the COB. Measurements have suffered from large uncertainties due to the difficulty in assessing and subtracting these bright foreground sources of emission \citep{hauser_dwek}. 

Performing EBL measurements from the outer solar system where scattered light from the Sun is reduced is an attractive option \citep{Zemcov_PASP}. Even beyond the bright ZL, COB measurements are challenging and require careful characterization and removal of all foreground emission sources to ensure the residual isolates the COB. For any arbitrary image of the astrophysical sky made above the atmosphere of Earth, the total measured brightness can be expressed as the sum of several components:
\begin{gather}
\lambda I_{\lambda}^{\mathrm{meas}} = \nonumber\\
\lambda I_{\lambda}^{\mathrm{*}} + \lambda I_{\lambda}^{\mathrm{ISL}} + \lambda I_{\lambda}^{\mathrm{DGL}} + \lambda I_{\lambda}^{\mathrm{IPD}} + \lambda I_{\lambda}^{\mathrm{inst}} + \epsilon\lambda I_{\lambda}^{\mathrm{COB}} ,
\label{eq:cob}
\end{gather}
where ``meas'' denotes the measured brightness of a sky image, ``*'' denotes the brightness of resolved stars, ``ISL'' denotes the brightness of the integrated starlight (ISL), including faint stars and the extended point spread function (PSF) of masked stars, ``DGL'' denotes the brightness of the diffuse galactic light (DGL) scattered by dust in the interstellar medium of the Milky Way, ``IPD'' denotes the brightness of light scattered by interplanetary dust (IPD) in the solar system, which is thought to be small at large ($>$ 10 AU) distances from the Sun, ``inst'' denotes any brightness caused by the instrument itself, ``COB'' denotes the brightness of the COB, and $\epsilon$ is a factor accounting for galactic extinction. 
Due to the faintness of the COB, a small error in the estimation of any of these components can produce large errors in its measured value.

The COB has been measured using a variety of instruments and methods from the vicinity of Earth. Photometric measurements include the ``dark cloud'' method, a differential measurement where the intensity of a high galactic latitude opaque Milky Way dust nebula is compared to the intensity of a nearby dust-free surrounding area. If the ISL can be accounted for, the difference between the dark cloud and surrounding region is a measurement of the EBL \citep{mattila_1990, Mattila_cob, dark_cloud}. Observation of the $\gamma$-ray emission from high-energy blazars offers a second method that takes advantage of the extinction of high-energy photons through the production of electron-positron pairs via interactions with EBL photons. In this method, the measured spectra of blazars is compared to the predicted spectra and the extinction from the EBL is estimated \citep{hess,fermilat_gamma,magic_gamma,gev_tev_gamma}. Direct number counts of galaxies offer a third method that provides a lower limit to the COB \citep{conselice}. Galaxy counts have been performed many times using deep integrations with a variety of facilities \citep[e.g.][]{Driver2016} and now have achieved $\sim 1$ \nw\ uncertainties across the optical.

The most direct way to measure the COB is through absolute photometry. In this method, estimates for the different terms of Eq.~\ref{eq:cob} are subtracted from the observed sky brightness, and the residual is associated with the COB. However, this method depends strongly on the ability to accurately remove the foreground emission, and attempts near Earth have yielded disparate results \citep{cooray_review}. From vantage points in the distant solar system where the foregrounds are smaller, the COB has been measured with data from Pioneers 10 and 11 (\citealt{Pioneer_1,pioneer_2} but see \citealt{Matsumoto2018}) and \nh\ \citep{nature,Lauer,lauer_2022}. Most recently, the measurements made with the Long-Range Reconnaissance Imager (LORRI) have assessed the COB with small statistical uncertainty in a broad band covering 440 to 870 nm at a pivot wavelength of $\bar{\lambda} = 655$ nm for a flat-spectrum source. Early work generated upper limits consistent with the expected light from galaxies \citep{nature}, but more recent measurements incorporating significantly more data in better-selected regions have yielded results about a factor of two brighter than the expected integrated galactic light (IGL; \citealt{Lauer,lauer_2022}). These results, if correct, have profound implications for the diffuse photon background at optical wavelengths, and combined with measurements at near-IR wavelengths \citep[e.g.][]{ciber,skysurf_2} may point to major problems with our accountancy of the electromagnetic products of structure formation in the universe.

In this paper, we present a new analysis of the COB drawn from all publicly available LORRI data as of mid-2022. In Section 
\ref{sec:data_selection}, we describe the LORRI data products used for our measurement and our data selection process. In Section \ref{sec:data_analysis}, we detail our data analysis pipeline and calibration procedure. In Section \ref{sec:foregrounds}, we discuss astrophysical foreground characterization and subtraction. In Section \ref{sec:error_analysis}, we develop our error budget and characterize the sources of uncertainty in our measurement. In Section \ref{sec:results}, we present our measurement in the context of previous work and discuss implications for future studies. Our calibrated and masked data products will be archived on the Planetary Data System for future public use. Additional details of this analysis are presented in \citet{Symons2022}. 

\vspace{30pt}
\section{Data Set}
 \label{sec:data_selection}
 
In this Section, we describe the nature of the data, the data selection process, and cuts applied to the available data sets to yield our scientific sample. 

\subsection{Input Data Characteristics}
\label{sec:nh}

\nh\ is NASA's first mission to survey the Pluto system and Kuiper Belt \citep{nh_first}. Launched in January 2006, \nh\ performed a flyby of Jupiter in 2007 as it traveled to the outer solar system. It completed its primary mission objective, a survey of Pluto, in 2015 \citep{nh_pluto}. After being approved for the Kuiper Belt Extended Mission (KEM; \citealt{KEM}), \nh\ performed a flyby of Arrokoth, a Kuiper Belt Object (KBO), in January 2019. \nh\ was recently approved for a second mission extension through 2025 as it continues to traverse the Kuiper Belt on its way out of the solar system. The LORRI instrument onboard \nh\ \citep{cheng_2008} is a 20.8 cm Ritchey-Chr\'etien telescope with a clear filter, broad optical passband (approximately 440 -- 870 nm) and a $0^{\circ}.29 \times 0^{\circ}.29$ field of view (FOV). It operates in both a 1 $\times$ 1 binning mode with 1024 $\times$ 1024 pixels 
and a more sensitive 4 $\times$ 4 binning mode with on-chip binning to 256 $\times$ 256 effective pixels that we use for our measurement. In its 4 $\times$ 4 mode, the point source sensitivity in a 10 second exposure is $V$ = 17 \citep{cheng_2008,Conard,Morgan}. 

Since launch in 2006, LORRI has taken a total of 19,990 publicly-available exposures as of 2022 Aug.~17. Pre-processed LORRI data are served from the Planetary Data System (PDS) as FITS files comprising intensity and error images, as well as metadata containing information about the observation taken and spacecraft status at the observing time. LORRI data are pre-processed by the LORRI instrument team to return science-grade images in raw units (DN). Because we later calibrate these to surface brightness units, we will refer to the pre-processed LORRI exposures as ``raw'' and our final calibrated products as ``calibrated.'' The LORRI pre-processing pipeline performs: a bias subtraction from in-flight dark images to correct pixel-to-pixel variations; smear removal to correct charge transfer effects in the CCD on bright objects; and finally flat-fielding using responsivity corrections obtained during ground testing. This results in the final raw exposure in DN \citep{cheng_2008}. 

\subsection{Survey Selection}
\label{sec:phase}

Because we are performing archival data analysis, not every LORRI exposure is a good candidate for measuring the COB.
Six data deliveries are available in the PDS Small Bodies Node. The data we consider in our analysis include: 

\textit{Post Launch}  The post-launch checkout data were taken from 2006 Feb.~24 -- 2006 Oct.~18 and include instrument commissioning tests and calibration data. There are a total of 1,235 exposures, including a set of bias images taken before LORRI's aperture door was opened on 2006 Aug. 29 \citep{lor_launch}. While we did not find any usable science exposures in this set, we do use the bias images to compare dark current before and after the aperture was uncovered. This set of dark images contains 359 exposures in the 4 $\times$ 4 binned mode taken from 2006 Apr.~23 -- 2006 May 3. 

\textit{Jupiter Encounter}  The Jupiter encounter data were acquired from 2007 Jan.~8 -- 2007 Jun.~11. There are 1,114 exposures including observations of the Jovian atmosphere, features, and ring system, the Galilean moons, and several smaller moons \citep{lor_jupiter}. Additionally, LORRI's optical scattering was characterized using these data \citep{Cheng2010}. We do not derive any of our science data from this phase, but we do use a set of six exposures of Callirrhoe, a small, $\simmod$10 km radius minor outer moon of Jupiter observed on 2007 Jan.~10 in order to test LORRI's operations on Pluto's moons pre-encounter. This field is designated Ghost 1 and discussed further in Section \ref{sec:ghostmask}. 

\textit{Pluto Cruise}  The Pluto cruise phase data were acquired from 2007 Sep.~29 -- 2014 Jul.~26. While the spacecraft spent a significant amount of time in hibernation during this period, the set includes 984 exposures taken during various check-outs in preparation for the Pluto encounter. Science observation targets included several KBOs as well as the planets Jupiter, Uranus, and Neptune \citep{lor_cruise}. The science fields of interest taken during this phase are called PC1 -- PC4, and these were previously analyzed to result in the COB measurement described in \citet{nature}. 

\textit{Pluto Encounter}  The Pluto encounter data were taken from 2015 Jan.~25 -- 2016 Jul.~16. This set of 6,773 exposures constitutes the bulk of the observations that fulfilled \nh' primary mission. The majority of the exposures are observations of Pluto and its moons taken during approach, the encounter, and departure from the Pluto system. There are also KBO observations and calibration tests \citep{lor_pluto}. Our science fields from this phase include PE1 -- PE4, which also make up the testing set used for pipeline development. This set contains 135 exposures of KBOs taken from 2016 Apr.~7 -- 2016 Jul.~13. 

\textit{KEM Cruise}  The KEM cruise phase data were taken from 2017 Jan.~28 -- 2017 Dec.~6. This phase has 1,863 exposures including observations of KBOs, calibration tests, and observations taken during the approach to Arrokoth \citep{lor_kemcruise}. Our science fields taken from this phase include KC1 -- KC4, a set of 174 exposures of KBOs taken from 2017 Sep.~21 -- 2017 Nov.~1. 

\textit{Arrokoth Encounter} The Arrokoth encounter data were acquired from 2018 Aug.~16 -- 2020 Apr.~23 and downlinked before 2020 May 1. Additional data taken during this time period that were downlinked after 2020 May 1 will be publicly available in a future release. This set of 8,021 exposures includes observations of Arrokoth, various KBOs, Pluto, Triton, and interplanetary dust \citep{lor_kem}. Our science fields from this set include AE1 -- 7, a set of high galactic latitude, low galactic foreground exposures previously analyzed by \cite{Lauer}, which comprise a set of 194 exposures acquired from 2018 Aug.~20 -- 2019 Sep.~4.

\subsection{Data Cuts}
\label{sec:reg}

Starting from the full collection of 19,990 exposures, we first exclude all data with exposure time $<$ 5 seconds as very short exposures do not have sufficient COB signal-to-noise ratio (S/N) to accurately assess subtle noise or instrumental features that may be present. Additionally, we wish to balance S/N with maintaining the largest possible data set. The remaining exposures are all acquired in LORRI's 4 $\times$ 4 binning mode. The dark exposures taken early in the mission, while useful for examining dark current, are also not useful for measuring the COB. The optical design of LORRI causes scattered light from baffle illumination due to low solar elongation angle (SEA, \citealt{Cheng2010}) to make some exposures unsuitable \citep{Lauer}. 
As a result, we exclude all exposures with SEA $<$ 90$\degree$, although as explored further in Section \ref{sec:jackknife}, extending this cut to SEA $<$ 105$\degree$ has little effect on the final measurement. The remaining light exposures are then astrometrically registered using \url{http://astrometry.net} \citep{astrometry} in order to associate right ascension ($\alpha$) and declination ($\delta$) for each pixel in a given exposure. We find that a small fraction of exposures are not able to be registered due to pointing drift or a defect in image quality that prevents accurate detection of point sources, so these are cut from the data set. All images surviving these cuts are visually inspected and classified based on the presence of bright objects (including images of the geography of Pluto) and obvious image-space defects. The number of exposures excluded for each of these reasons is given in Table \ref{tab:data_cuts} along with the fraction of the total available exposures and the total viable exposures remaining after all data cuts. 

\begin{deluxetable}{lcc}[htbp!]
\tabletypesize{\scriptsize}
\tablecaption{Data cuts made to total available LORRI data as both number of exposures cut and fraction of the total that this represents. The cuts include exposure time, astrometric registration, exposures containing Pluto or its moons, dark exposures taken before the LORRI aperture cover was opened (although these are used to estimate dark current), galactic latitude, solar elongation angle, pointing drift, irregular exposures, and the camera's power-on effect. We also list the total number of available LORRI exposures and the total remaining science exposures after all cuts have been completed. \label{tab:data_cuts}}
\tablehead{\colhead{\textbf{Type of Data Cut}} & \colhead{\textbf{\# of Exposures}} & \colhead{\textbf{Fraction of Total}}}
\centering
\startdata
    \textbf{Total Available} & 19,990 & 100\% \\
    \textbf{Exposure Time Cut} & 10,613 & 53\% \\
    \textbf{Registration Cut} & 504 & 2.5\% \\
    \textbf{Pluto Cut} & 1,405 & 7\% \\
    \textbf{Dark Image Cut} & 359 & 1.8\% \\
    \textbf{\boldmath$b$ Cut} & 4,223 & 21\% \\
    \textbf{SEA Cut} & 1,305 & 6.5\% \\
    \textbf{Pointing Drift Cut} & 246 & 1.2\% \\
    \textbf{Irregular Image Cut} & 10 & 0.05\% \\
    \textbf{Camera Power-On Cut} & 796 & 4\% \\ \hline \hline
    \textbf{Total Remaining} & 529 & 2.6\%
\enddata
\end{deluxetable}

The Milky Way is bright at optical wavelengths and so we concentrate on exposures at mid-to-high galactic latitude. This also excludes observations of Pluto and Arrokoth that were all taken within a few degrees of the galactic plane, which mitigates several foregrounds that complicate the measurement. At lower latitudes, the increased density of stars means that a greater fraction of the exposure will need to be masked, greatly reducing the number of background pixels that contribute to a measurement. Additionally, ISL and DGL are also much brighter at lower latitudes due to greater concentrations of stars and dust. The DGL in particular does not scale linearly with thermal emission in the optically thick regime \citep{leinert}. We therefore exclude any exposures at $b$ $<$ 30$\degree$ to avoid unassessed systematics in our DGL scaling, resulting in our second largest cut of 21\% of the total available data. 

When \nh\ is tracking KBOs, sequential exposures of the same target occasionally exhibit significant ($\sim 1^{\circ}$) drift over the course of several minutes. Because we average together multiple exposures of the same field later in our analysis, fields with $\geq$ $0^{\circ}.5$ of movement from exposure to exposure cannot be easily combined. We exclude 1.2\% of the complete data set to avoid these issues. 

A very small number of exposures (10 out of the data remaining from all previous cuts) display irregularities when compared with the bulk of the data. These exposures have extremely negative surface brightness, containing almost entirely negative pixel values in raw units. Since the surface brightness reported by the detector is unphysical, these exposures likely suffer from some kind of electronic irregularity. The exposures taken sequentially before and after those affected do not display the same issue and the cause is unknown, but we suspect transient cosmic ray upsets of the detector electronics. As these few exposures are true outliers with non-physical data values, we exclude them. 

\begin{figure*}[htb!]
    \centering
    \subfloat{{\includegraphics[width=8cm]{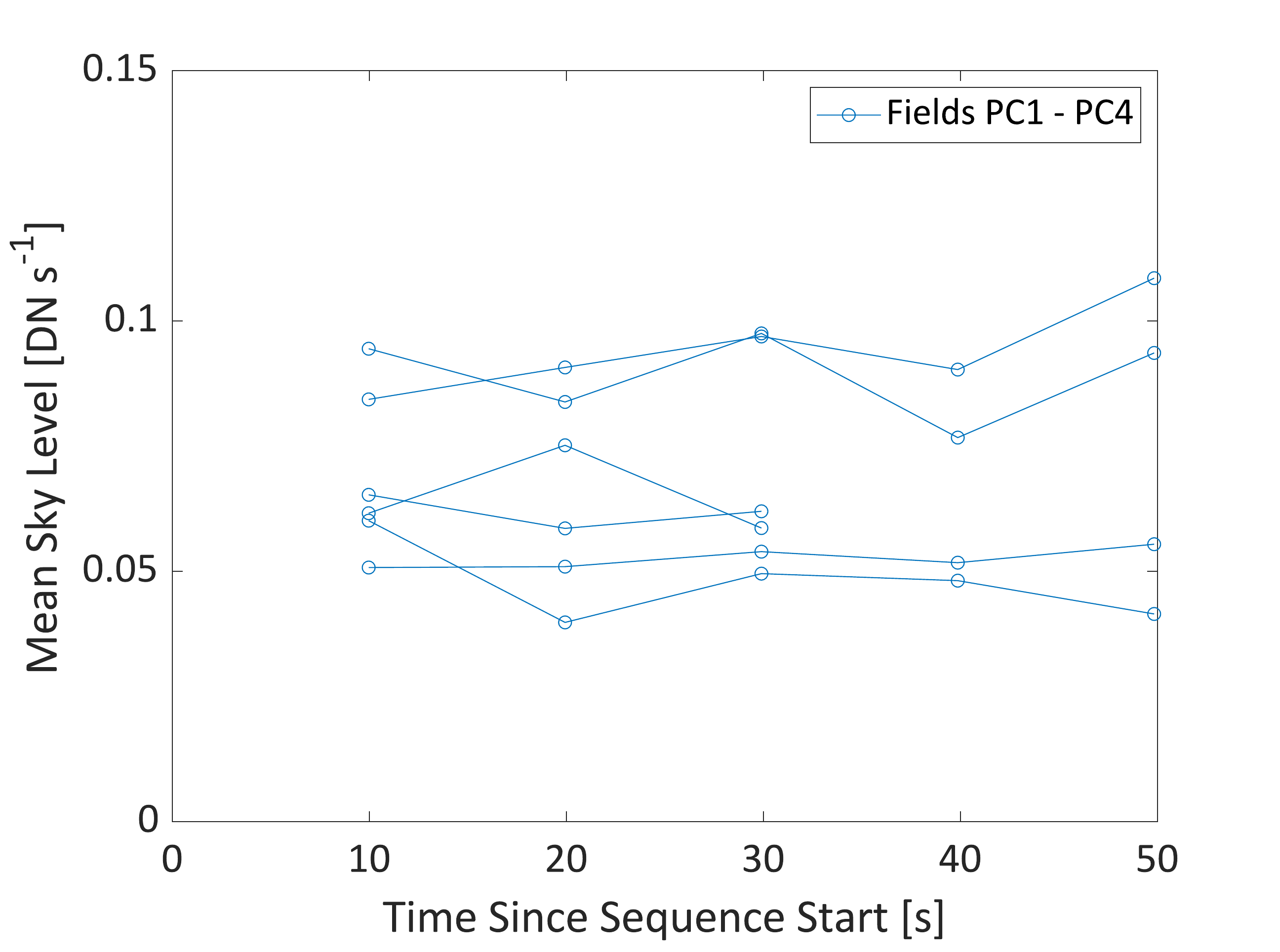} }}%
    \subfloat{{\includegraphics[width=8cm]{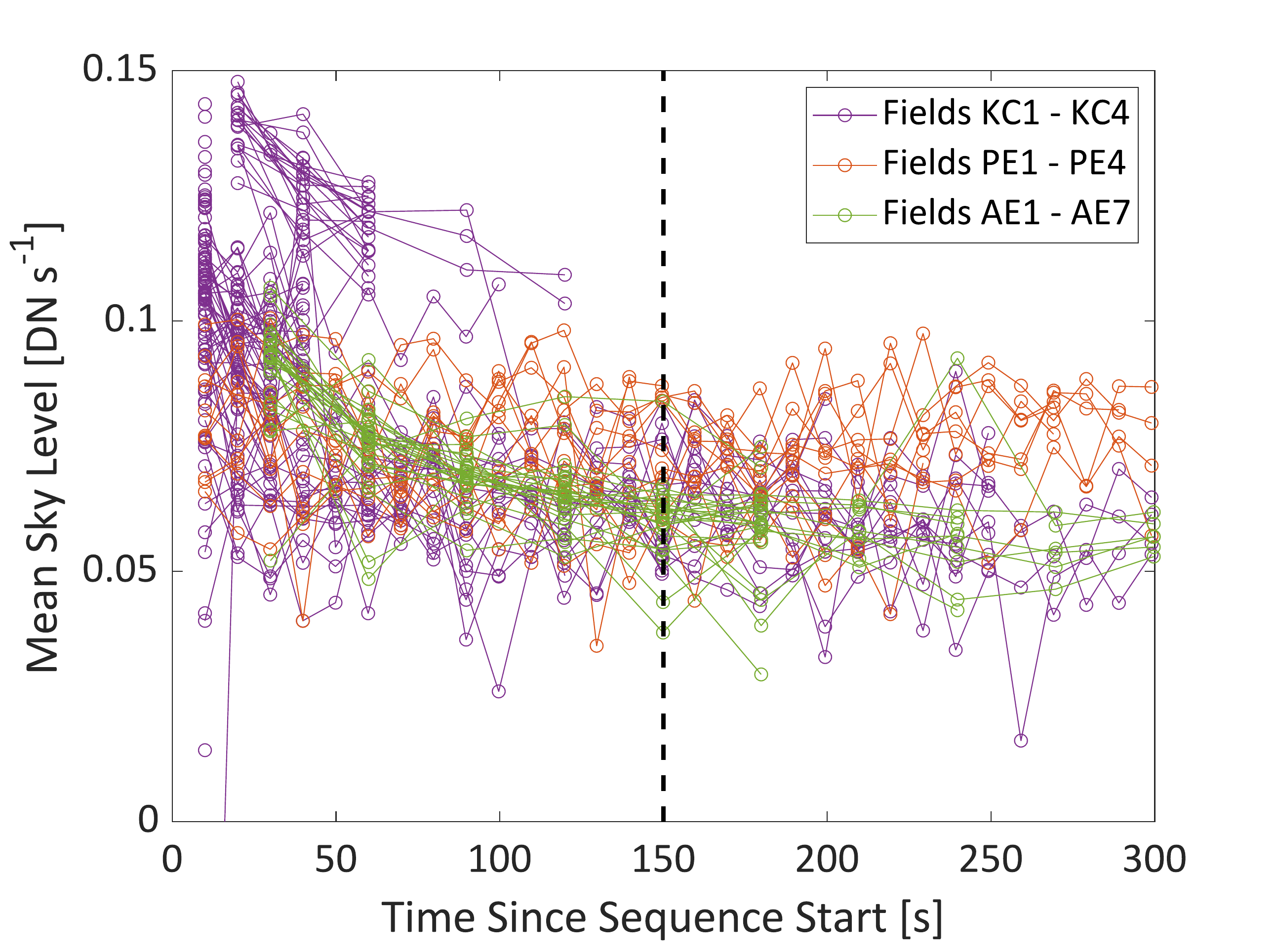} }}%
    \caption{Left: A comparison of mean sky level per observation sequence for fields PC1 -- PC4. Each sequence is shown as a separate line. No drop-off in mean sky level is detected for any sequence in these fields. Right: The same comparison for fields KC1 -- KC4 (purple), fields PE1 -- PE4 (orange), and fields AE1 -- AE7 (green). Here, a noticeable decay in the absolute brightness of the image is seen up to 150 seconds (dashed line) of data per sequence. We choose to exclude data taken before 150 seconds of observing time has elapsed. This also effectively excludes the population of data clustered around 0.13 DN s$^{-1}$, which is anomalous compared to the rest of the set. }%
    \label{fig:sequence}%
\end{figure*}

\citet{Lauer,lauer_2022} investigate an effect where exposures taken after the LORRI camera is first powered on exhibit significantly higher background sky levels that drop off over a period of 150 seconds after camera activation. This effect is likely an electrical or thermal transient that corrupts reads following a power cycle of the detector, and the cause is unknown. Previous analyses exclude the first 150 seconds of data taken after camera power-on as anomalous. We explored this issue for all data remaining after the previously described cuts by calculating the mean sky level in DN s$^{-1}$ of our masked exposures (masking procedures to be described in Section \ref{sec:mask}). LORRI data are divided into observation sequences of multiple exposures of the same target. We compared the mean brightness for all exposures from the same sequence for up to 400 seconds of data, where each observing sequence is assumed to begin with camera power-on. This is not necessarily true of all sequences, but serves as a proxy to analyze this effect.
Our comparison of image brightness after observing sequence start for all sequences in our data set is shown in Figure \ref{fig:sequence}. The PC fields contain at most 50 seconds of data, and do not display any noticeable drop-off in mean sky level. Therefore, we elect not to exclude any part of this data set beyond the cuts that have already been made. The KC and AE fields all demonstrate a drop-off through 150 seconds of data, so we choose to exclude the first 150 seconds from each of these sets, resulting in a reduction of 4\% of the complete data set. We investigate the systematic error associated with this choice in Section \ref{sec:camcut_results}.

\vspace{-10pt}
\subsection{Data Used in this Analysis}
\label{sec:final}

The data surviving these cuts form the set used for scientific analysis, as summarized in Table \ref{tab:fields}. Our pipeline has been designed for analysis against a training data set, and the final analysis is performed blind on the combination of the training set and a large data set we call the science set. Here, we describe these data sets, as well as the ancillary data sets used in developing our analysis procedures but not used to constrain the COB directly. 

The training data set is comprised of science-quality fields, mostly acquired earlier in time and thus closer to the Sun, which are used to develop our data analysis pipeline and associated procedures. This set of 303 exposures comes from exposures on four distinct fields and comprises almost an hour of integration time; we denote these PE1 -- PE4. 

\begin{figure*}[htb!]
    \centering
    \subfloat{{\includegraphics[width=8.5cm]{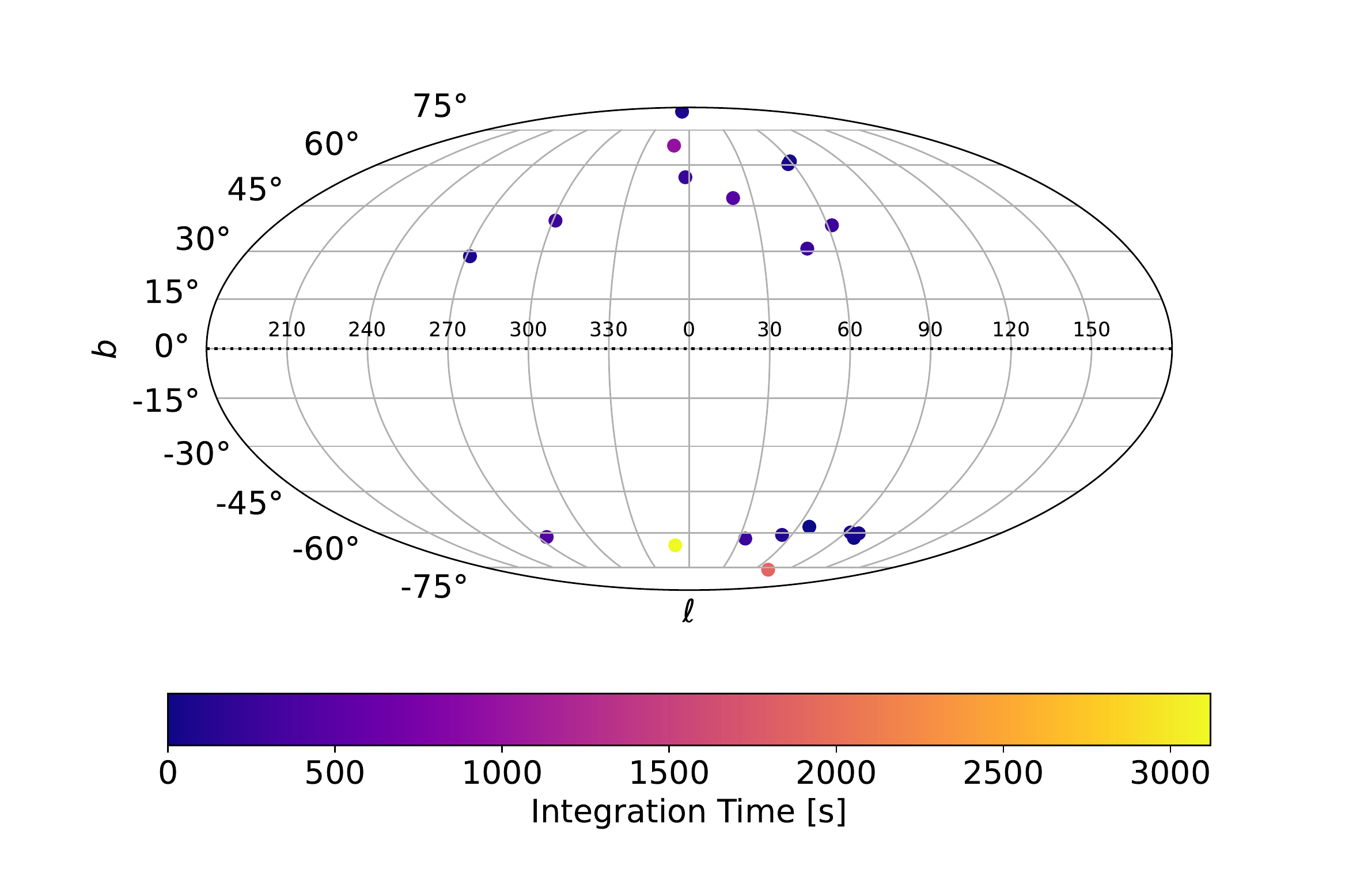} }}%
    \subfloat{{\includegraphics[width=8.5cm]{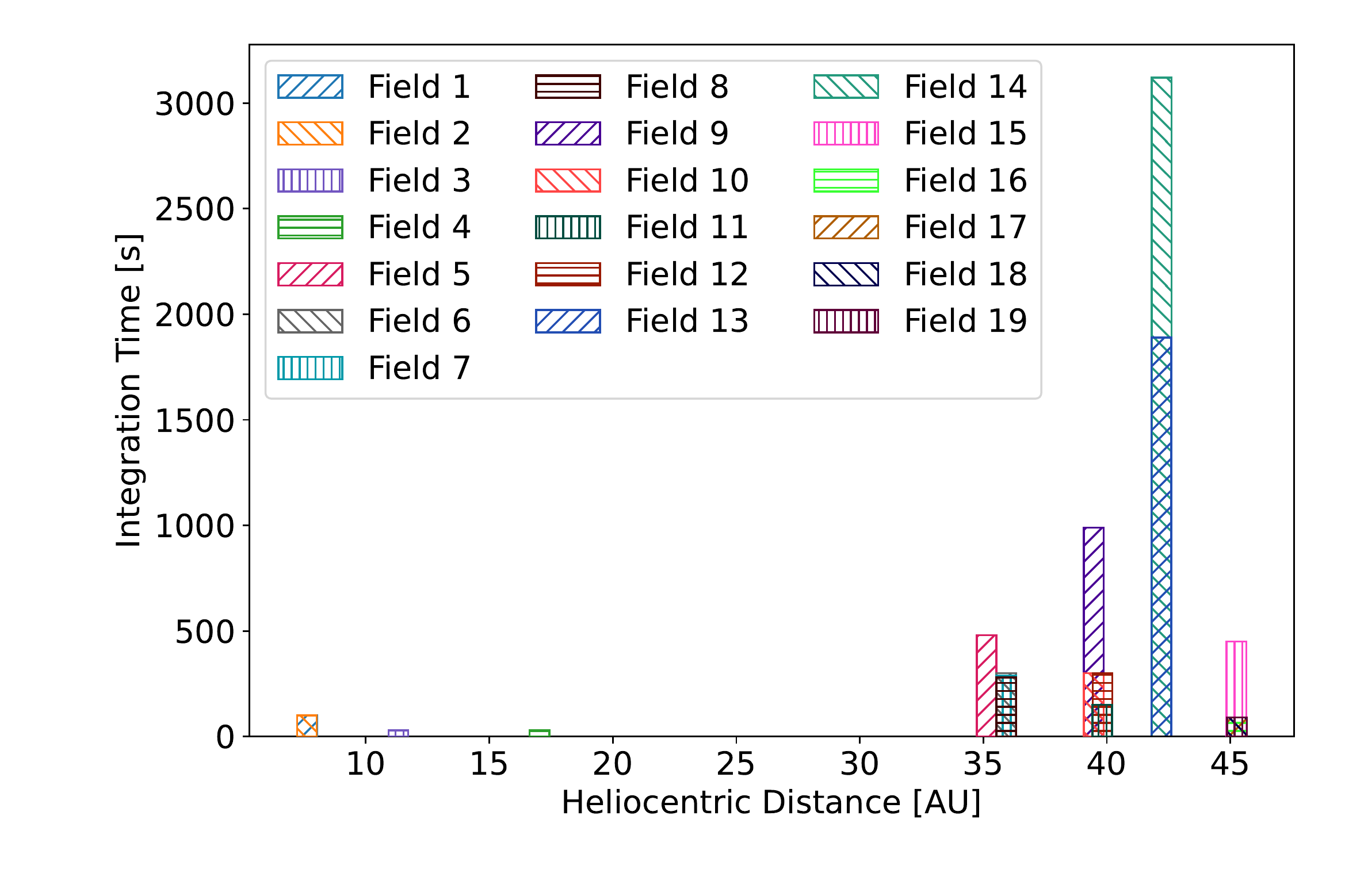} }}%
    \caption{Left: Galactic coordinates of science fields color-coded by total integration time per field. Right: Heliocentric distance of each science field. The height of each bar indicates the total integration time per field. }%
    \label{fig:field_loc}%
\end{figure*}

\begin{figure*}[htb!]
\centering
\noindent\includegraphics[width=\textwidth]{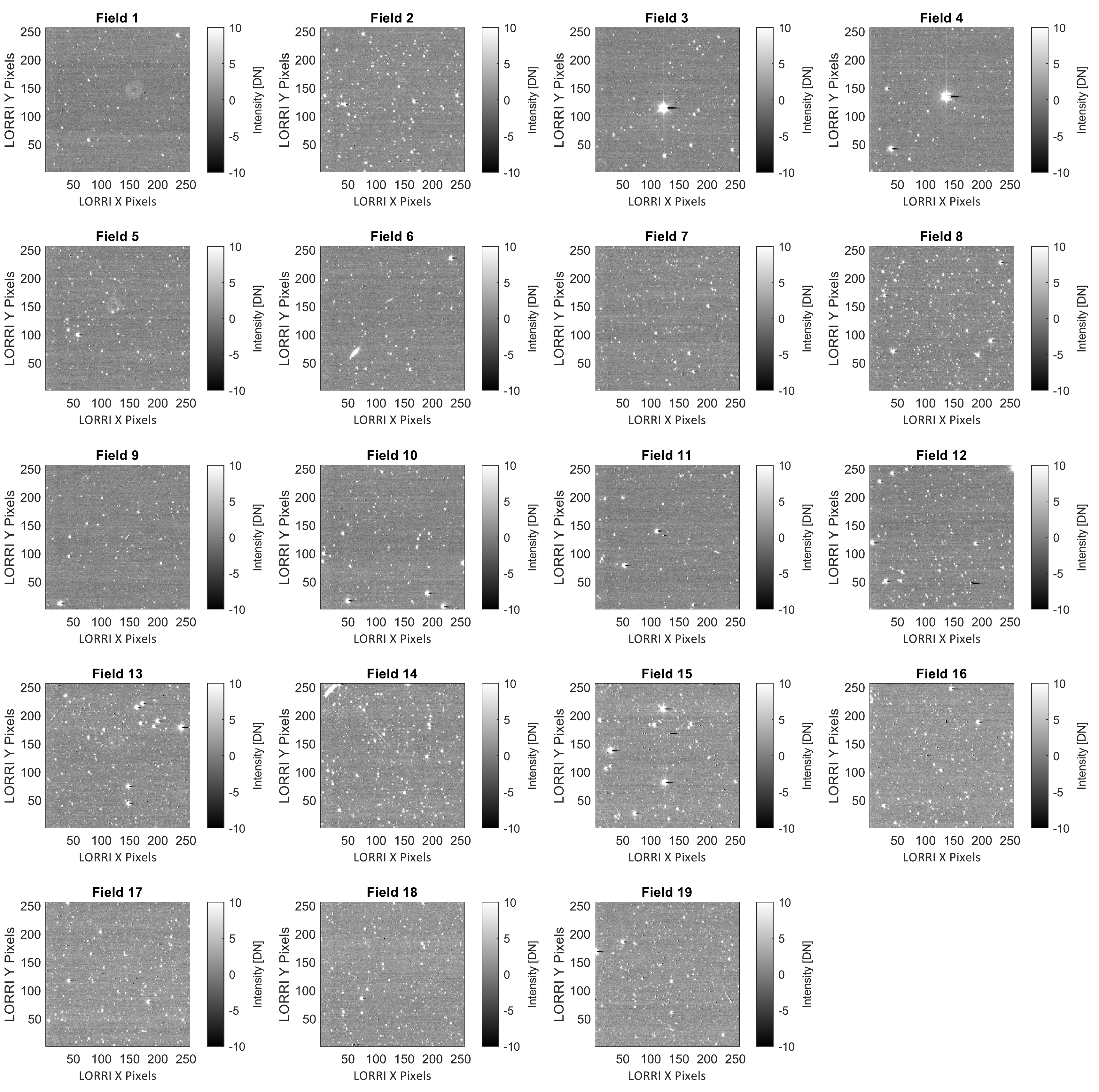}
\caption{All science fields input to our pipeline in raw units. For each of the 19 science fields contributing to our measurement of the COB, we show one example exposure in DN. The field numbers match those assigned in Table \ref{tab:fields}; Fields 5 through 8 comprise the training data set we use before unblinding the analysis. An optical ghost is faintly visible in Field 1, and Neptune is visible as the bright source in Fields 3 and 4. Field 6 has two bright foreground galaxies that will also be masked.}\label{fig:allfields_raw}
\end{figure*}

Our final list of 19 science fields is selected from the full set of available data, and is summarized in Table \ref{tab:fields}. This set includes 11 fields previously analyzed by \cite{nature} and \cite{Lauer}, which we re-analyze, as well as eight new fields not previously analyzed (PE1 -- PE4 and KC1 -- KC4). This full set represents 9,170 seconds (2.5 hours) of total integration time and includes observations spanning 12 years in time over a heliocentric distance of 8 -- 45 AU. Figure \ref{fig:field_loc} shows the galactic positions scattered near the galactic poles and the heliocentric distance of each field by total integration time and Figure \ref{fig:allfields_raw} shows a single raw example exposure of each of the 19 fields. 

A set of fields used solely in the development of the analysis methods is the ghost training set. This set includes fields with exposures that contain visible optical ghosts. The exposures in this set were specially selected to characterize LORRI's optical ghosting and develop ways to mitigate its contribution to the background. The set contains 125 exposures from four different fields, including fields PC1, PE1, and PE4 from the science field set. These fields are summarized in Table \ref{tab:ghost_fields}. Field Ghost 1 is the only field that does not also appear in the science set. Only a subset of exposures from fields PE1 and PE4 were used in the ghost training set as those were the only exposures with visible ghosts. 

\begin{deluxetable*}{cccccccccc}[htb!]
\tabletypesize{\normalsize}
\tablecaption{Description of 19 LORRI fields, comprising 529 images, used to measure the COB in this analysis. Fields PC1 -- PC4 were analyzed as part of \citet{nature}, fields AE1 -- AE7 were analyzed as part of \citet{Lauer}, and fields PE1 -- PE4 and KC1 -- KC4 have not yet appeared in publications.\label{tab:fields}}
\tablehead{
\colhead{\textbf{Field}} &
  \colhead{\textbf{Field}} &
  \colhead{\textbf{\boldmath$\alpha$ (J2000)}} &
  \colhead{\textbf{\boldmath$\delta$ (J2000)}} &
  \colhead{\textbf{\boldmath$\ell$}} &
  \colhead{\textbf{\boldmath$b$}} &
  \colhead{\textbf{\boldmath$N_{\rm exps}$}} &
  \colhead{\textbf{Exp.~per}} &
  \colhead{\textbf{Nominal}} &
  \colhead{\textbf{Obs.}} \\
\colhead{\textbf{Number}} &
  \colhead{\textbf{Name}} &
  \colhead{\textbf{hh:mm:ss}} &
  \colhead{\textbf{dd:mm:ss}} &
  \colhead{\textbf{(\boldmath$\degree$)}} &
  \colhead{\textbf{(\boldmath$\degree$)}} &
  \colhead{\textbf{}} &
  \colhead{\textbf{Image}} &
  \colhead{\textbf{Target}} &
  \colhead{\textbf{Date}}}
  \startdata
1  & PC1  & 13:04:03.83 & 23:56:56.04  & 345.41 & 85.74  & 10  & 10s     & Haumea     & 10/06/07  \\
2  & PC2  & 10:47:37.50 & -26:47:02.14 & 271.45 & 28.41  & 10  & 10s     & Chariklo   & 10/06/07  \\
3  & PC3  & 23:04:26.69 & -07:07:11.33 & 66.27  & -57.69 & 3   & 10s     & Neptune    & 10/16/08 \\
4  & PC4  & 00:07:12.40 & -01:15:04.85 & 98.81  & -62.03 & 3   & 10s     & Neptune    & 06/23/10  \\\hline
5  & PE1  & 15:40:44.90 & 12:15:59.01  & 20.89  & 47.72  & 28  & 10s     & 1994 JR1   & 04/07/16   \\
6  & PE2  & 14:43:10.25 & 04:47:32.43  & 357.91 & 55.25  & 30  & 10s     & Quaoar     & 07/13/16  \\
7  & PE3  & 12:45:23.39 & -22:49:46.60 & 301.11 & 40.02  & 29  & 10s     & Ixion      & 07/13/16  \\
8  & PE4  & 17:19:09.79 & 25:54:03.80  & 48.34  & 30.86  & 48  & 10s     & MS4        & 07/13/16  \\
9  & KC1  & 13:56:06.49 & 11:03:36.23 & 349.46  & 67.87 & 99  & 10s      & 2014 OE394 & 09/21/17  \\
10 & KC2  & 22:49:45.65 & -23:25:56.76 & 33.78  & -62.32 & 30  & 10s     & 2011 HJ103 & 09/21/17  \\ 
11 & KC3  & 23:00:01.78 & -13:53:31.14 & 54.33  & -60.76 & 15  & 10s     & 2011 HJ103 & 10/31/17 \\ 
12 & KC4  & 16:55:14.76 & 38:23:01.04  & 61.88  & 38.45  & 30  & 10s     & MS4        & 11/01/17  \\\hline
13 & AE1   & 00:07:06.96 & -17:46:40.80 & 73.08  & -76.15 & 63  & 30s     & 2014 OE394 & 08/20/18  \\
14 & AE2   & 23:12:14.66 & -41:38:09.60 & 350.96 & -65.06 & 104 & 30s     & 2014 OJ394 & 08/22/18  \\
15 & AE3   & 02:13:37.66 & -50:45:10.44 & 275.02 & -61.69 & 15  & 30s     & n3c61f     & 09/01/18   \\
16 & AE4   & 23:52:58.27 & -00:31:05.88 & 92.71  & -59.91 & 3   & 30s     & ZL         & 09/04/19   \\
17 & AE5   & 00:03:13.58 & 00:17:29.40  & 98.06  & -60.23 & 3   & 30s     & ZL         & 09/04/19   \\
18 & AE6   & 14:59:57.00 & 36:13:59.16  & 59.51  & 61.34  & 3   & 30s     & ZL         & 09/04/19   \\
19 & AE7   & 15:05:56.76 & 35:17:52.44  & 57.26  & 60.26  & 3   & 30s     & ZL         & 09/04/19   \\ \hline
\enddata
\end{deluxetable*}


\begin{deluxetable*}{ccccccccc}[htb!]
\tablecaption{Description of fields that make up the ghost training set used to characterize optical ghosting for masking and subtraction of diffuse ghost intensity. Field Ghost 1 does not appear in the science data set.\label{tab:ghost_fields}}
\tablehead{
  \colhead{\textbf{Field}} &
  \colhead{\textbf{\boldmath$\alpha$ (J2000)}} &
  \colhead{\textbf{\boldmath$\delta$ (J2000)}} &
  \colhead{\textbf{\boldmath$\ell$}} &
  \colhead{\textbf{\boldmath$b$}} &
  \colhead{\textbf{\boldmath$N_{\rm exps}$}} &
  \colhead{\textbf{Exp.~per}} &
  \colhead{\textbf{Nominal}} &
  \colhead{\textbf{Obs.}} \\
  \colhead{\textbf{Name}} &
  \colhead{\textbf{hh:mm:ss}} &
  \colhead{\textbf{dd:mm:ss}} &
  \colhead{\textbf{(\boldmath$\degree$)}} &
  \colhead{\textbf{(\boldmath$\degree$)}} &
  \colhead{\textbf{}} &
  \colhead{\textbf{Image}} &
  \colhead{\textbf{Target}} &
  \colhead{\textbf{Date}}}
  \startdata
Ghost 1   & 13:04:04.80 &  23:57:00.00 & 345.52 & 85.73  & 6   & 10s     & Callirrhoe   & 01/10/07  \\
PC1  & 13:04:03.83 & 23:56:56.04  & 345.41 & 85.74  & 10  & 10s     & Haumea     & 10/06/07  \\
PE1  & 15:40:44.90 & 12:15:59.01  & 20.89  & 47.72  & 79  & 10s     & 1994 JR1   & 04/07/16   \\
PE4  & 17:19:09.79 & 25:54:03.80  & 48.34  & 30.86  & 30  & 10s     & MS4        & 07/13/16  \\\hline
\enddata
\end{deluxetable*}

\vspace{-40pt}
\section{Data Processing and Calibration}
\label{sec:data_analysis}

\begin{figure*}[htb!]
\centering
\noindent\includegraphics[width=0.9\textwidth]{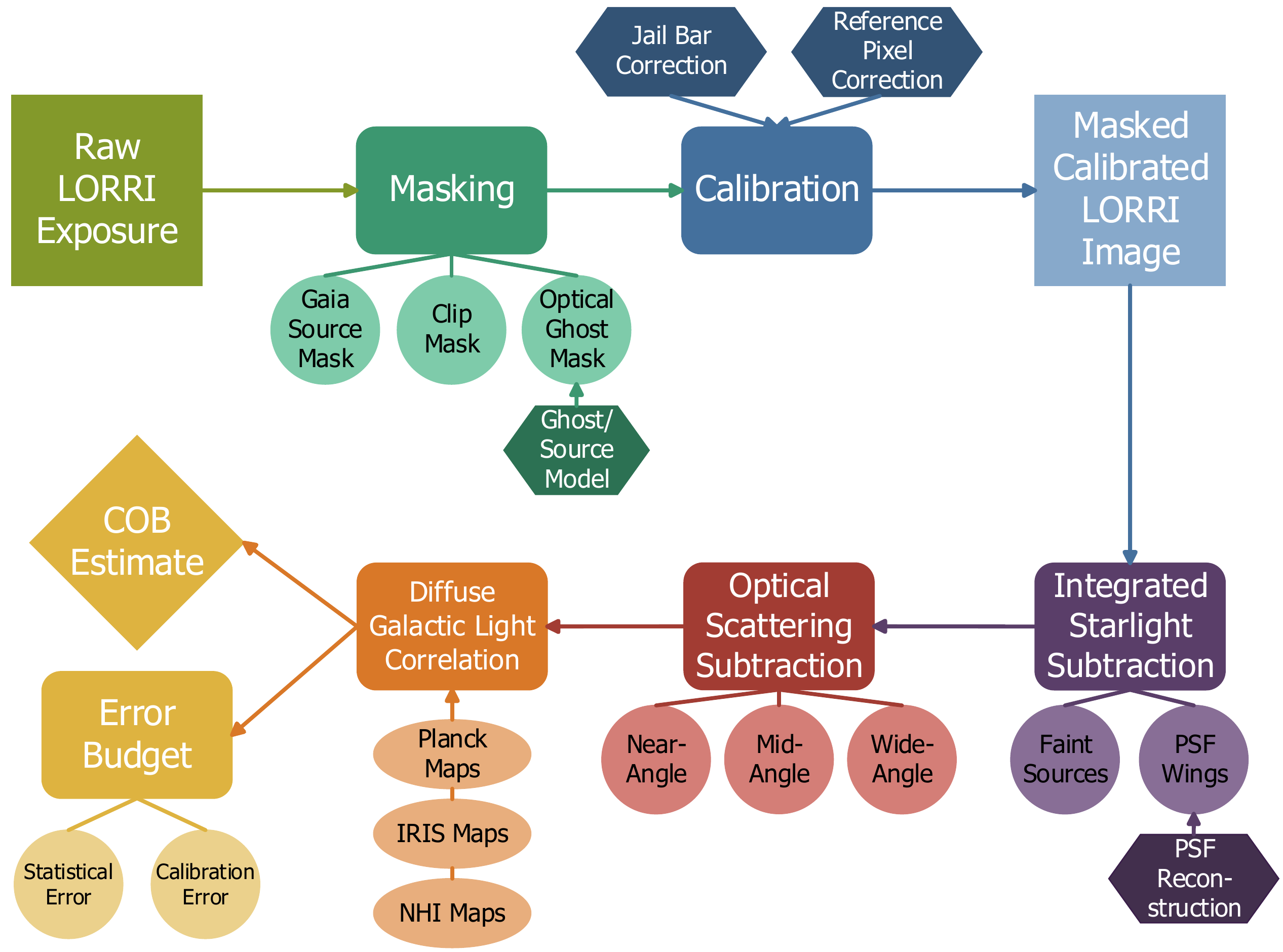}
\caption{Flowchart illustrating the modules and sequence of the data analysis pipeline, starting from pre-processed LORRI exposures through final COB and error budget estimation. Intermediate steps include characterization and subtraction of astrophysical foreground components. The data processing (upper family of boxes) is discussed in this Section, the foreground compensation that leads to the COB estimate (lower family of boxes) is discussed in Section \ref{sec:foregrounds}, and our development of the overall the error budget is discussed in Section \ref{sec:error_analysis}. }\label{fig:flowchart}
\end{figure*}

Our pipeline is trained against a subset of the data and then deployed against the full set listed in Table \ref{tab:fields}. First, we develop masks for foreground sources including bright stars that can be masked via catalog information. Next, we correct the data for a few subtle effects that can greatly affect the final data values after calibration. The first is a detector defect causing an offset in alternating columns in a ``jail bar'' pattern, and the second is an adjustment to the LORRI pre-processing pipeline's method of compensating for dark current. Following these corrections, we calibrate the images to astronomical intensity units. 
We assess the astrophysical foregrounds that can be directly subtracted from each calibrated image in the next Section. 
The overall flow of the pipeline, including our assessment of the astrophysical foregrounds discussed in Section \ref{sec:foregrounds}, is illustrated in Figure \ref{fig:flowchart}. 

\subsection{Masking}
\label{sec:mask}
\begin{figure*}[htb!]
    \centering
    \subfloat{{\includegraphics[width=7.2cm]{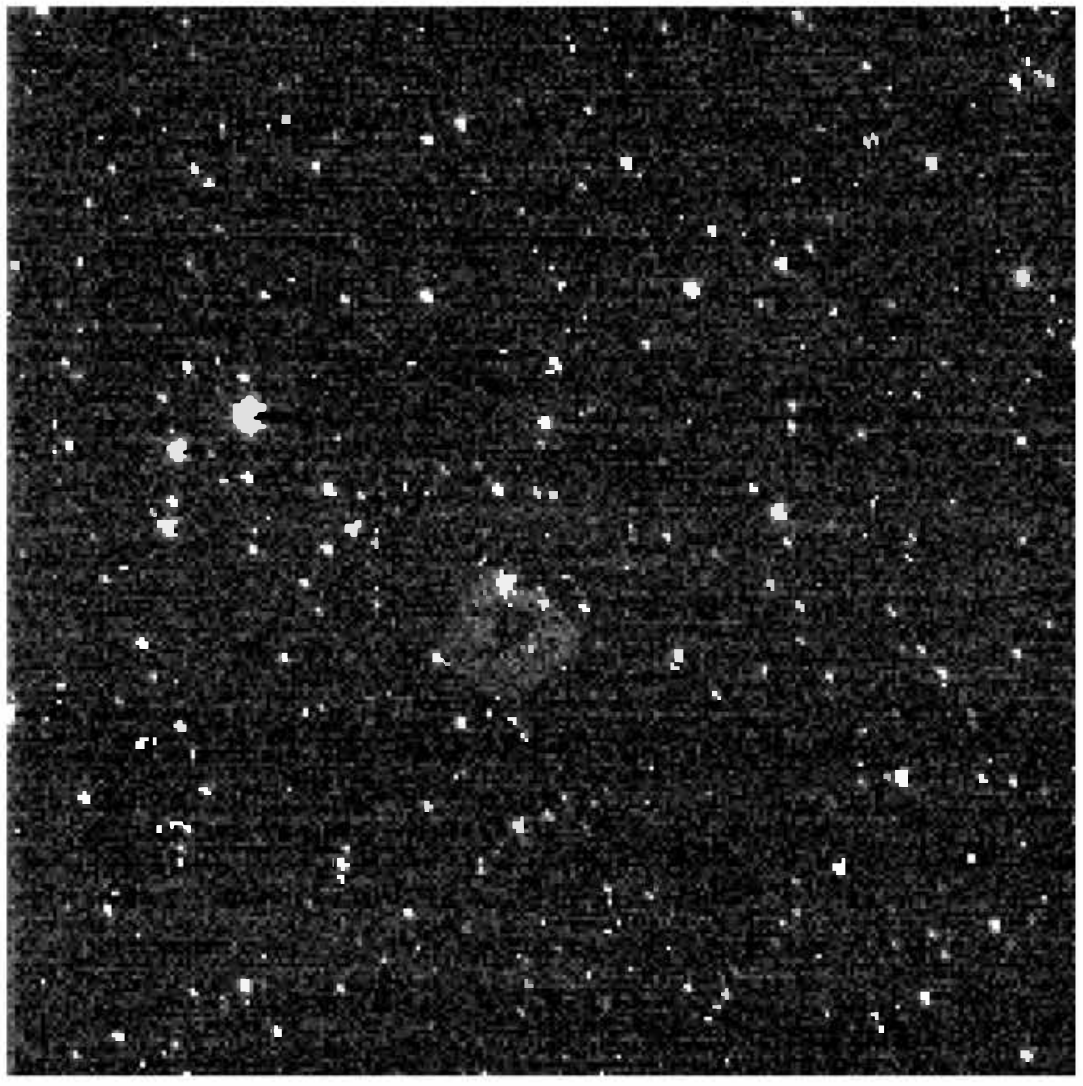} }}%
    \qquad
    \subfloat{{\includegraphics[width=7.2cm]{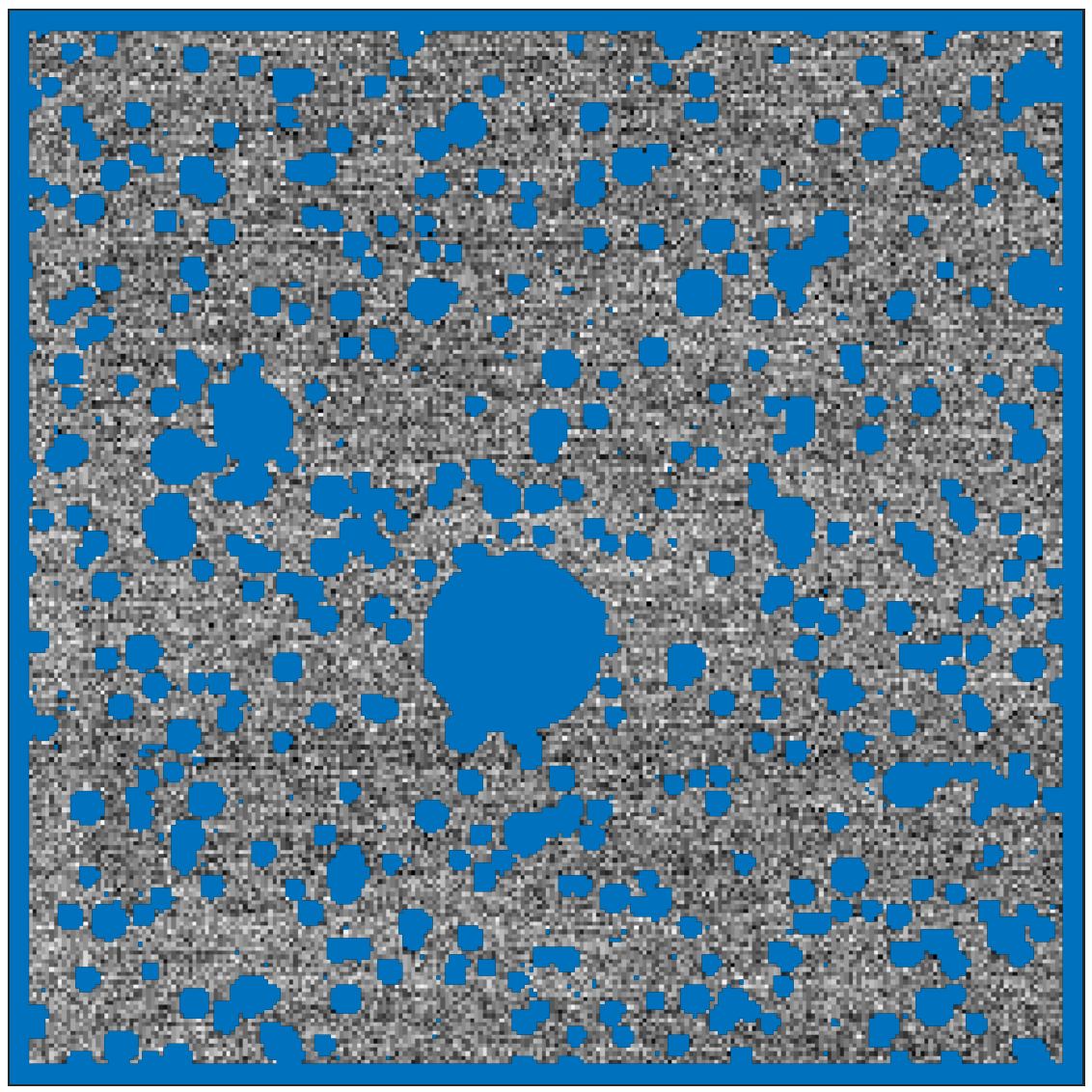} }}%
    \caption{Left: An example unmasked LORRI exposure after pre-processing. Right: The same exposure after all masks have been applied. The large, circular mask near the center of the exposure masks an optical ghost caused by an off-axis bright star, faintly visible in the unmasked exposure.}%
    \label{fig:mask}%
\end{figure*}

The first pipeline task is to perform various types of masking wherein a map of pixels designated for exclusion from the analysis is developed. The most prominent foreground component of any exposure is resolved stars, which are masked via catalog reference. The mask that removes optical ghosting due to bright sources just off-field is then calculated. Masking of charge-transfer artifacts caused by detector readout of over-saturated stars is then applied. Next, manual masking of detector defects and resolved or solar system sources is applied. Lastly, other hot pixels that remain unmasked by the previous procedures are masked using clip masking. An example exposure before and after all masks are applied is demonstrated in Figure \ref{fig:mask}. 

\subsubsection{Star Masking}
\label{sec:starmask}
To accurately subtract the contribution from bright stars, $\lambda I_{\lambda}^{*}$ in Eq.\ref{eq:cob}, we have developed a procedure for masking bright sources using the \gaia\ Data Release 2 (DR2) catalog \citep{gaia_mission,gaia}. From \gaia\ DR2, we return all sources that fall within a given exposure based on astrometric registration. We calculate the color correction between the two bandpasses, $\Delta m$, using the ratio of the integrated, scaled bandpasses. 

\begin{figure*}[htb!]
\centering
\noindent\includegraphics[width=4in]{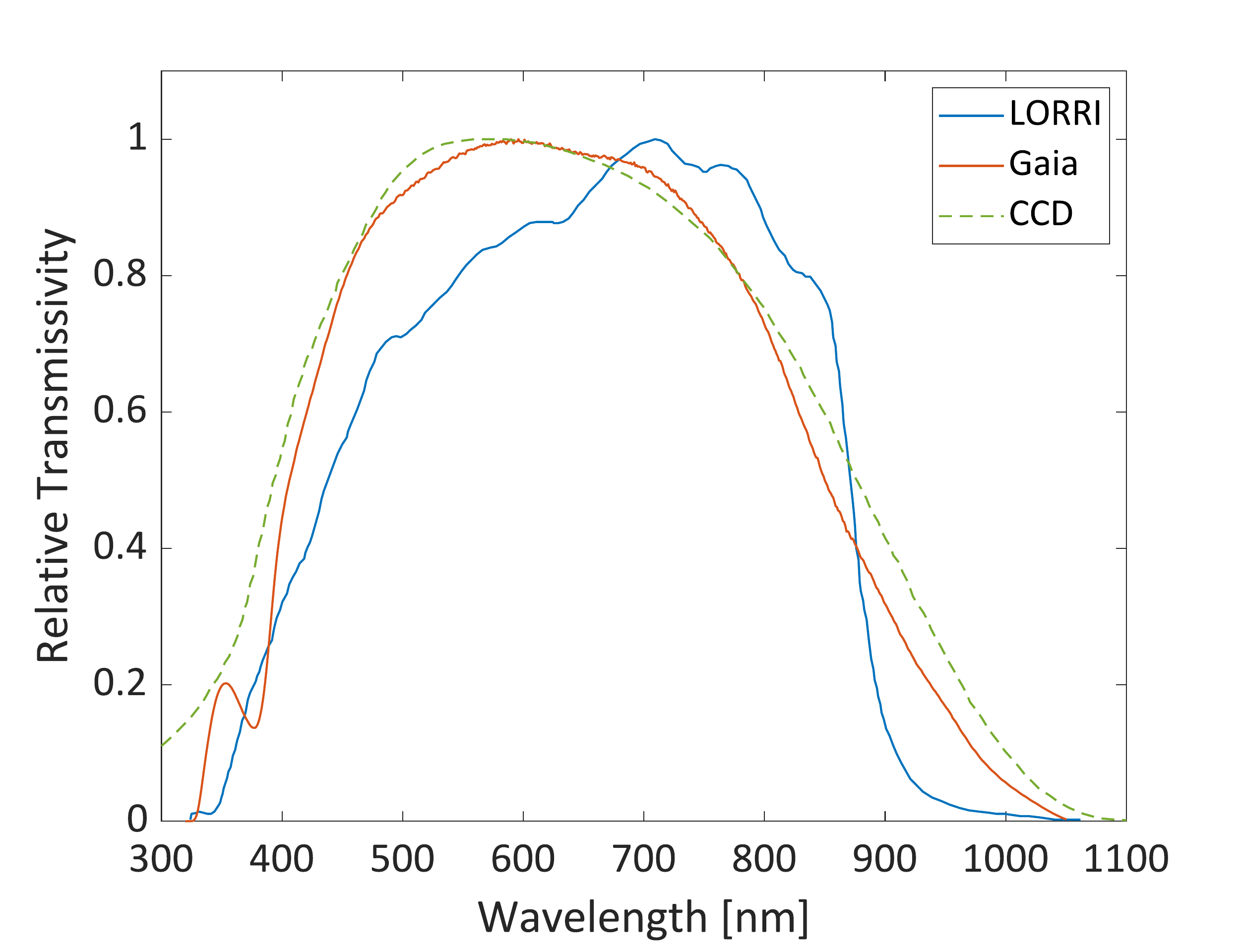}
\caption{Comparison of LORRI and \gaia\ bandpasses. The \gaia\ bandpass, shown in orange, extends from 330 to 1050 nm \citep{gaia_passband}. The LORRI bandpass, in blue, has a range of 360 to 910 nm \citep{weaver2020}. Though they have slightly different spectral sensitivity, to approximately flat-spectrum sources like DGL and the COB \gaia\ $G$ magnitudes are very similar to LORRI magnitudes \citep{Symons2022}. The LORRI CCD's intrinsic response is shown as the green dashed line; modulo the free normalization, the difference between this and the blue line is the transmissivity of the LORRI optics.}\label{fig:bandpass}
\end{figure*}

This gives $\Delta m$ = -0.0323. Because the bandpasses of LORRI and the \gaia\ $G$-band are almost identical (Figure \ref{fig:bandpass}), we are able to use \gaia\ magnitudes directly in our masking algorithm. Using these magnitudes, we mask to a radius in the image that is weighted by the magnitude of each source,
\begin{equation}
\label{eq:mask}
r = 2.5\Big(\frac{m_{\mathrm{max}}}{m}\Big)^{2} ,
\end{equation}
where $m_{\mathrm{max}}$ is the faintest magnitude that can be reliably masked, $m$ is the magnitude of each source, and $r$ is the mask radius in pixels. 
To determine $m_{\mathrm{max}}$, we compare the surface brightness from sources in the \gaia\ DR2 catalog to the expected total surface brightness in each magnitude bin for 10 TRILEGAL simulations \citep{trilegal} per LORRI field. 
We find that \gaia\ matches the TRILEGAL expectation of the total ISL in our fields to $m_{G}$ $\simmod$21, so set $m_{\mathrm{max}} = 21$ and mask all sources down to this magnitude.

\gaia\ DR2 does not differentiate between stars and galaxies, so we use a catalog developed by \cite{gaia_gal} that identifies galaxies in \gaia\ DR2 in order to prevent masking galaxies that contribute to the COB signal. We use this second catalog to remove sources identified as possible galaxies from the masking process by matching potential galaxies in both catalogs using their DR2 identifiers and excluding them from the star mask. We explore the uncertainty from this catalog's purity in Section \ref{sec:error_analysis}. 

\vspace{10pt}
\subsubsection{Static and Manual Masking}
\label{sec:statmanmask}

Next, we mask out of every exposure those pixels that are obviously problematic to future processing steps. This static mask includes the outermost five pixel ``rind'' of each exposure. At this stage we also mask solar system objects, such as planets, via their coordinates at the time of observation and expected intensity. This typically removes pixels near the center of the frame, as many of our science observations targeted solar system objects of various types (see Table \ref{tab:fields}). Finally, we manually mask two resolved foreground galaxies in field PE2. Although galaxies source the COB, the local and bright galaxies that appear resolved in a LORRI exposure do not contribute to the diffuse background of such an exposure and would bias our measurement. 

\subsubsection{Optical Ghost Masking}
\label{sec:ghostmask}
\begin{figure*}[htb!]
    \centering
    \subfloat{{\includegraphics[width=7.4cm]{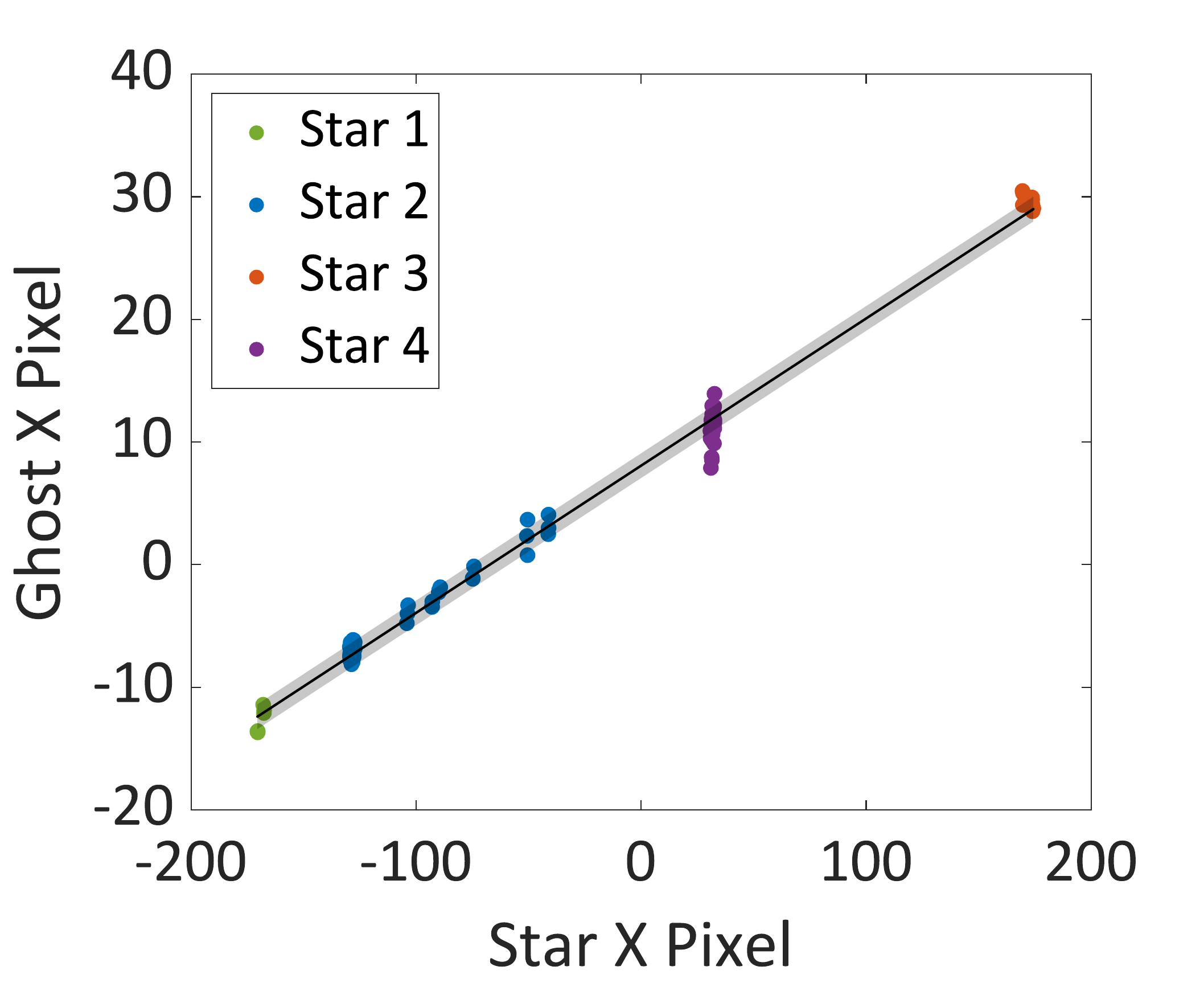} }}%
    \qquad
    \subfloat{{\includegraphics[width=7.4cm]{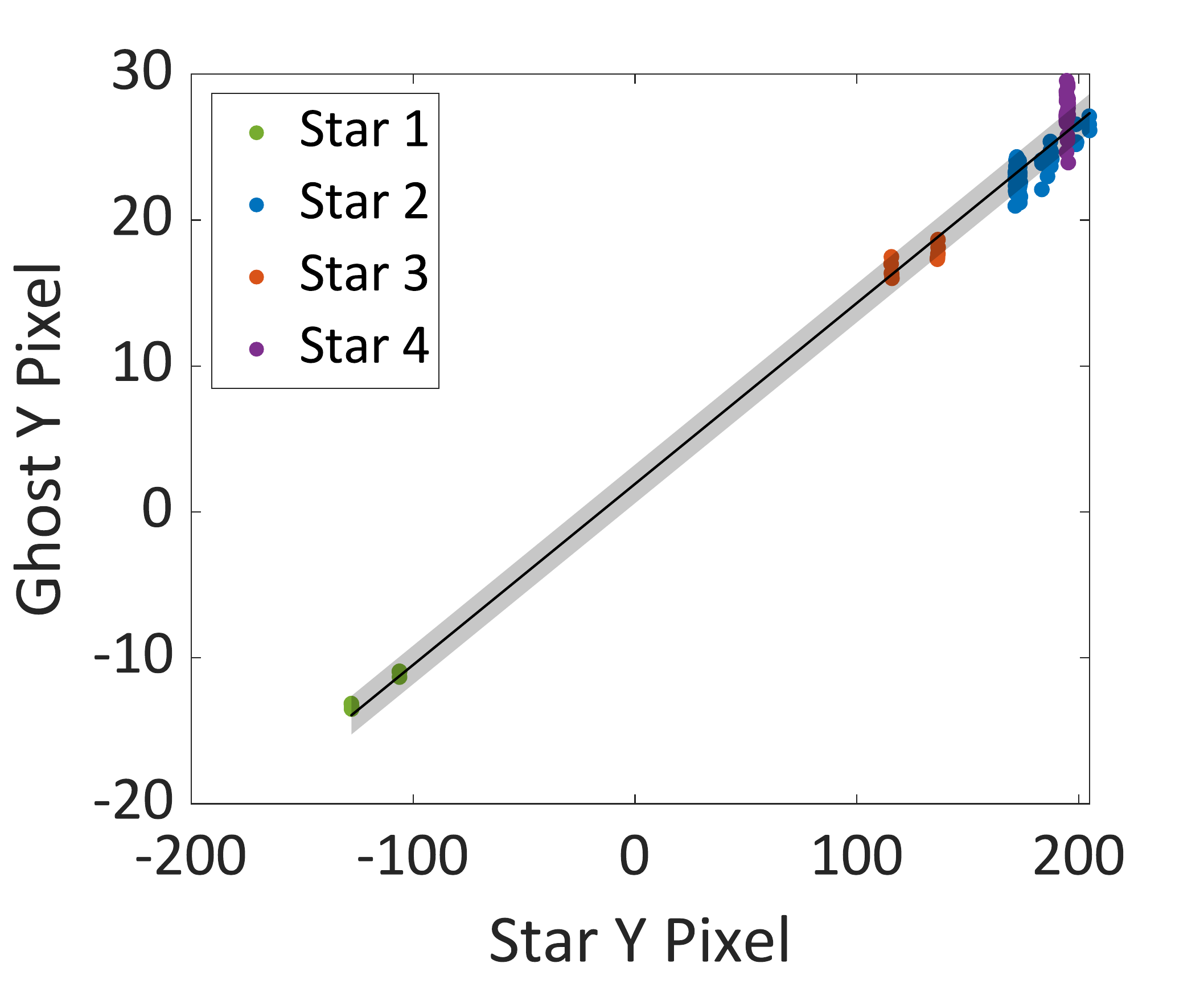} }}%
    \caption{From the training set of ghosts, coordinates were recorded for each ghost and the star causing the ghost. The black lines give linear fits between the ghost and star pixel coordinates in $x$ and $y$, which are successfully used to predict the locations of ghosts for masking. The grey shaded regions give the RMS error on the fits.}%
    \label{fig:coords}%
\end{figure*}

LORRI has known optical ghosting caused by direct illumination of the camera lenses by sources that are up to $0^{\circ}.37$ from the center of the FOV \citep{cheng_2008,Cheng2010}. Using the \gaia\ DR2 catalog, we were able to identify potential bright stars in this region as the source of each ghost. Successive LORRI exposures often display slight pointing shifts that allow us to track the location of candidate stars and ghosts over time. This allowed us to develop a geometric model relating the location of a star and the ghost it causes, illustrated in Figure \ref{fig:coords}. Details about the model construction can be found in \citet{Symons2022}.

We use this model to predict the location of a ghost when an exposure has a $m_{G} <$ 8 star within $0^{\circ}.37$ of the FOV center and automatically mask a radius of 21.5 pixels, which was the maximum radius necessary to mask all ghosts in the training set.

\subsubsection{Line Masking}
\label{sec:linemask}
The brightest stars in an exposure saturate the detector response and can leave charge-transfer artifacts when the detector is read out. These artifacts typically appear as extremely negative pixel values in the read direction following a bright source. We automatically mask the row in which the center of a star is located from the central pixel of the star to the right-hand edge of the exposure for any star with $m$ $<$ 13. This limit was empirically determined based on visual observations of charge-transfer artifacts.

\subsubsection{Clip Masking}
\label{sec:clipmask}
The final mask applied is clip masking, in which any pixels with values greater than n-$\sigma$ from the mean of the unmasked pixels are masked, which excludes pixels suffering from transient effects like cosmic rays. We tested multiple $\sigma$-levels for our entire testing set of exposures to arrive at the choice of 3$\sigma$, which we apply in several iterations. 

\subsection{Jail Bar Correction}
\label{sec:jailbar}
\begin{figure*}[htb!]
\centering
    \subfloat[][Before correction]{{\includegraphics[width=5.15cm]{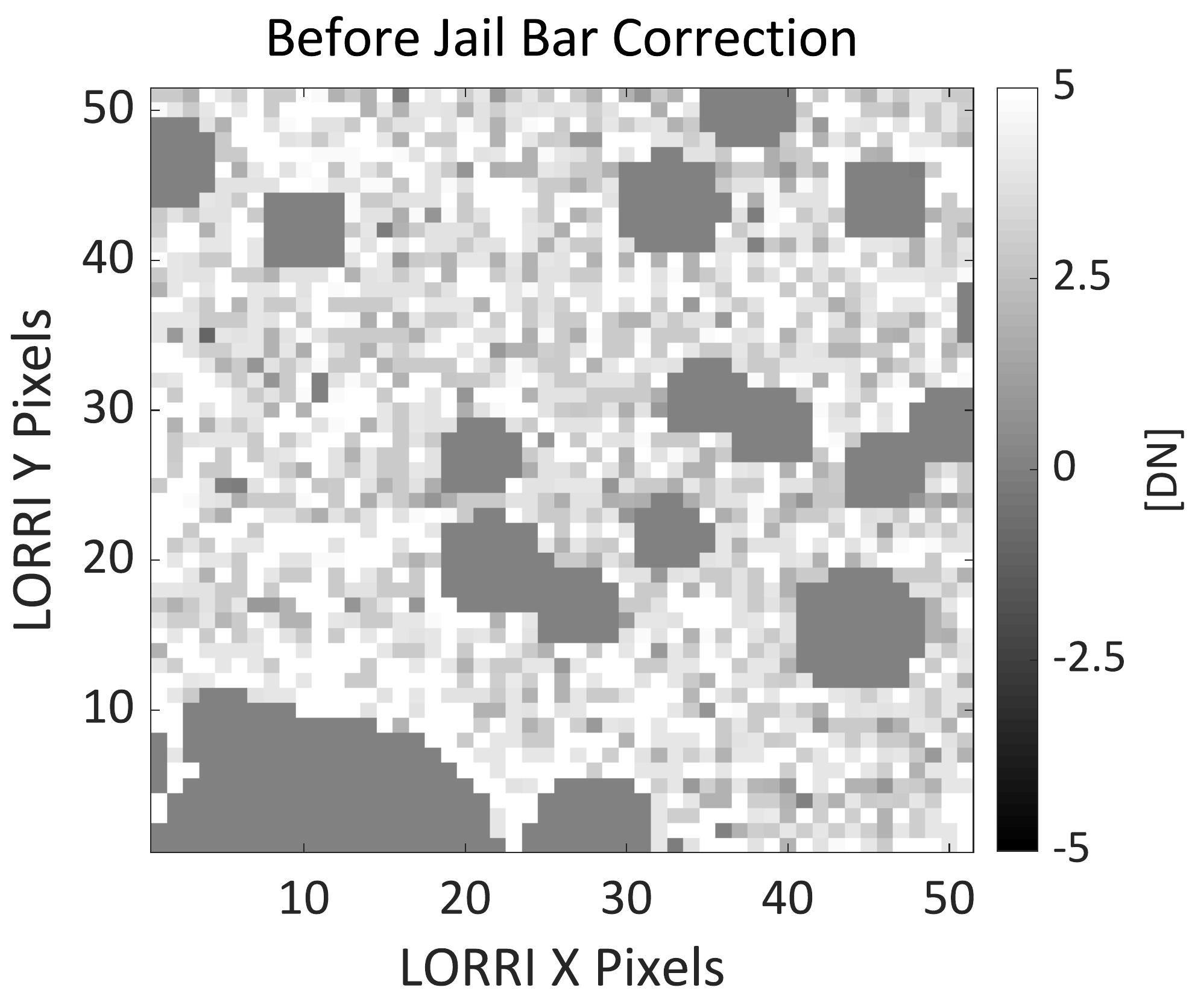} }}%
    \subfloat[][After correction]{{\includegraphics[width=5.15cm]{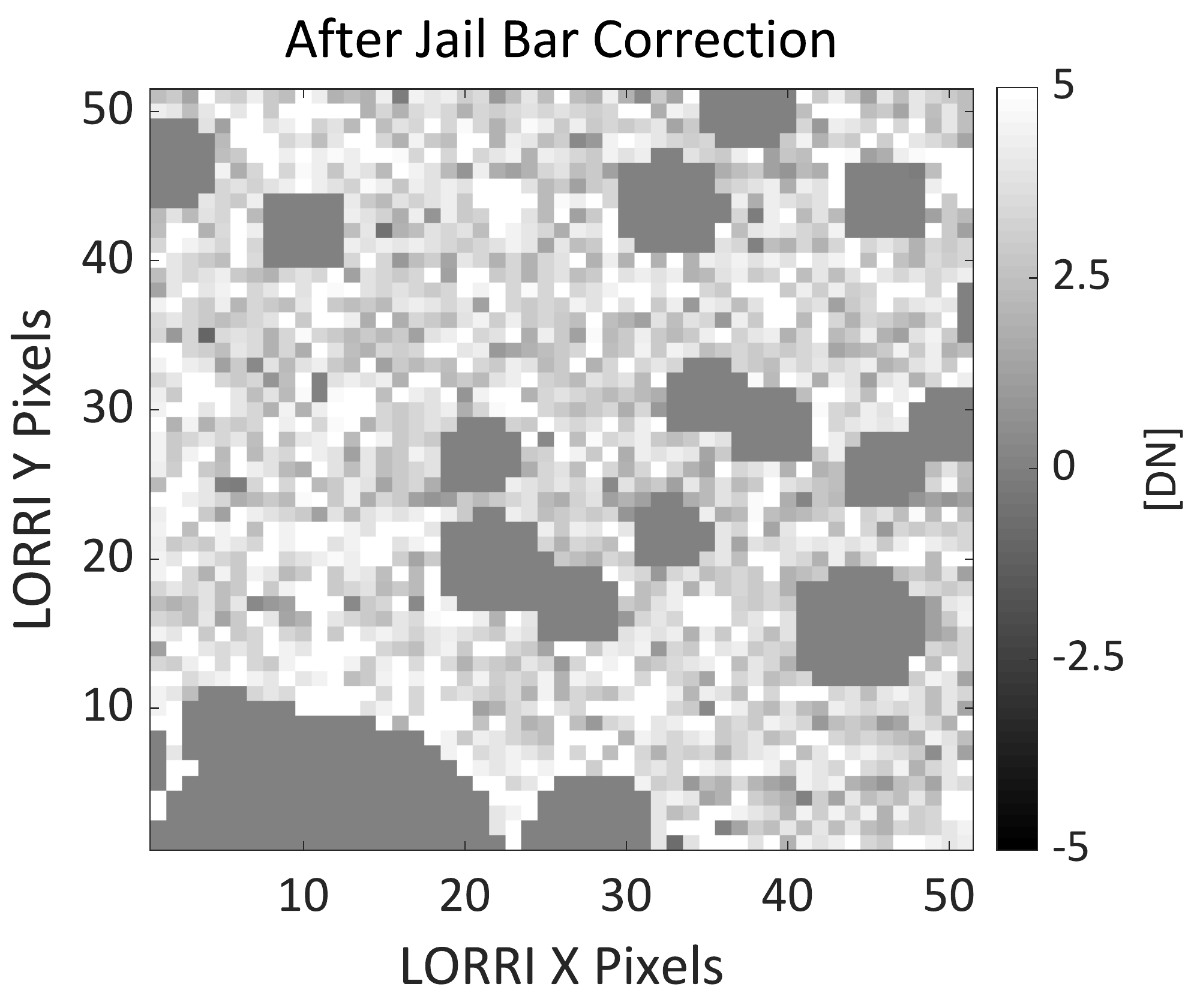} }}%
    \subfloat[][Before - After]{{\includegraphics[width=5.15cm]{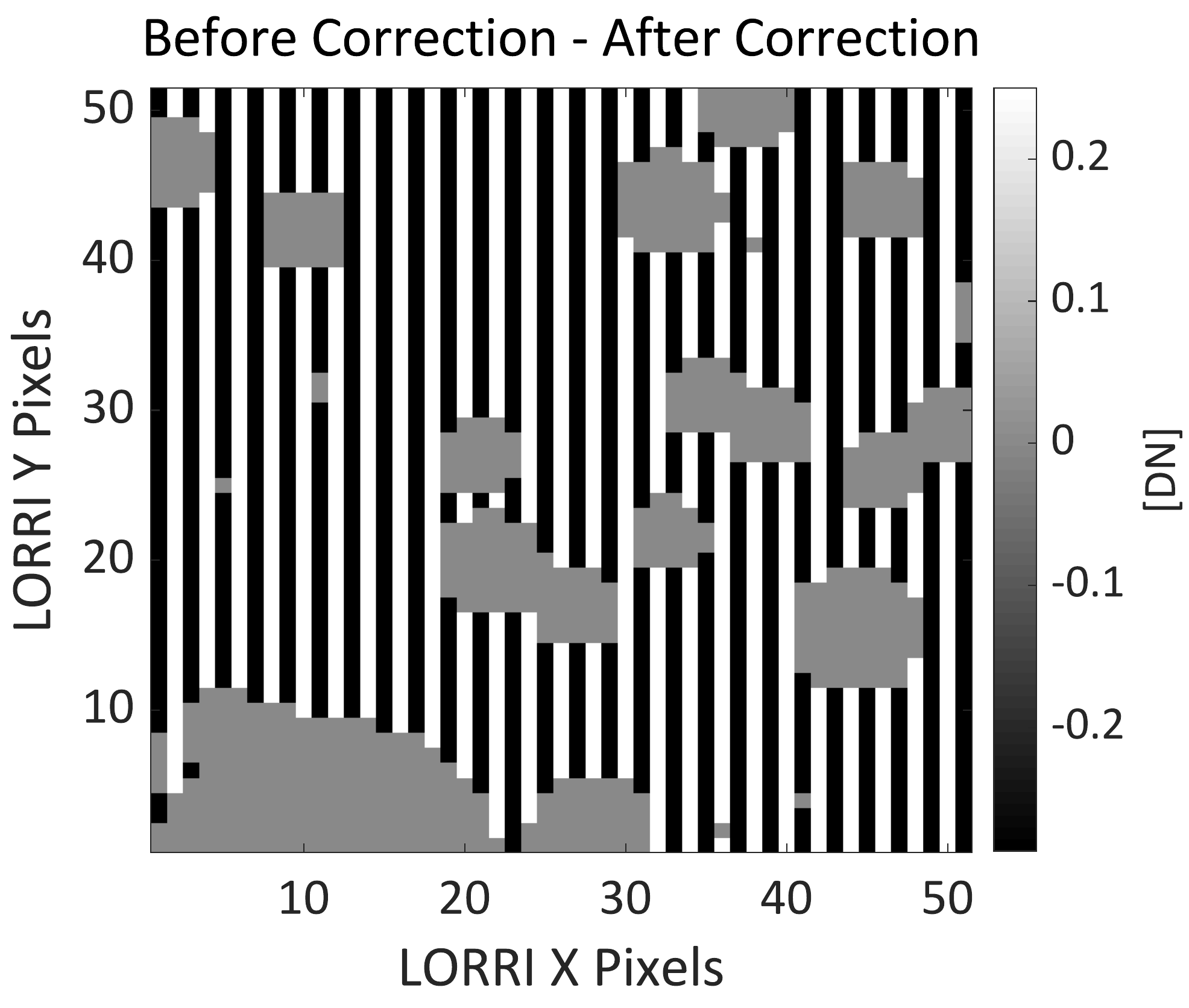} }}%
\caption{A 50 $\times$ 50 pixel stamp image of the same single LORRI exposure (a) before the jail bar correction is applied and (b) after the correction is applied. The color stretch is 10 DN with masked pixels appearing as 0 DN, but the effect is so subtle as to not be visible. In (c) we show the difference between (a) and (b) with a color stretch of 0.5 DN and shifted negative 0.25 DN for clarity. This demonstrates how this highly subtle effect must be carefully corrected to obtain accurate background sky values. \label{fig:jaildiff}}
\end{figure*}
Recently, \cite{weaver2020} and \cite{Lauer} pointed out a LORRI detector defect of unknown origin that causes an excess or deficit of 0.5 DN in alternating columns in a ``jail bar'' pattern. This effect is demonstrated for a portion of a single exposure in Figure \ref{fig:jaildiff}. To correct for this effect, we take the difference of every pair of even and odd columns in an exposure and observe a mean deviation of either $+0.5$ or $-0.5$ DN per exposure. We have determined that if the offset is positive, the correction must be subtracted from the even columns, and if the offset is negative, the correction must be added. We subtract or add as appropriate the absolute value of the mean column difference to the even columns. 

\subsection{Reference Pixel Correction}
\label{sec:refcorr}
\begin{figure*}[htb!]
\centering
\includegraphics[width=4.5in]{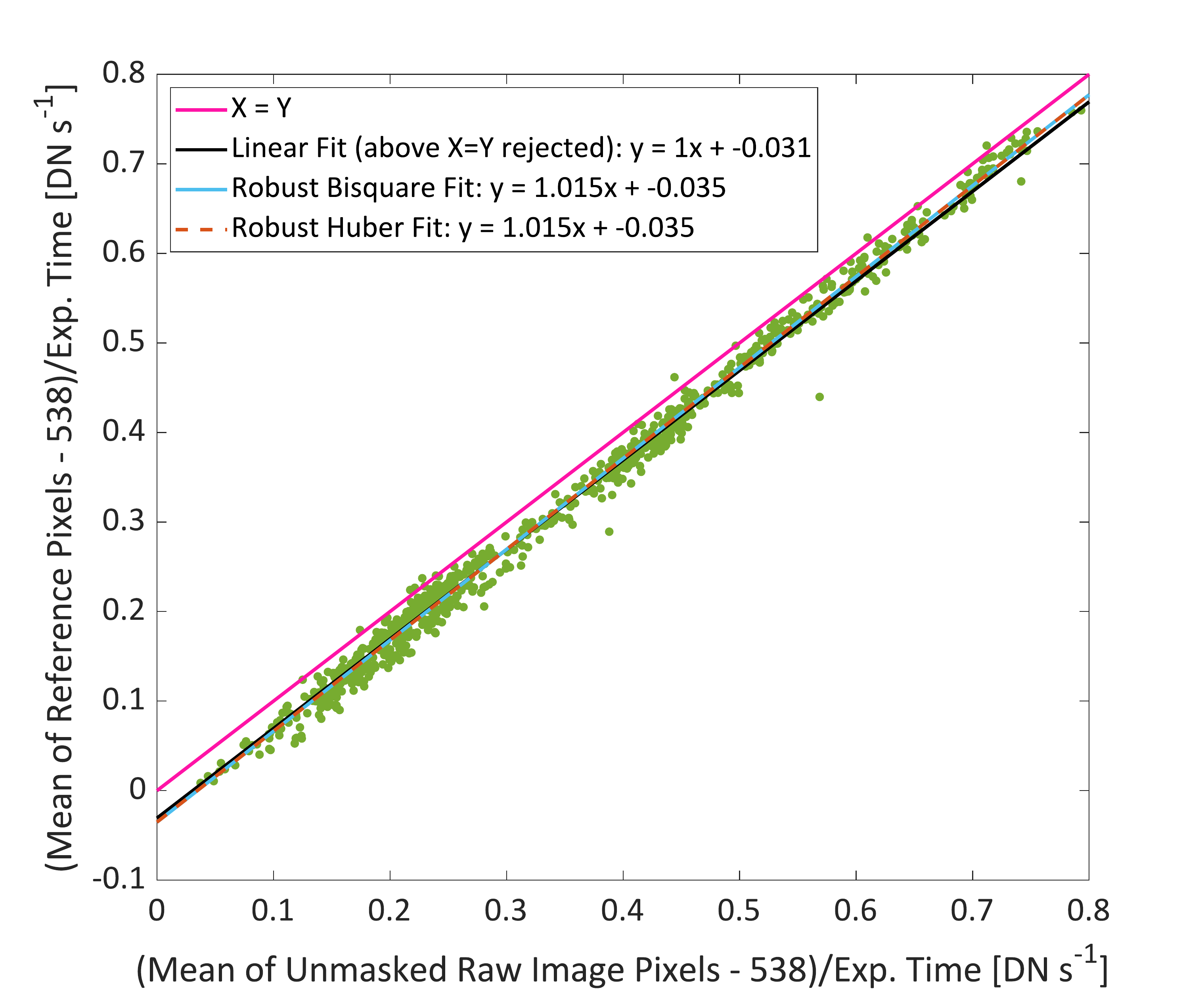}
\caption{LORRI reference pixel offset. Here, we compare the mean of the reference pixels to the mean of the raw exposure pixels for all testing exposures. A constant value of 538 DN is subtracted from both for clarity, and then the means are normalized for different exposure times. The pink line indicates the line along which X = Y, indicating that most of the data have a negative offset from this expected relationship. The black line indicates a linear fit rejecting all values above the pink line, the teal line indicates a robust regression with bisquare weights, and the dashed orange line indicates a robust regression with Huber weights. We select the bisquare-weighted regression to calculate the offset needed to correct the reference column data, 0.035 DN s$^{-1}$. \label{fig:reffit}}
\end{figure*}
LORRI's detector contains four reference columns that are shielded from incoming light with a metal shade to provide a real-time measure of the active pixel bias and dark current levels \citep{cheng_2008}. In 4 $\times$ 4 binning mode, this translates to a single reference column located on the right side of the detector. As part of LORRI's pre-processing pipeline prior to 2020 July 30, the median of the reference column is subtracted from the raw data \citep{soc}. However, \citet{nature} determined that the median is often skewed due to cosmic rays or defective pixels and that a $\sigma$-clipped mean gives a more stable correction that does not produce correlation with the final image mean. Following this procedure, we undo the median subtraction and instead subtract the mean of the reference column after pixels with values $>$ 3$\sigma$ from the mean have been rejected over a series of two iterations:
\begin{equation}
D_{\mathrm{c}} = D_{\mathrm{r}} + R_{\mathrm{m}} - R_{\mathrm{\sigma}} ,
\label{eq:refsub}
\end{equation}
where $D_{\mathrm{c}}$ is the corrected exposure data, $D_{\mathrm{r}}$ is the raw exposure data, $R_{\mathrm{m}}$ is the median of the data in the reference column, and $R_{\mathrm{\sigma}}$ is the $\sigma$-clipped mean of the data in the reference column.
After 2020 July 30, the LORRI pre-processing pipeline was changed by the instrument team to use a different measure of the reference column. First, valid pixels are determined to be those that are not classified as missing with values between 530 -- 560 DN. If no valid pixels are present, the bias is calculated from the FPU board temperature based on ground calibration. If there are valid pixels, a robust mean is taken ignoring outliers beyond a specific range of empirically-determined DN (LORRI collaboration, private communication 2022). Without knowledge of this range, we are unable to reproduce the robust mean for all data and instead use the same $\sigma$-clipped mean after undoing the robust mean using a recorded value from the header.

We then compare the $\sigma$-clipped mean of the reference column to the mean of the unmasked raw exposure pixels for the entire testing set, shown in Figure \ref{fig:reffit}. If the bias column tracks the light detected in the array, we would expect an intensity of 0 DN in the reference column to be equivalent to an intensity of 0 DN in the raw data. Instead, we find that the reference column has a slight negative offset when compared to the raw data. Therefore, subtracting any bias based purely on the reference column values will result in an oversubtraction. \cite{lauer_2022} recently discovered an analog-to-digital conversion error that causes the mean bias level of the reference column to be 0.02 DN too low. Though it is not clear precisely what the cause of these effects are, nor how these observations are related at the hardware level, the important point here is that the reference pixels have a slightly different zero-point than the light pixels and that this effect must be corrected.

In order to compensate, we first subtract an arbitrary 538 DN from both the reference column mean and the raw exposure mean to reduce the numerical values of both families of pixels to near zero. This reduces the importance of the covariance between the slope and offset when we determine the relationship between the two to determine the offset. We then normalize all data points by dividing by the appropriate exposure time to convert to DN s$^{-1}$ before applying any fits to the data. 

\citet{Symons2022} details a variety of tests we performed to determine the best fitting algorithm to relate the light and dark pixels. We use a robust regression with bisquare weighting, which yields an offset of -0.035 DN s$^{-1}$ that must be subtracted from the correction to compensate for the reference column data. Our new correction becomes
\begin{equation}
D_{\mathrm{c}} = D_{\mathrm{r}} + R_{\mathrm{x}} - R_{\mathrm{\sigma}} + (0.035\cdot t_{\mathrm{E}})\ [\mathrm{DN}] ,
\label{eq:refsubmod}
\end{equation}
where $R_{\mathrm{x}}$ is either the median of the reference column for older data with no recorded bias measurement ($R_{\mathrm{m}}$) or that which is recorded in the header ($R_{\mathrm{b}}$). The reference correction is multiplied by the appropriate exposure time, $t_{\mathrm{E}}$. The 0.02 DN correction applied by \citet{lauer_2022} is included in the correction we apply.

The reference correction naturally removes any dark current in the detectors \citep{cheng_2008}, which should be negligible at the temperatures at which the CCD was operated following the Jupiter encounter \citep{ccds, nature}. 
As detailed in \citet{Symons2022}, the CCD temperature has continued to decrease as \nh\ moves away from the Sun, so we do not expect a dynamic contribution that is not already accounted for by the reference pixel correction. 

\subsection{Conversion to Surface Brightness}
\label{sec:sbconv}
After these corrections are made to the raw data in DN, we calibrate to surface brightness units in \nw\ using the following conversion that we derived to be straightforward and reproducible:
\begin{equation}
\lambda I_{\lambda} = \left(\frac{\alpha f_{0} 10^{-0.4 m_{0}}}{t_{\mathrm{E}} \Omega_{\mathrm{beam}}} \right) I_{\mathrm{raw}} ,
\label{eq:cal}
\end{equation}
where: $I_{\mathrm{raw}}$ is the raw LORRI exposure flux in DN; $f_{0} = 3050 \,$ Jy is the zero-point of Vega in the LORRI bandpass; $m_{0}$ is the empirically determined zero-point magnitude; $t_{\mathrm{E}}$ is the exposure time; $\Omega_{\mathrm{beam}}$ is the solid angle of the beam; and $\alpha$ is the conversion from Jy to \nw\ \citep{Symons2022}. The solid angle of the beam, $\Omega_{\mathrm{beam}}$, is computed as $\Omega_{\mathrm{beam}} = \Omega_{\mathrm{PSF}}\cdot \mathrm{pix}_{\mathrm{size}}^{2}$ where $\Omega_{\mathrm{PSF}}$ = 2.64 pix$^{2}$ is the total point source solid angle determined via source stacking \citep{nature}, and $\mathrm{pix}_{\mathrm{size}}$ is the LORRI 4 $\times$ 4 binned pixel width of 1.98$\times$10$^{-5}$ rad. The zero-point magnitude, $m_{0}$, is derived in the LORRI ($R_{\mathrm{L}}$) band from the Johnson $V$-band zero-point ($m_{V}$ = 18.88; \citealt{weaver2020}) as follows. Given that a source's magnitude ($m$) in any bandpass $i$ is calculated from its flux ($f$) and zero-point in magnitudes (ZP) via
\begin{equation}
m_{i} = -2.5\ \log_{10}(f_{i}) + \mathrm{ZP}_{i} ,
\end{equation}
the difference between a magnitude in $V$-band ($m_{V}$) and $R_{\mathrm{L}}$-band ($m_{R_{\mathrm{L}}}$) is determined from:
\begin{gather}
m_{V} - m_{R_{\mathrm{L}}} = \nonumber\\
-2.5\ \log_{10}(f_{V}) + \mathrm{ZP}_{V} +\ 2.5\ \log_{10}(f_{R_{\mathrm{L}}}) - \mathrm{ZP}_{R_{\mathrm{L}}} .
\end{gather}
We compute the $R_{L}$-band flux zero-point to be $m_{R_{\mathrm{L}}}$ = 0.046 by interpolating the magnitude of Vega in the $U$, $B$, $V$, $R$, $I$, and $J$ bands (covering 360 -- 1250 nm; \citealt{Megessier_1995}), giving $m_{V} - m_{R_{\mathrm{L}}} = -0.016$. Given knowledge that ZP$_{V}$ is 18.88, $f_{V}$ (the zero-point of Vega in $V$-band) is 3636 Jy, and $f_{R_{\mathrm{L}}}$ (the zero-point of Vega in LORRI's band) is 3050 Jy, we calculate ZP$_{R_{\mathrm{L}}}$ to be
\begin{gather}
\mathrm{ZP}_{R_{\mathrm{L}}} = \nonumber\\
-2.5\ \log_{10}(f_{V}) + \mathrm{ZP}_{V} +\ 2.5\ \log_{10}(f_{R_{\mathrm{L}}}) -\ m_{V} +\ m_{R_{\mathrm{L}}} \nonumber\\
= 18.71 .
\end{gather}

\begin{figure*}[htb!]
\centering
\noindent\includegraphics[width=\textwidth]{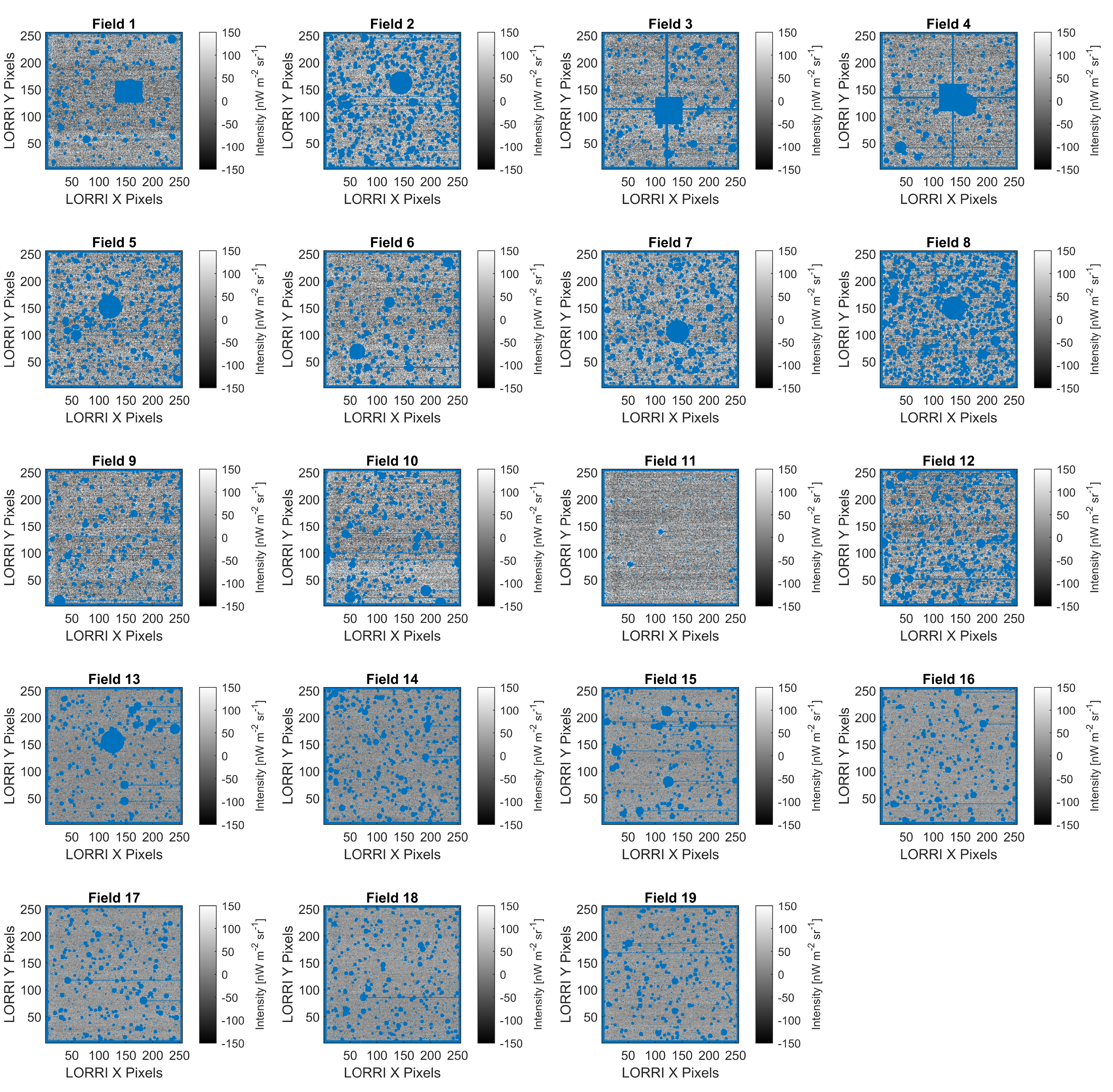}
\caption{All science fields calibrated to surface brightness including image masks. For each science field, we show an example calibrated, masked image with masked pixels in blue. These images have been calibrated to \nw. Most masked objects are stars, but the largest masks are for optical ghosts. Additionally, Field 6 contains two masked foreground galaxies. Fields 13 -- 19 appear less noisy because they are 30 second exposures while all others are 10 second exposures.}\label{fig:allfields_cal}
\end{figure*}

This flux zero-point gives a total conversion factor of 475.45 $\frac{\mathrm{nW\ m^{-2}\ sr^{-1}}}{\mathrm{DN}\ \mathrm{s}^{-1}}$. When a raw exposure is multiplied by this factor, the resulting calibrated image is in surface brightness units of \nw, allowing unmasked pixels to be used to calculate diffuse brightness of the image background. Examples of the final reduced, masked and calibrated images in each of the 19 science fields are shown in Figure \ref{fig:allfields_cal}. The mean and 1-$\sigma$ error for each field are listed in Table \ref{tab:lildiff}. Additionally, we make our calibrated images with masks available on PDS.

\begin{deluxetable}{cccc}[htb!]
\tabletypesize{\scriptsize}
\tablecaption{The calibrated, masked image mean ($\lambda I_{\lambda}^{\mathrm{diff}}$) calculated per-field as the mean of all images of a given field in both DN s$^{-1}$ and \nw. This includes all calibration corrections. The 1-$\sigma$ error, $\delta\lambda I_{\lambda}^{\mathrm{diff}}$, is calculated as the standard deviation of all image means for each field.\label{tab:lildiff}}
\tablehead{
\colhead{\textbf{Field \#}} & \colhead{\textbf{$\lambda I_{\lambda}^{\mathrm{diff}}$} [DN s$^{-1}$]} &\colhead{\textbf{$\lambda I_{\lambda}^{\mathrm{diff}}$} [\nw]}
& \colhead{\textbf{$\delta\lambda I_{\lambda}^{\mathrm{diff}}$} [\nw]} }
\startdata
\textbf{Field 1} & 0.050 & 23.86 & 2.89 \\
\textbf{Field 2} & 0.092 & 43.60 & 4.19 \\
\textbf{Field 3} & 0.062 & 29.46 & 1.59 \\
\textbf{Field 4} & 0.065 & 30.97 & 4.19 \\
\textbf{Field 5} & 0.074 & 35.17 & 4.20 \\
\textbf{Field 6} & 0.061 & 28.93 & 4.12 \\
\textbf{Field 7} & 0.082 & 38.81 & 3.35 \\
\textbf{Field 8} & 0.075 & 35.73 & 3.58 \\
\textbf{Field 9} & 0.055 & 26.14 & 7.16 \\
\textbf{Field 10} & 0.066 & 31.48 & 4.61 \\
\textbf{Field 11} & 0.070 & 33.39 & 4.09 \\
\textbf{Field 12} & 0.061 & 28.78 & 4.47 \\
\textbf{Field 13} & 0.057 & 26.88 & 4.55 \\
\textbf{Field 14} & 0.054 & 25.44 & 2.70 \\
\textbf{Field 15} & 0.059 & 28.09 & 5.03 \\
\textbf{Field 16} & 0.062 & 29.59 & 1.17 \\
\textbf{Field 17} & 0.062 & 29.33 & 0.50 \\
\textbf{Field 18} & 0.051 & 24.33 & 3.97 \\
\textbf{Field 19} & 0.056 & 26.53 & 2.08
\enddata
\end{deluxetable}

\section{Astrophysical Foreground Corrections}
\label{sec:foregrounds}

After converting our raw exposures to calibrated images, we estimate and account for the per-image contribution from several diffuse astrophysical foregrounds in order to measure the COB. These foregrounds include the ISL, multiple sources of diffuse optical scattering, the DGL, galactic extinction, and light from IPD. 

\vspace{30pt}
\subsection{Integrated Starlight}
\label{sec:isl}

The brightest sky component in the LORRI images is starlight. A large fraction of this component is removed by source masking, but there is still residual stellar emission from faint sources below the masking threshold and the wings of the PSF. Accordingly, we decompose the term describing remaining starlight into $\lambda I_{\lambda}^{\mathrm{ISL}}$ = $\lambda I_{\lambda}^{\mathrm{faint}}$ + $\lambda I_{\lambda}^{\mathrm{PSF}}$, where $\lambda I_{\lambda}^{\mathrm{faint}}$ includes contributions from unmasked sources with $m_{G} >$ 21, and $\lambda I_{\lambda}^{\mathrm{PSF}}$ includes the unmasked extended PSF response for our masked bright sources. 

For the populations of faint stars below the masking limit, we use the TRILEGAL model \citep{trilegal} to generate a simulated star catalog for each LORRI field to $m$ = 32 in the $G$ band over a 0.0841 square degree area. To probe the variation in the surface brightness from such sources,  we generate ten independent TRILEGAL simulations for each field. For all $N$ sources with $m >$ 21 in each field's simulation, we calculate $\lambda I_{\lambda}^{\mathrm{faint}}$ as the mean of the summed surface brightness from the simulated sources over the ten-member ensemble.

In order to determine to contribution from the extended, unmasked PSF response of resolved sources, we first need to reconstruct LORRI's PSF. We have developed an algorithm for PSF reconstruction that combines computationally simple techniques in a way that is robust to noise and other complicating factors, detailed in \cite{Symons}. Using this estimated PSF, we construct a noiseless simulated image for each LORRI exposure with sources from the \gaia\ DR2 catalog. Point sources convolved with the PSF are placed in their known coordinates within the mock image, the previously determined mask for that exposure is applied, and the mean of the remaining unmasked pixels is taken as the contribution from the extended PSF, $\lambda I_{\lambda}^{\mathrm{PSF}}$. 

\subsection{Optical Scattering Contributions}
\label{sec:scatt}

\begin{figure*}[htb!]
\centering
\includegraphics[width=4in]{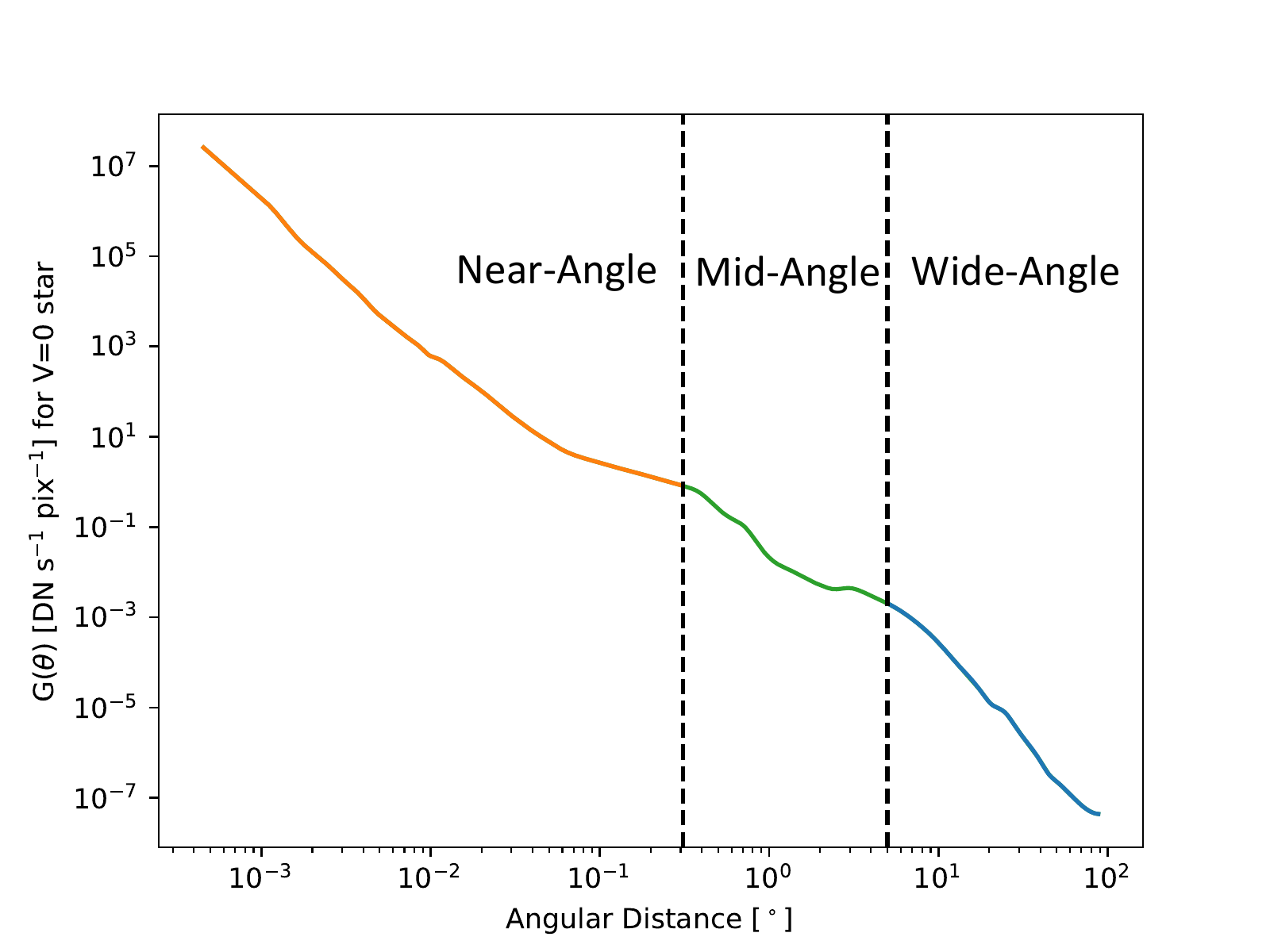}
\caption{The extended response function of LORRI \citep{Lauer}. The function within the LORRI FOV was calculated from in-flight measurements of stars, while the function beyond the FOV was calculated using a combination of in-flight measurements of scattered sunlight and pre-flight testing. We use this function to determine the amount of scattered light that is detected by LORRI from all sources out to the measured extent of 88$\degree$. The orange section represents what we define to be near-angle scattering ($\leq$ $0^{\circ}.31$). The green represents mid-angle scattering ($0^{\circ}.31 < \theta \leq 5^{\circ}$), and the blue represents wide-angle scattering ($>$ 5$\degree$). \label{fig:extpsf}}
\end{figure*}

LORRI experiences significant optical scattering from off-axis sources. While bright sources cause optical ghosting that has been characterized \citep{Cheng2010}, more recent studies of LORRI's extended response function have shown that all sources may cause significant scattering out to 45$\degree$ from the center of the FOV, and possibly beyond \citep{Lauer}. At the levels of the uncertainty in our COB measurement, this is an important component that must be removed, which we account as part of $\lambda I_{\lambda_{\rm inst}}$ in Eq.~\ref{eq:cob}. We define three regimes over which this scattering is calculated: near-angles where diffuse optical ghost intensity exists from all sources; mid-angles at $0^{\circ}.31 < \theta \leq 5^{\circ}$ where light from sources illuminating the baffle scatters into the optical path; and far-angles out to 88$\degree$ where the full extent of LORRI's extended response contributes surface brightness. Though we estimate the scattered contributions in each regime differently, we can combine the extended response function to a point source at an off-axis angle $\theta$ in each regime into a single function called $G(\theta)$ 
\citep{tsumura_ciber}. $G(\theta)$ is normalized to DN s$^{-1}$ pix$^{-1}$ for a $V$ = 0 star, and is illustrated in Figure \ref{fig:extpsf}. In the following Sections we detail the construction of this gain function and how it is used to estimate the scattered contribution to the diffuse surface brightness in our science data set.

\subsubsection{Near-Angle Scattering}
\label{sec:diffghost}

\begin{figure*}[htb!]
\centering
\includegraphics[width=4in]{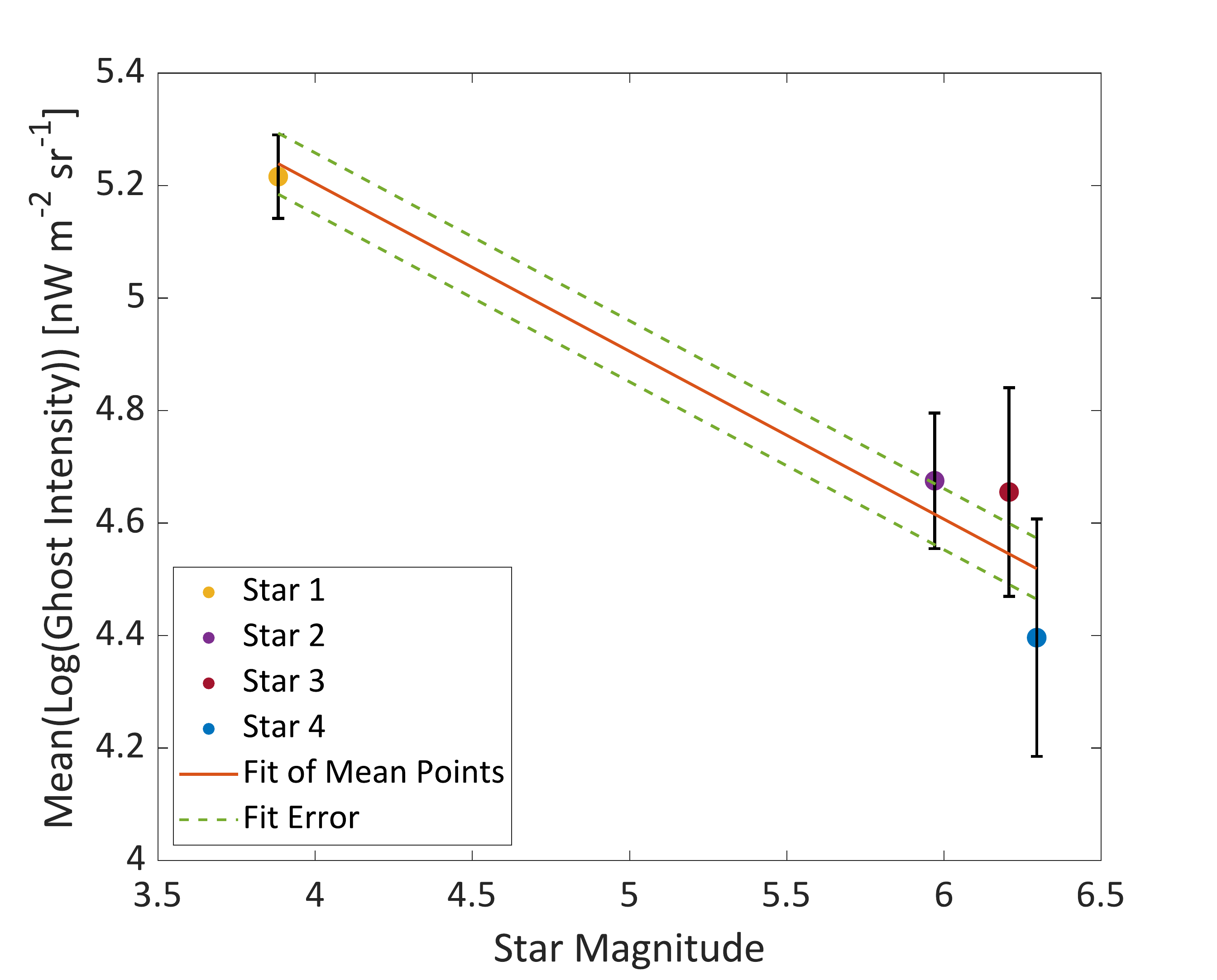}
\caption{The fitted relationship between mean ghost intensity $\lambda I_{\lambda}^{\mathrm{ghost}}$ and $m_{G}$ of the star causing the ghosts in our training set. Each star is given a color-coded point, with black error bars indicating the standard deviation of all ghost intensities generated by that star. The orange line gives the linear fit between the points, with green dashed lines indicating the standard error on the fit. \label{fig:fit}}
\end{figure*}

In addition to the ghosts that cause obvious image-space artifacts, all stars within the region of space that directly illuminates the LORRI lens relay introduce additional diffuse brightness into the image region where ghosts are known to appear. To avoid masking that entire region of the exposure (approximately the central third), we develop a relationship between star magnitude and expected ghost intensity so that this additional diffuse foreground contribution may be subtracted from the exposure using the ghost training set described in Section \ref{sec:final} and the geometric relation discussed in Section \ref{sec:ghostmask}. For each ghost in the training set, intensity is estimated by taking the mean of the background-subtracted unmasked pixels within the ghost radius, calculated as the mean of the non-ghost unmasked pixels.
This gives the most probable intensity of the ghost, which is then multiplied by the number of pixels within the ghost radius, yielding the ghost intensity, ${\lambda I_{\lambda}^{\mathrm{ghost}} = \lambda I_{\lambda}^{\mathrm{M}}\cdot N_{\mathrm{pix}}}$, where $\lambda I_{\lambda}^{\mathrm{M}}$ is the mean value for the ghost and $N_{\mathrm{pix}}$ is the number of pixels. This intensity is then related to the flux of the star causing the ghost, as shown in Figure \ref{fig:fit}. Additional details about this model and the validations we performed can be found in \citet{Symons2022}.

\begin{figure*}[htb!]
\centering
\includegraphics[height=2.5in]{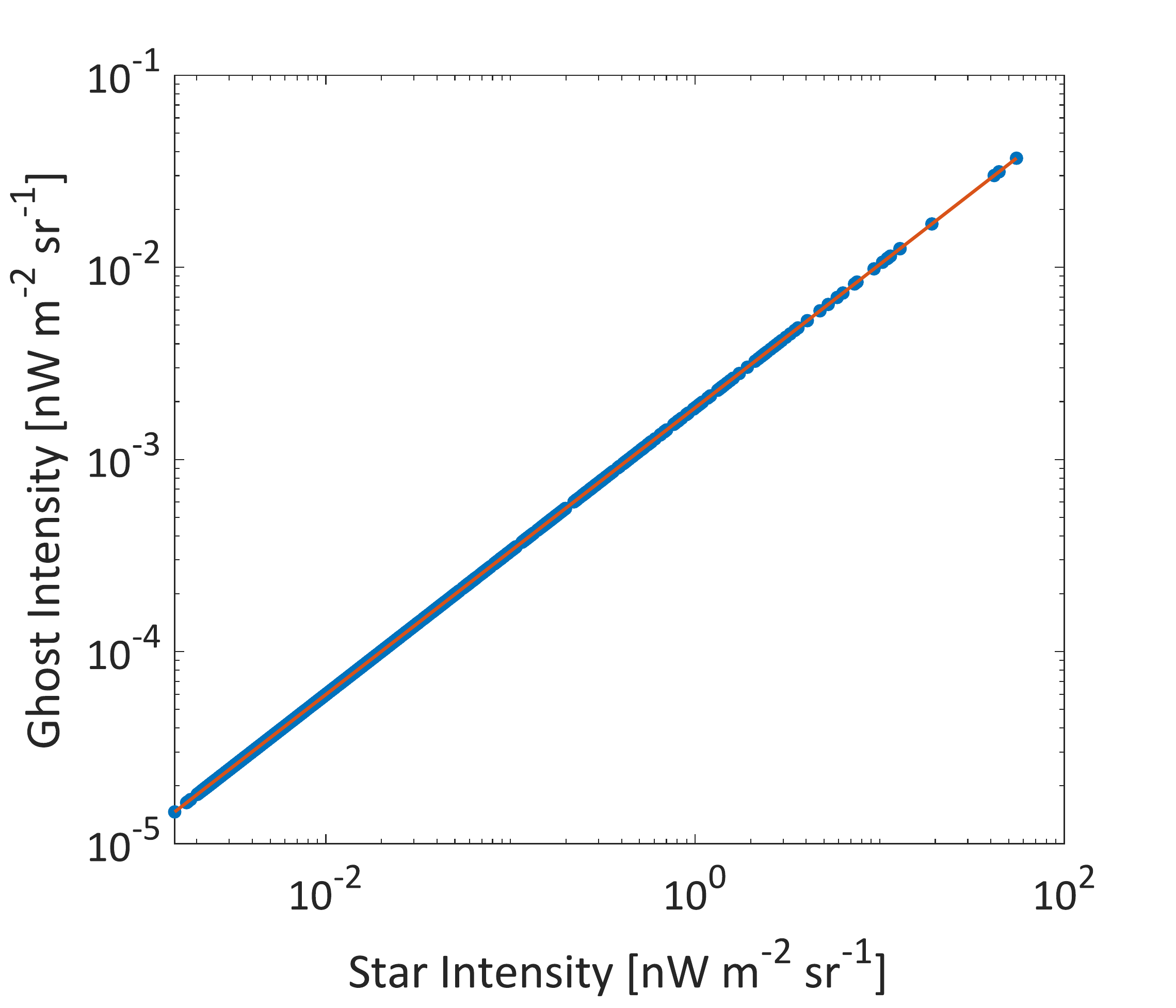}
\includegraphics[height=2.5in]{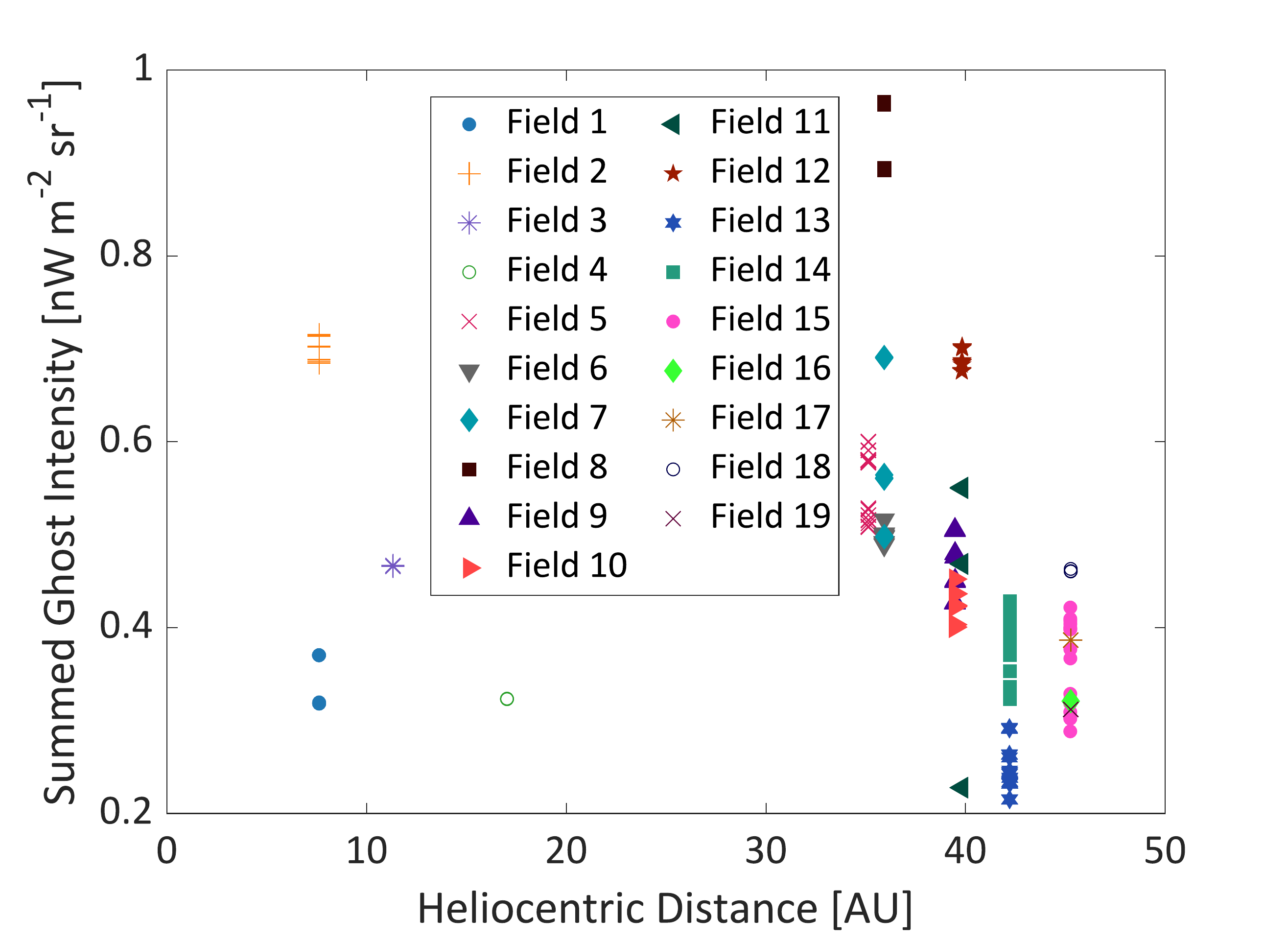}
\caption{Left: Predicted ghost intensities compared to star intensity for all stars in range to cause a ghost in one exposure. Using this linear relationship (orange line), we estimate diffuse ghost intensity that must be subtracted per-exposure for each star. The sum of all ghost intensities associated with all stars is the quantity explored in the right panel. Right: Summed diffuse ghost intensity calculated for all science exposures as a function of heliocentric distance. Points are color-coded with varying symbols by field number. Differing populations of stars near each field cause a natural variation in the total diffuse ghost intensity. \label{fig:pred}}
\end{figure*}

With this model relating the geometry and intensity of the near-angle scattering, we can predict the surface brightness of each source falling in the scattering region. For each science exposure, a list of all stars that meet the distance criteria to cause ghosts is created. For each star in this list, the predicted ghost intensity is calculated via the model, illustrated for a single field in the left panel of Figure \ref{fig:pred}. For each exposure, these intensities are summed to form the total ghost intensity $\lambda I_{\lambda}^{\mathrm{ghost}}$. As an example, $\lambda I_{\lambda}^{\mathrm{ghost}} = 0.58$ \nw\ for the exposure shown in Figure \ref{fig:pred}. When this estimation is repeated for all science exposures, the summed ghost intensity ranges from 0.21 to 0.97 \nw, as shown in the right panel in Figure \ref{fig:pred}. We subtract this quantity from $\lambda I_{\lambda}^{\mathrm{meas}}$ to correct for the diffuse optical ghosting. The contribution of this geometric model to $G(\theta)$ represented as an azimuthal average is shown in Figure \ref{fig:extpsf}. 

\subsubsection{Mid-Angle Scattering}
\label{sec:medscatt}

Beyond the region where sources directly illuminate the lens relay ($0^{\circ}.31$), the LORRI extended response function has been determined by \cite{Lauer} and is shown in Figure \ref{fig:extpsf}. 
At intermediate angles, we estimate the expected scattered intensity using this function and the \gaia\ DR2 catalog to estimate the scattered intensity from individual sources in $0^{\circ}.31 < \theta \leq 5^{\circ}$.
For each catalog source in this range, we compute the surface brightness that would be coupled to the detector through the response function, and sum the intensities to determine the mid-angle scattering contribution per exposure, $\lambda I_{\lambda}^{\mathrm{scatt_{m}}}$. 

\subsubsection{Wide-Angle Scattering}
\label{sec:widescatt}

At angles $>$ 5$\degree$, we estimate the ISL brightness by combining the wide-angle part of $G(\theta)$ shown in Figure \ref{fig:extpsf} with an all-sky ISL map \citep{gaia_isl}. The map is in HEALpix format \citep{healpix} with $N_{\mathrm{side}}$ = 64 and gives $G$-band luminosity in W m$^{-2}$ sr$^{-1}$ for each $\simmod$55' pixel. We convert this to the equivalent flux of Vega, and then sum map pixels into 40 linearly-spaced radial bins spanning 5$\degree$ to 88$\degree$. 
The number of bins and bin spacing were empirically optimized to minimize the effect of binning choices. The total scattered intensity due to each bin is calculated as the sum of the product of the binned ISL flux and $G(\theta)$, which yields the total intensity contribution from wide-angle scattering, $\lambda I_{\lambda}^{\mathrm{scatt_{w}}}$. The parameter describing total combined off-axis scattering is then defined to be ${\lambda I_{\lambda}^{\mathrm{scatt}} = \lambda I_{\lambda}^{\mathrm{scatt_{m}}} + \lambda I_{\lambda}^{\mathrm{scatt_{w}}}}$. We carry $\lambda I_{\lambda}^{\mathrm{ghost}}$ that captures the intensity from near-angle scattering as a separate quantity forming part of $\lambda I_{\lambda}^{\mathrm{inst}}$. 

\vspace{10pt}
\subsection{Diffuse Galactic Light Correction}
\label{sec:dgl}
DGL is a significant diffuse contribution to the overall surface brightness in an exposure, and is expected to be comparable in amplitude to the COB at high galactic latitudes. 
Our large and diverse set of science fields permits us to apply two distinct methods to estimate the DGL. In the first, we use thermal dust emission templates and a coupling constant to estimate the optical contribution from DGL, based on the procedure described in \cite{nature}. In the second method, we do not directly subtract the DGL but instead correlate a measure of the sum of COB and DGL with the template, effectively avoiding the large uncertainty associated with the DGL coupling parameter \citep{arendt,dirbe1}. This is the first time this direct-fit method has been applied to LORRI data, and since it solves for the COB brightness and the DGL coupling directly, it is the preferred method for our measurement of the COB intensity. 

\subsubsection{DGL Template Generation}
\label{sec:dgltemplate}

To calculate templates for the spatial structure of the DGL, we begin by computing a spatial template for the emission based on three different analyses that combine similar data in different ways: the Planck component-separated dust maps \citep{planck}; the Improved Reprocessing of the IRAS Survey (IRIS) maps \citep{iris}; and a combined IRIS and Schlegel, Finkbeiner, and Davis (SFD; \citealt{sfd}) map \citep{iris_sfd} that combines IRIS at scales $<$ 30' and SFD at scales $>$ 30'. In all cases the FIR point sources have been removed. The extragalactic CIB intensity is not included in the Planck and IRIS/SFD templates, but we subtract 0.48 MJy sr$^{-1}$ from the IRIS maps to account for it \citep{Dole}.

For each template, we compute the spatial emission in each LORRI field at a reference wavelength of $100 \, \mu$m. The expected surface brightness of the DGL in each exposure can be calculated via
\begin{equation}
\lambda I_{\lambda}^{\mathrm{DGL}}(\lambda, \ell, b) = \nu \langle I_{\nu}(100\mu m, \ell, b)\rangle\cdot \bar{c}_{\lambda}\cdot d(b) ,
\end{equation}
where $\nu \langle I_{\nu}(100\mu m)\rangle$ is the mean 100 $\mu$m intensity over the field in MJy sr$^{-1}$ at wavelength $\lambda$ and galactic coordinates $(\ell, b)$, $\bar{c}_{\lambda}$ is a bandpass-weighted scaling factor between the optical and FIR, and $d(b)$ is a geometric function that modifies $\bar{c}_{\lambda}$ \citep{nature}. We note other works often parametrize the scaling as $\nu b_{\lambda}$ (unrelated to galactic latitude $b$; see \citealt{dirbe5,dirbe6}) with units \nw / MJy sr$^{-1}$. While $\bar{c}_{\lambda}$ carries the same dimensions as $\nu b_{\lambda}$, it includes the geometric factor $d(b)$ and $\nu b_{\lambda}$ does not, so the two quantities are not directly comparable. To provide quantities with like units, we introduce the parameter $\nu \beta_{\lambda}$ = 30 $\cdot \bar{c}_{\lambda}$ and present estimates for the values of $\nu \beta_{\lambda}$ and $\nu b_{\lambda}$ in Section \ref{sec:results}.

\begin{figure*}[htbp!]
    \centering
    \subfloat{{\includegraphics[width=7.75cm]{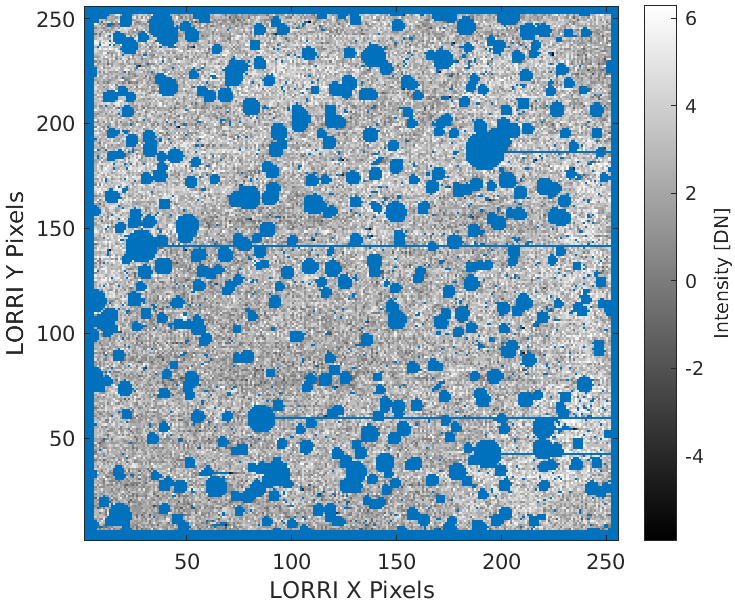} }}%
    \subfloat{{\includegraphics[width=7.75cm]{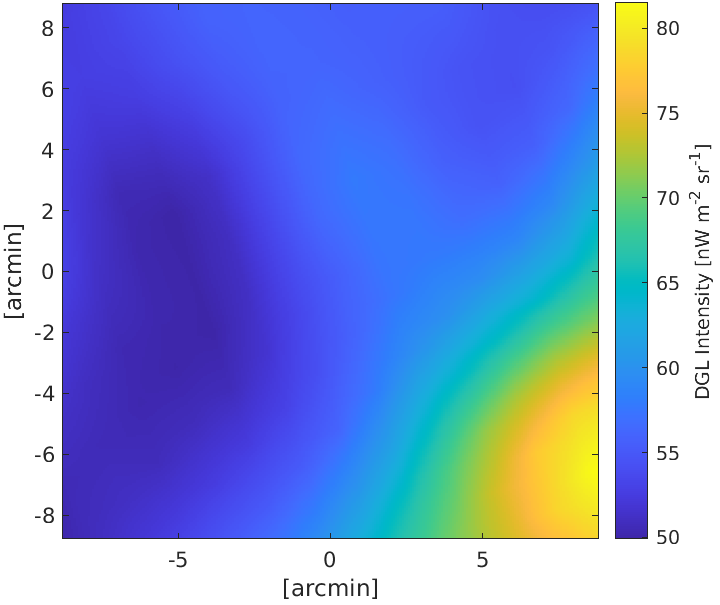} }}%
\caption{Left: A masked example LORRI exposure taken at $b$ = $44^{\circ}.8$ on 2019 Mar.~9 with observation ID 0414427168. While this image is not one our science fields, it provides an example of visible DGL structure in a LORRI observation. Right: The DGL for the same field calculated using Planck thermal dust emission maps \citep{planck} assuming a fixed $\bar{c}_{\lambda}$ = 0.491. The resulting mean intensity of the DGL emission in this field is $\lambda I_{\lambda}^{\rm DGL} = 57.3$ \nw. \label{fig:dgl}}
\end{figure*}

The parameter $d(b)$ is computed as
\begin{equation}
\label{eq:db}
d(b) = d_{0}(1 - 1.1g\sqrt{\sin{|b|}}) ,
\end{equation}
where $d_{0}$ = 1.76 is computed by normalizing $d(b)$ at $b$ = 25$\degree$ \citep{lillie}, and the asymmetry factor of the scattering phase function $g$ \citep{jura} is computed by taking a bandpass-weighted mean of a model for the high-latitude DGL \citep{draine} to yield $g$ = 0.61. In Figure \ref{fig:dgl}, we demonstrate an example of the DGL using a fixed value of $\bar{c}_{\lambda}$ for a single LORRI field compared to the masked exposure for the same field. 
We also demonstrate the numerical differences for the predictions for the different spatial templates for the example field PE1  
in Table \ref{tab:100m_dgl}. In this example all other model parameters are fixed, so the different DGL predictions are due entirely to differences in the input templates.

\begin{deluxetable*}{ccccc}[htbp!]
\tabletypesize{\normalsize}
\tablecaption{Comparison between 100 $\mu$m emission and DGL intensity for spatial templates for LORRI field PE1. For each of three spatial templates, we calculate a comparison between mean 100 $\mu$m emission, $\nu I_{\nu}^{\mathrm{100\mu m}}$, and mean DGL intensity, $\lambda I_{\lambda}^{\mathrm{DGL}}$. We also list the associated uncertainties in these quantities due to the error in the spatial templates and the parameters $\bar{c}_{\lambda}$ and $d(b)$. We note that the slightly larger $\nu I_{\nu}^{\mathrm{100\mu m}}$ predicted by the IRIS template may be sourced by residual ZL that has not been removed, while the other templates have had more careful removal of ZL \citep{planck}. This effect is included in our error budget. The relatively small differences in 100 $\mu$m emission from the various templates propagate to larger differences in DGL intensity because of the nature of the scaling relationship.\label{tab:100m_dgl}}
\tablehead{
\colhead{\textbf{Template}} & \colhead{$\nu I_{\nu}^{\mathrm{100\mu m}}$ [MJy sr$^{-1}$]} & \colhead{$\delta \nu I_{\nu}^{\mathrm{100\mu m}}$ [MJy sr$^{-1}$]} & \colhead{$\lambda I_{\lambda}^{\mathrm{DGL}}$ [\nw]} & \colhead{$\delta \lambda I_{\lambda}^{\mathrm{DGL}}$ [\nw]} }
\startdata
\textbf{IRIS} & 2.53 & 0.03 & 23.74 & 8.69 \\
\textbf{IRIS/SFD} & 2.31 & 0.03 & 19.87 & 7.27 \\
\textbf{Planck} & 2.32 & 0.09 & 19.57 & 7.16
\enddata
\end{deluxetable*}

\vspace{-20pt}
\subsubsection{Method 1: Direct DGL Subtraction}
\label{sec:dgl_direct}

Our direct subtraction of DGL to isolate the COB proceeds by choosing a particular scaling $\bar{c}_{\lambda}$ between the FIR and optical intensity and then subtracting the scaled template from each image. We fit measurements \citep{ienaka} to a model \citep{zda04} of the $\frac{I_{\nu}(\mathrm{optical})}{I_{\nu}(100 \mu \mathrm{m})}$ scaling to arrive at $\bar{c_{\lambda}}$ = 0.491. We then calculate $\lambda I_{\lambda}^{\mathrm{COB}}$ on a per-image basis by rearranging Eq.~\ref{eq:cob} as:
\begin{gather}
\lambda I_{\lambda}^{\mathrm{COB}} = \nonumber\\
\epsilon \cdot (\lambda I_{\lambda}^{\mathrm{diff}} - \lambda I_{\lambda}^{\mathrm{ghost}} - \lambda I_{\lambda}^{\mathrm{ISL}} - \lambda I_{\lambda}^{\mathrm{scatt}} - \lambda I_{\lambda}^{\mathrm{DGL}}) .
\end{gather}
To combine measurements from multiple images of a single field, we take the mean of $\lambda I_{\lambda}^{\mathrm{COB}}$ for all images of the same field. Finally, the mean over all of our science fields yields a combined measurement of $\lambda I_{\lambda}^{\mathrm{COB}}$, which we refer to as our ``direct-subtraction'' COB measurement. We perform this process separately for our three spatial templates of the DGL, IRIS, IRIS/SFD, and Planck, arriving at a unique $\lambda I_{\lambda}^{\mathrm{COB}}$ for each template.

\subsubsection{Method 2: DGL Correlation Estimation}

To account for the DGL via correlation with 100 $\mu$m emission, we calculate $\lambda I_{\lambda}^{\mathrm{EBL+DGL}}$ on a per-image basis via
\begin{equation}
\lambda I_{\lambda}^{\mathrm{EBL+DGL}} = \lambda I_{\lambda}^{\mathrm{diff}} - \lambda I_{\lambda}^{\mathrm{ghost}} - \lambda I_{\lambda}^{\mathrm{ISL}} - \lambda I_{\lambda}^{\mathrm{scatt}}.
\end{equation}
To combine measurements from multiple images of a single field, we again compute the mean of this quantity over the exposures of that field.

To estimate the DGL scaling, we perform a linear fit of $\lambda I_{\lambda}^{\mathrm{EBL+DGL}}$ to the independent parameter $(d(b) \cdot \nu I_{\nu}^{\mathrm{100\mu {\rm m}}})$ via
\begin{equation}
\label{eq:cobest}
 \lambda I_{\lambda}^{\mathrm{EBL+DGL}} = \nu \beta_{\lambda} \cdot(d(b) \cdot \nu I_{\nu}^{\mathrm{100\mu m}}) + \lambda I_{\lambda}^{\mathrm{COB}},
\end{equation} 
where $\nu \beta_{\lambda}$ is the slope and $\lambda I_{\lambda}^{\mathrm{COB}}$ the offset of the fit.

\subsection{Correction for Galactic Extinction}
\label{sec:ext}

Galactic extinction is the absorption of extragalactic photons by dust in the interstellar medium of the Milky Way. The extinction templates are therefore constructed in a very similar fashion to the DGL estimates. For each exposure, we compute extinction in magnitudes, $\Delta m_{b}$, \citep{sfd_reddening} using an SFD all-sky map \citep{sfd} of galactic reddening, $E(B-V)$, assuming the Landolt $R$ filter ($\lambda_{\mathrm{eff}}$ = 642.78 nm) with $R_{V}$ = 3.1, which gives $A_{b}$ = 2.169, from:
\begin{equation}
\Delta m_{b} = E(B-V) A_{b} .
\end{equation}
We then calculate the extinction flux correction as:
\begin{equation}
\epsilon = \frac{f_{\mathrm{c}}}{f_{\mathrm{uc}}} = \frac{1}{10^{-0.4\Delta m_{b}}} ,
\end{equation}
where $f_{\mathrm{uc}}$ is the uncorrected flux and $f_{\mathrm{c}}$ is the corrected flux. 

\subsection{IPD Foreground Light Estimation}

Light scattering from IPD generated from asteroidal collisions and cometary outgassing is a challenging foreground signal when conducting observations in the inner solar system (e.g., $r<5$ AU); however, IPD foregrounds are generally estimated to be negligible in the outer solar system. 
We can use the interplanetary dust model of \citet{Poppe_2019b} to predict $\lambda I_{\lambda}^{\rm{IPD}}$ along each LORRI line of sight.
Across all LORRI observations used, the modeled $\lambda I_{\lambda}^{\rm{IPD}}$ varies from 0.056 nW m$^{-2}$ sr$^{-1}$ to 0.51 nW m$^{-2}$ sr$^{-1}$, which is at least an order of magnitude smaller than the expected COB brightness. We investigate this component further in Section \ref{sec:ipd_results}. 

\subsection{Final COB Estimates}
\label{sec:cob_est}

To compute our best estimate of the COB intensity, we fit Eq.~\ref{eq:cobest} for the parameter $\lambda I_{\lambda}^{\mathrm{COB}}$, which is the best combined measurement of the COB intensity referred to as the ``correlative COB measurement.'' The fit is weighted by the overall uncertainty in the field surface brightness, which is derived 
in Section \ref{sec:error_analysis}. The fit is adjusted for extinction using an iterative method that minimizes $\chi^{2}$ \citep{num_rec}. As a first step, we establish a design matrix for our fit containing the $d(b)\cdot \nu I_{\nu}^{\mathrm{100\mu m}}$ values for each of the 19 fields. We define our weights to be:
\begin{gather}
N = \nonumber\\
\frac{1}{(\delta\lambda I_{\lambda}^{\mathrm{EBL+DGL}})^{2} + (\nu \beta_{\lambda})^{2}\cdot (\delta d(b)\cdot \nu I_{\nu}^{\mathrm{100\mu m}})^{2}}
\end{gather}
using an initial guess for $\nu \beta_{\lambda}$ of 7 \nw/MJy sr$^{-1}$, where the $\delta$ values are the error bars on the various quantities (see Section \ref{sec:error_analysis}). The Normal fitting method using this design matrix and uncertainty weight then 
yields $\nu \beta_{\lambda}$ and an estimate for the COB before it is adjusted for extinction. 

We then repeat the fitting procedure to perform the adjustment for galactic extinction. Our new design matrix contains the $d(b)\cdot \nu I_{\nu}^{\mathrm{100\mu m}}$ values and $
\epsilon = \frac{f_{\mathrm{c}}}{f_{\mathrm{uc}}}$ for each field. The new weights are set to:
\begin{gather}
N' = \nonumber\\
\frac{1}{(\delta\lambda I_{\lambda}^{\mathrm{EBL+DGL}})^{2} + (\nu \beta_{\lambda})^{2}\cdot (\delta d(b)\cdot \nu I_{\nu}^{\mathrm{100\mu m}})^{2}} ,
\end{gather}
where $\nu b_{\lambda}$ is now the previous best-fit value and the same error values are used. We again use the Normal method to obtain an extinction-adjusted slope and $\lambda I_{\lambda}^{\mathrm{COB}}$. We do not find that additional iterations of this method change the fit parameters appreciably.

In addition to thermal dust templates, we also include a spatial template based on NHI column density as measured by the HI4PI survey \citep{hi4pi}. Because DGL is correlated with NHI column density \citep{Toller1981}, this provides a robust check of our COB intensity based on an independent physical tracer. The method proceeds as for the thermal dust templates, with $d(b)\cdot \mathrm{NHI}$ replacing $d(b)\cdot \nu I_{\nu}^{\mathrm{100\mu m}}$.

\section{Error Analysis}
\label{sec:error_analysis}

\begin{figure*}[htbp!]
\centering
\noindent\includegraphics[width=\textwidth]{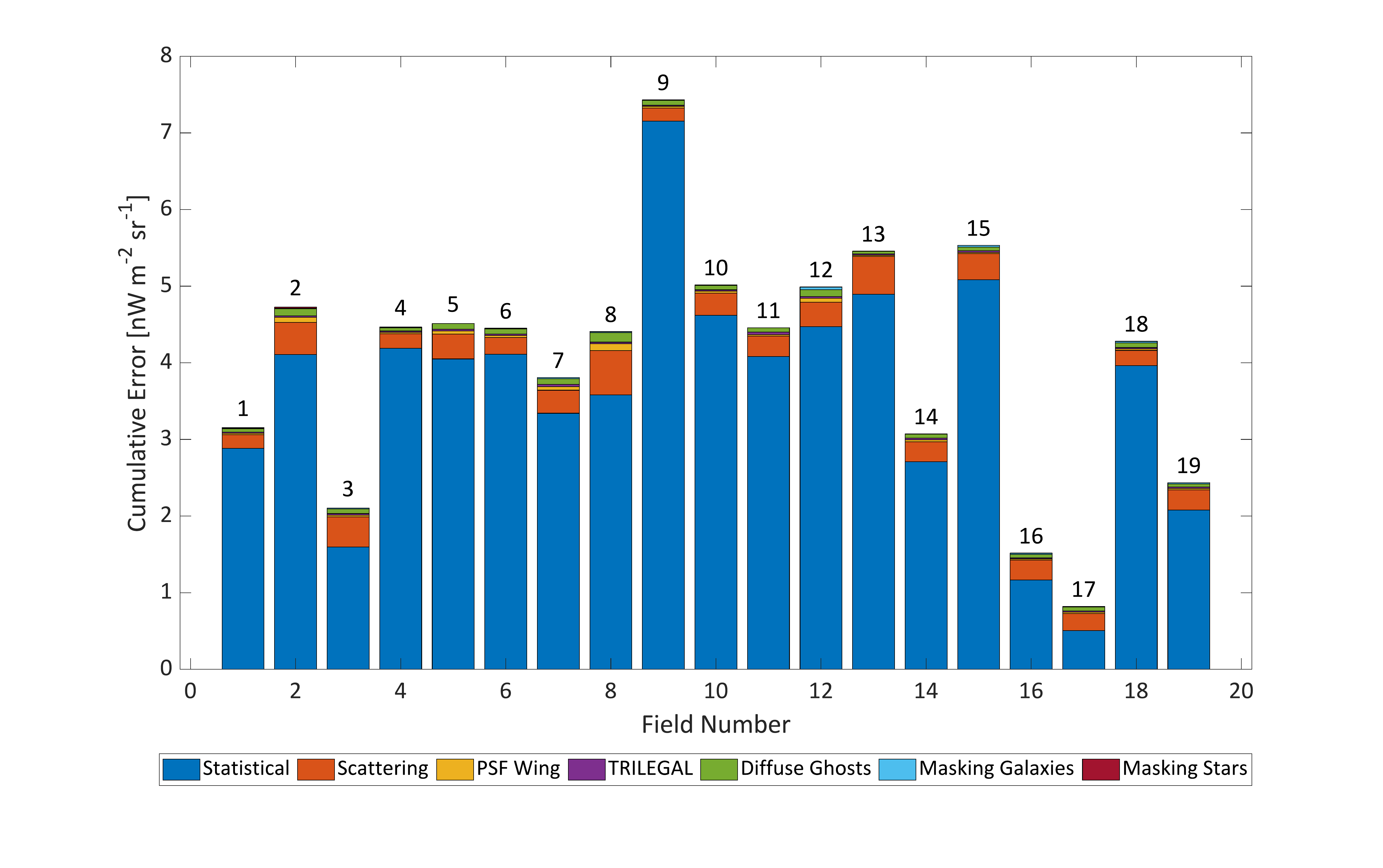}
\caption{For each of the LORRI science fields, we divide $\delta \lambda I_{\lambda}^{\mathrm{EBL+DGL}}$ into its constituent sources of error. The largest source of error for all fields is statistical, followed by the uncertainty on optical scattering. }\label{fig:field_errs}
\end{figure*}

The errors in our measurement of $\lambda I_{\lambda}^{\mathrm{COB}}$ include calibration uncertainty, systematic uncertainty in both the instrument and estimation of astrophysical foregrounds, and statistical uncertainty. The total uncertainty budget is given in Table \ref{tab:budget} and summarized in Figure \ref{fig:field_errs}. 

\begin{deluxetable*}{c|ccc}[htbp!]
\tablecaption{Total error budget given as mean values for all fields combined. The first column gives the type of error, the second column the error's source, the third the quantity for which the error provides uncertainty, and the fourth the uncertainty in that quantity. Errors marked with (*) are included in $\delta \lambda I_{\lambda}^{\mathrm{EBL+DGL}}$.\label{tab:budget}}
\tablehead{
\colhead{\textbf{Error Type}} & \colhead{\textbf{Source}} & \colhead{\textbf{Quantity}} & \colhead{\textbf{Error [nW m\boldmath$^{-2}$ sr$^{-1}$]}}}
\startdata
\multirow{2}{*}{\textbf{Instrumental}} & Dark Current & $\lambda I_\lambda^\mathrm{inst}$ & -0.36 \\ 
\cline{2-4}
 & Diffuse Ghosts & $\lambda I_\lambda^\mathrm{ghost}$ & (+0.062, -0.055)* \\ 
\hline
\multirow{2}{*}{\textbf{Calibration}} & Photometric Calibration & $\lambda I_\lambda^\mathrm{diff}$ & $\pm$ 0.61 \\ 
\cline{2-4}
 & Solid Angle of Beam & $\lambda I_\lambda^\mathrm{diff}$ & $\pm$ 1.21 \\ 
\hline
\multirow{11}{*}{\textbf{Astrophysical}} & IPD & $\lambda I_\lambda^\mathrm{IPD}$ & (+1.90, -0.02) \\ 
\cline{2-4}
 & Masking Galaxies & $\lambda I_\lambda^\mathrm{diff}$ & $\pm$ 0.01* \\ 
\cline{2-4}
 & Masking Stars & $\lambda I_\lambda^\mathrm{diff}$ & $\pm$ 0.002* \\ 
\cline{2-4}
 & PSF Wings & $\lambda I_\lambda^\mathrm{PSF}$ & $\pm$ 0.004* \\ 
\cline{2-4}
 & TRILEGAL Simulations & $\lambda I_\lambda^\mathrm{faint}$ & $\pm$ 0.019* \\ 
\cline{2-4}
 & Mid-Angle Scattering & $\lambda I_\lambda^\mathrm{scatt_{m}}$ & $\pm$ 0.240 \\ 
\cline{2-4}
 & Wide-Angle Scattering & $\lambda I_\lambda^\mathrm{scatt_{w}}$ & $\pm$ 0.059 \\ 
\cline{2-4}
 & Total Scattering & $\lambda I_\lambda^\mathrm{scatt}$ & $\pm$ 0.299* \\ 
\cline{2-4}
 & DGL - IRIS & $\lambda I_\lambda^\mathrm{DGL}$ & $\pm$ 8.58 \\ 
\cline{2-4}
 & DGL - IRIS/SFD & $\lambda I_\lambda^\mathrm{DGL}$ & $\pm$ 6.65 \\ 
\cline{2-4}
 & DGL - Planck & $\lambda I_\lambda^\mathrm{DGL}$ & $\pm$ 6.49 \\ 
\hline
\hline
\multirow{2}{*}{\textbf{Total}} & Calibration Error & $\lambda I_\lambda^\mathrm{COB}$ & $\pm$ 1.36 \\ 
\cline{2-4}
 & Statistical Error & $\lambda I_\lambda^\mathrm{COB}$ & $\pm$ 1.23
\enddata
\end{deluxetable*}

\vspace{-20pt}
\subsection{Instrumental Errors}
\label{sec:inst_err}
Instrumental errors include those sources of uncertainty that are primarily associated with the LORRI instrument itself. These include uncertainty in the estimation of the dark current and diffuse optical ghosting (near-angle scattering). 

\subsubsection{Dark Current}
\label{sec:dc}
Dark current is assessed via LORRI's reference pixels. While the reference pixels are identical to the photo-responsive pixels, the metal shade that shields them from light may cause up to a 20\% reduction in the measured dark current due to electromagnetic coupling between the shade and the pixels \citep{nature}. We estimate the mean dark current for all science exposures and then calculate 20\% of that value to be the uncertainty in the dark current, which is 0.36 \nw. As this error would cause an over-compensation in the reference pixel correction, the resulting error on $\lambda I_{\lambda}^{\mathrm{inst}}$ is only in the negative direction. 

\subsubsection{Near-Angle Scattering}
\label{sec:diffghost_err}
We use a model to predict diffuse ghost intensity based on the magnitude of the star causing the ghost (Section \ref{sec:diffghost}). The dominant source of error in this estimation is the error on the linear fit used to predict ghost intensity, which gives the error associated with the diffuse ghost intensity per star, $\delta \lambda I_{\lambda}^{\mathrm{ghost}}$ (see Figure \ref{fig:fit}). We calculate the upward-going error as $\delta \lambda I_{\lambda}^{\mathrm{ghost},+}$ and the downward-going error as $\delta \lambda I_{\lambda}^{\mathrm{ghost},-}$ for every star within $0^{\circ}.31$ of the center of each science exposure. Finally, just as $\lambda I_{\lambda}^{\mathrm{ghost}}$ is summed for all stars in a given exposure, $\delta \lambda I_{\lambda}^{\mathrm{ghost}}$ for all stars is also summed:
\begin{equation}
\delta \lambda I_{\lambda}^{\mathrm{ghost}} = \sum_{i=1}^{N_{\mathrm{stars}}}(\delta \lambda I_{\lambda}^{\mathrm{ghost},i}) ,
\end{equation}
where $N_{\mathrm{stars}}$ is the total number of stars for the exposure. The process is repeated for both $\delta \lambda I_{\lambda}^{\mathrm{ghost},+}$ and $\delta \lambda I_{\lambda}^{\mathrm{ghost},-}$ to yield the total positive and negative error on $\lambda I_{\lambda}^{\mathrm{ghost}}$ per exposure. These quantities are averaged for all exposures of a given field to obtain the overall per-field uncertainty due to the scattering model. 

\subsection{Calibration Errors}
\label{sec:cal_err}
Calibration errors are those errors associated with our photometric calibration of LORRI exposures from raw units to surface brightness in \nw\ with a defined zero-level. These include uncertainty in the photometric calibration zero-point and the solid angle of the beam.

The photometric zero-point of LORRI was recently recalibrated by \cite{weaver2020} to be 18.88 in $V$-band with a $\simmod$2\% 1$\sigma$ accuracy for a solar-type SED. We convert this zero-point into the $R_{\mathrm{L}}$ band (Section \ref{sec:sbconv}), which carries a negligible error compared to the overall photometric accuracy. We apply this uncertainty as a $\pm$ 2\% error on $\lambda I_{\lambda}^{\mathrm{diff}}$. 

The error on the beam solid angle was assessed by \cite{nature} via half-half jackknife tests on PSF stacking to be $\pm$ 4\%, which propagates to a 4\% error on $\lambda I_{\lambda}^{\mathrm{diff}}$.

\subsection{Astrophysical Errors}
\label{sec:astro_err}
Astrophysical errors include any source of uncertainty associated with the estimation and subtraction of astrophysical foregrounds. 

\subsubsection{Masking Stars}
\label{sec:stars}
The dominant source of uncertainty in masking stars is the size of the mask, which depends directly on the magnitude of each source via Equation \ref{eq:mask}. We use the \gaia-reported $G$-band magnitude error, $\delta m_{G}$, to estimate the error from varying the size of the masks. We compute the error on each source's magnitude as a random Gaussian with width that matches $\delta m_{G}$. 
These new error-adjusted magnitudes create a new star mask for each LORRI exposure that is then propagated through the entire data analysis pipeline and compared with the original $\lambda I_{\lambda}^{\mathrm{diff}}$. The difference between these quantities gives the error associated with the star mask for each exposure, $\delta \lambda I_{\lambda}^{\mathrm{star}}$. The mean of this error for all exposures of a given field gives the same error for that field. The error on the absolute calibration of \gaia\ is negligible in comparison to the individual source magnitude errors.

\subsubsection{Masking Galaxies}
\label{sec:gals}
In the process of excluding potential galaxies from the \gaia\ DR2 catalog, some galaxies may be incorrectly identified as stars and masked, just as some stars may be incorrectly identified as galaxies and unintentionally left unmasked. The purity of the \gaia\ galaxy catalog is 71.3\% \citep{gaia_gal}, meaning that of the galaxies identified in the catalog, only 71.3\% of them can be expected to be correct identifications. 

To explore this source of error, we create 100 randomized versions of the galaxy catalog for each field, each of which selects only 71.3\% of the available galaxies for masking. This simulates the effect of only a random subset of the possible galaxies being correctly identified. For each LORRI exposure we generate 100 new masks using the 100 different galaxy catalogs. These new masked images are then processed through the pipeline and a new $\lambda I_{\lambda}^{\mathrm{diff}}$ is calculated for each. Again, the difference between the original $\lambda I_{\lambda}^{\mathrm{diff}}$ and the error-adjusted version is taken. Because we have 100 simulations per exposure, we take the mean difference as the error for a given exposure:
\begin{equation}
\delta \lambda I_{\lambda}^{\mathrm{gal}} = \frac{\sum_{i=1}^{N_{\mathrm{sim}}}(|\lambda I_{\lambda}^{\mathrm{diff}}-\lambda I_{\lambda}^{\mathrm{err}_{i}}|)}{N_{\mathrm{sim}}} ,
\end{equation}
where $\delta \lambda I_{\lambda}^{\mathrm{gal}}$ is the per-exposure error due to incorrectly masking galaxies, $\lambda I_{\lambda}^{\mathrm{diff}}$ is the non-error-adjusted value, $\lambda I_{\lambda}^{\mathrm{err}_{i}}$ is the $i$th error-adjusted $\lambda I_{\lambda}^{\mathrm{diff}}$, and there are $N_{\mathrm{sim}}$ = 100 total simulations. The mean of $\delta \lambda I_{\lambda}^{\mathrm{gal}}$ for all exposures of a given field is taken to be the $\delta \lambda I_{\lambda}^{\mathrm{gal}}$ for that field. 

\subsubsection{PSF Wings}
\label{sec:psf}
In calculating $\lambda I_{\lambda}^{\mathrm{PSF}}$, we use catalog-simulated images with masks determined from \gaia\ DR2. The primary source of uncertainty in this calculation is the reported \gaia\ DR2 $\delta m_{G}$, which affects the radii of the star masks as well as the summed source fluxes in each simulated image. To assess this error, we vary the magnitude of each source by $\delta m_{G}$ and generate new simulated images and new masks to recalculate $\lambda I_{\lambda}^{\mathrm{PSF}}$ for each science exposure. We then take the difference between $\lambda I_{\lambda}^{\mathrm{PSF}}$ and its error-adjusted version to be the per-exposure $\delta \lambda I_{\lambda}^{\mathrm{PSF}}$. 

\subsubsection{TRILEGAL Simulations}
\label{sec:trilegal}
The TRILEGAL simulation draws from a statistical model to generate a catalog of sources in each field, and each realization has a slightly different number of sources in a given magnitude range. To account for this variation, we compute the standard deviation of $\lambda I_{\lambda}^{\mathrm{faint}}$ over 10 simulations of each field, which yields the error $\delta \lambda I_{\lambda}^{\mathrm{faint}}$ for each exposure. 

\subsubsection{Mid-Angle Scattering}

Mid-angle scattering is calculated using the \gaia\ DR2 catalog. The two most prominent sources of uncertainty in this estimation are error in computing the flux of each source and error in the extended response function itself. 

The error in the calculation of each source's flux is derived from the error in the \gaia\ $G$-band zero-point, which is $m_{G_{0}} \pm \delta m_{G_{0}}$ = 25.6885 $\pm$ 0.0018 \citep{gaia}. We estimate the corresponding error in the flux zero-point to be 0.17\% of each source's flux \citep{Symons2022}. 
The total uncertainty for all sources for a given exposure, $\delta\lambda I_{\lambda}^{f}$, is then the sum of the uncertainties for individual sources. 

The extended response function has an uncertainty in its amplitude of 10\% \citep{Lauer}, which we apply as an fixed positive or negative uncertainty to $G(\theta)$ when computing the mid-angle scattering term. 
The total uncertainty for all sources in one exposure, $\delta\lambda I_{\lambda}^{g}$,  is the sum of all individual sources' response to the modified $G(\theta)$. Since these are uncorrelated errors, the total uncertainty associated with the mid-angle scattering, $\delta \lambda I_{\lambda}^{\mathrm{scatt_{m}}}$, is then the quadrature sum of these two sources of error:
\begin{equation}
\delta \lambda I_{\lambda}^{\mathrm{scatt_{m}}} = \big[(\delta\lambda I_{\lambda}^{f})^{2} + (\delta\lambda I_{\lambda}^{g})^{2}\big]^{1/2} .
\end{equation}

\subsubsection{Wide-Angle Scattering}

The diffuse contribution from wide-angle scattering, $\lambda I_{\lambda}^{\mathrm{scatt_{w}}}$, has uncertainties due to the calibration of intensity in the all-sky map as well as the extended response function. Because the all-sky ISL map used to determine flux is derived from \gaia, the uncertainty on the \gaia\ zero-point is again the ultimate source of the intensity error. The error on this parameter is calculated by varying the ISL map by this factor in both the positive and negative directions and computing the difference with the fiducial value, yielding the total intensity error term $\delta\lambda I_{\lambda}^{\mathrm{f}}$. The 10\% uncertainty in the amplitude of the extended response function is calculated in a similar fashion, yielding the error term $\delta\lambda I_{\lambda}^{\mathrm{g}}$. 
These two errors are then combined as uncorrelated uncertainties:
\begin{equation}
\delta \lambda I_{\lambda}^{\mathrm{scatt_{w}}} = \big[(\delta\lambda I_{\lambda}^{\mathrm{f}})^{2} + (\delta\lambda I_{\lambda}^{\mathrm{g}})^{2}\big]^{1/2} .
\end{equation}
The total error we quote on optical scattering, $\delta \lambda I_{\lambda}^{\mathrm{scatt}}$, is the combination of the mid-angle and wide-angle scattering uncertainties:
\begin{equation}
\delta \lambda I_{\lambda}^{\mathrm{scatt}} = \delta\lambda I_{\lambda}^{\mathrm{scatt_{m}}} + \delta\lambda I_{\lambda}^{\mathrm{scatt_{w}}} .
\end{equation}

\subsubsection{DGL Estimation}
\label{sec:dgl_err}
The error on $\nu I_{\nu}^{\mathrm{100\mu m}}$, $\delta \nu I_{\nu}^{\mathrm{100\mu m}}$, is calculated differently for the three spatial templates. For the IRIS and IRIS/SFD templates, this error is based on the root mean square noise of the IRIS map, 0.06 MJy sr$^{-1}$ \citep{iris,iris_sfd}. We scale this from the solid angle of the IRIS beam to the solid angle of a LORRI exposure:
\begin{gather}
\delta \nu I_{\nu}^{\mathrm{IRIS}} = \nonumber\\
\frac{0.06\ \mathrm{[MJy\ sr^{-1}]}}{\big[(1.13\cdot4.3^2)/(17.4^2)\big]^{1/2}} = 0.23\ \mathrm{[MJy\ sr^{-1}]} ,
\end{gather}
where the IRIS beam FWHM is 4.3' for a two-dimensional Gaussian beam and the LORRI exposure width is 17.4'. 

For the Planck template, because $\nu I_{\nu}^{\mathrm{100\mu m}}$ depends on $\tau$, $\beta$, and $T$, $\delta \nu I_{\nu}^{\mathrm{100\mu m}}$ will also depend on these parameters and their uncertainties. The Planck map provides individual error maps for each parameter. Because the parameters are codependent, we varied all parameters separately by a Gaussian function of their given errors such that each parameter is modified randomly up to the full value of the error. We did this for 100 trials per parameter, resulting in a mean $\nu I_{\nu}^{\mathrm{100\mu m}}$ for each parameter per trial. Then we calculated the error on $\nu I_{\nu}^{\mathrm{100\mu m}}$ associated with each parameter as the standard error on the mean:
\begin{gather}
\delta \nu I_{\nu}^{\mathrm{100\mu m},x} = 
\frac{\Big(\frac{\sum_{i}^{100}\big|\nu I_{\nu}^{\mathrm{100\mu m},x_{i}} - \langle\nu I_{\nu}^{\mathrm{100\mu m},x}\rangle\big|^{2}}{99}\Big)^{1/2}}{(100)^{1/2}};\nonumber\\ x=\{\tau,\beta,T\} .
\end{gather}
We found that when examining all LORRI test fields, $\delta \nu I_{\nu}^{\mathrm{100\mu m},\tau}$ was $\simmod$4$\times$ smaller in magnitude than $\delta \nu I_{\nu}^{\mathrm{100\mu m},\beta}$ and  $\delta \nu I_{\nu}^{\mathrm{100\mu m},T}$, which were of equivalent magnitude.
Because $\beta$ and $T$ are the dominant source of uncertainty, we calculate total error on $\nu I_{\nu}^{\mathrm{100\mu m}}$ for the Planck template as
\begin{equation}
\delta \nu I_{\nu}^{\mathrm{Planck}} = \Big[(\delta \nu I_{\nu}^{\mathrm{100\mu m},\beta})^{2} + (\delta \nu I_{\nu}^{\mathrm{100\mu m},T})^{2}\Big]^{1/2} .
\end{equation}

For the NHI spatial template, we compute the uncertainty as
\begin{equation}
\delta \mathrm{NHI} = \frac{5\sigma_{\mathrm{RMS}}}{5\sqrt{N_{\mathrm{beams}}}} = 4.45 \times 10^{17}\ [\mathrm{cm}^{-2}] ,
\end{equation}
where the 5$\sigma_{\mathrm{RMS}}$ = 43 mK \citep{hi4pi_err} and $N_{\mathrm{beams}}$ is the number of beams per LORRI exposure, which is 1.07 based on the HI4PI beam size of 16.2' \citep{hi4pi}. 

Because we scale $\nu I_{\nu}^{\mathrm{100\mu m}}$ and NHI by $d(b)$, we propagate their respective errors as
\begin{gather}
\delta \nu I_{\nu}^{\mathrm{100\mu m}}\cdot d(b) = 
\Big\{[\delta \nu I_{\nu}^{\mathrm{100\mu m}}\cdot d(b)]^{2} \nonumber\\
+ [(\delta g\cdot1.1\sqrt{\sin{|b|}})\cdot\nu I_{\nu}^{\mathrm{100\mu m}}]^{2}\Big\}^{1/2} ,
\end{gather}
where $\delta \nu I_{\nu}^{\mathrm{100\mu m}}$ is either $\delta \nu I_{\nu}^{\mathrm{IRIS}}$ or $\delta \nu I_{\nu}^{\mathrm{Planck}}$ as appropriate to match the source of $\nu I_{\nu}^{\mathrm{100\mu m}}$ (note that IRIS and IRIS/SFD have the same uncertainty). The error on $d(b)$, $\delta d(b)$, is $(\delta g\cdot1.1\sqrt{\sin{|b|}})$ (see Eq. \ref{eq:db}). For NHI, this becomes:
\begin{gather}
\delta \mathrm{NHI}\cdot d(b) = \nonumber\\
\Big\{[\delta \mathrm{NHI}\cdot d(b)]^{2} + [(\delta g\cdot1.1\sqrt{\sin{|b|}})\cdot\mathrm{NHI}]^{2}\Big\}^{1/2}
\end{gather}

The uncertainty on each direct measurement of the DGL is based on the errors associated with the model parameters $\nu \langle I_{\nu}(100\mu m)\rangle$, $\bar{c_{\lambda}}$, and $d(b)$. The error on $\nu \langle I_{\nu}(100\mu m)\rangle$ is calculated as:
\begin{equation}
\delta \lambda I_{\lambda}^{\mathrm{DGL},\nu} = [\bar{c_{\lambda}}\cdot d(b)]^{2}\cdot (\delta \nu I_{\nu})^{2}
\end{equation}
where $\delta \nu I_{\nu}$ is the uncertainty in the CIB subtraction, which is the dominant error. For the IRIS template, this error is 0.21 MJy sr$^{-1}$ \citep{Dole}. Because the Planck and IRIS/SFD templates are already CIB-subtracted, $\delta \nu I_{\nu}$ = 0 and $\delta \lambda I_{\lambda}^{\mathrm{DGL},\nu}$ = 0.

The error on $\bar{c_{\lambda}}$ is calculated as
\begin{equation}
\delta \lambda I_{\lambda}^{\mathrm{DGL},\bar{c_{\lambda}}} = [\nu \langle I_{\nu}(100\mu m)\rangle\cdot d(b)]^{2}\cdot (\delta \bar{c_{\lambda}})^{2} ,
\end{equation}
where $\delta \bar{c_{\lambda}}$ is the error on $\bar{c_{\lambda}}$, 0.129 \citep{ienaka}. 

The error on $d(b)$ is calculated as
\begin{gather}
\delta \lambda I_{\lambda}^{\mathrm{DGL},d(b)} = \nonumber\\
[\nu \langle I_{\nu}(100\mu m)\rangle \cdot \bar{c_{\lambda}}\cdot d_{0}\cdot 1.1\sqrt{\sin{|b|}})]^{2}\cdot [\delta g]^{2} ,
\end{gather}
where $\delta g$ is the error on $g$ and the remaining error on $d(b)$, 0.10 \citep{sano_2016}.

These errors are then combined to yield $\delta \lambda I_{\lambda}^{\mathrm{DGL}}$:
\begin{gather}
\delta \lambda I_{\lambda}^{\mathrm{DGL}} = \nonumber\\
\Big(\delta \lambda I_{\lambda}^{\mathrm{DGL},\nu} + \delta \lambda I_{\lambda}^{\mathrm{DGL},\bar{c_{\lambda}}} + \delta \lambda I_{\lambda}^{\mathrm{DGL},d(b)}\Big)^{1/2},
\end{gather}
which is the uncertainty on $\lambda I_{\lambda}^{\mathrm{DGL}}$ for any given LORRI exposure. 

For the correlative COB measurement, the fit is weighted by the error bars on both $\lambda I_{\lambda}^{\mathrm{EBL+DGL}}$ and $\nu I_{\nu}^{\mathrm{100\mu m}} \cdot d(b)$ or $\mathrm{NHI} \cdot d(b)$ as appropriate. The errors $\delta \nu I_{\nu}^{\mathrm{100\mu m}}\cdot d(b)$ and $\delta \mathrm{NHI}\cdot d(b)$ are discussed above. Here, we discuss the error on $\lambda I_{\lambda}^{\mathrm{EBL+DGL}}$, $\delta \lambda I_{\lambda}^{\mathrm{EBL+DGL}}$, which is a combination of truly random systematic errors and statistical error. The errors included in $\delta \lambda I_{\lambda}^{\mathrm{EBL+DGL}}$ are marked by (*) in Table \ref{tab:budget}. The systematic errors were introduced in the previous Sections. These errors are combined with statistical error as
\begin{gather}
\delta \lambda I_{\lambda}^{\mathrm{EBL+DGL}} = \nonumber\\ 
\big[(\delta \lambda I_{\lambda}^{\mathrm{ghost}})^{2} + (\delta \lambda I_{\lambda}^{\mathrm{star}})^{2} + (\delta \lambda I_{\lambda}^{\mathrm{gal}})^{2} + (\delta \lambda I_{\lambda}^{\mathrm{PSF}})^{2} \nonumber\\
+ (\delta \lambda I_{\lambda}^{\mathrm{faint}})^{2} + (\delta \lambda I_{\lambda}^{\mathrm{scatt}})^{2} + (\delta \lambda I_{\lambda}^{\mathrm{stat}})^{2}\big]^{1/2},
\end{gather}
where $\delta \lambda I_{\lambda}^{\mathrm{EBL+DGL}}$ is calculated for each LORRI field. The components of $\delta \lambda I_{\lambda}^{\mathrm{EBL+DGL}}$ for each science field are illustrated in Figure \ref{fig:field_errs}. 

Statistical error is derived from multiple independent measurements of the same field. This error encompasses different sources of random noise, such as photon noise, that are averaged down with increasing integration time. The statistical error on $\lambda I_{\lambda}^{\mathrm{EBL+DGL}}$ for each field is the standard deviation of the per-image $\lambda I_{\lambda}^{\mathrm{EBL+DGL}}$ (original calculation discussed in Section \ref{sec:cob_est}) for all images of that field:
\begin{gather}
\delta\lambda I_{\lambda}^{\mathrm{stat}} = \nonumber\\
\Bigg(\frac{\sum_{i}^{N}\Big|\lambda I_{\lambda}^{\mathrm{EBL+DGL_{\mathrm{img}}^{i}}} - \langle\lambda I_{\lambda}^{\mathrm{EBL+DGL_{\mathrm{img}}}}\rangle\Big|^{2}}{N_{\mathrm{img}}-1}\Bigg)^{1/2} ,
\label{eq:sem}
\end{gather}
where $N_{\mathrm{img}}$ is the number of images for a given field. This gives the per-field statistical error. We do not include any uncertainty due to our adjustment for galactic extinction as this is negligible compared to other sources of error. 

\subsection{Overall Error Budget}
\label{sec:budget}

The error budget includes all sources of uncertainty in $\lambda I_{\lambda}^{\mathrm{COB}}$, including instrumental, calibration, astrophysical, and statistical sources of error. Our budget, shown in Table \ref{tab:budget}, gives the total value of each error as the mean of that error over all science fields. We also indicate which quantity contributing to our $\lambda I_{\lambda}^{\mathrm{COB}}$ measurement the uncertainty modifies.

The total statistical error on $\lambda I_{\lambda}^{\mathrm{COB}}$ for any given spatial template is the error on the intercept of the fit, $\delta \lambda I_{\lambda}^{\mathrm{COB}}$. Our modeling errors are uncorrelated and carried as statistical errors, except $\delta \lambda I_{\lambda}^{\mathrm{inst}}$ due to dark current and $\delta \lambda I_{\lambda}^{\mathrm{IPD}}$, which cannot be properly assessed, and $\delta \lambda I_{\lambda}^{\mathrm{DGL}}$, which we do not directly subtract in our measurement. When all four templates are combined into a single measurement of the mean $\lambda I_{\lambda}^{\mathrm{COB}}$, the statistical errors are also combined via the mean. This effectively combines the statistical errors from the four independent COB measurements. 

The total calibration error on $\lambda I_{\lambda}^{\mathrm{COB}}$ is the quadrature sum of the two sources of calibration error. This is the combination of calibration error due to the photometric calibration of the zero-point and the solid angle of the beam. Ultimately, we quote the statistical/foreground and calibration uncertainties on our final COB measurement separately.

\section{Results \& Conclusions}
\label{sec:results}

Using the analysis methods presented above, we process our set of 19 science fields into a final measurement of the COB. 

\subsection{COB Estimation}
\label{sec:cob_results}

\begin{figure*}[htbp!]
    \centering
    \subfloat{{\includegraphics[width=7.5cm]{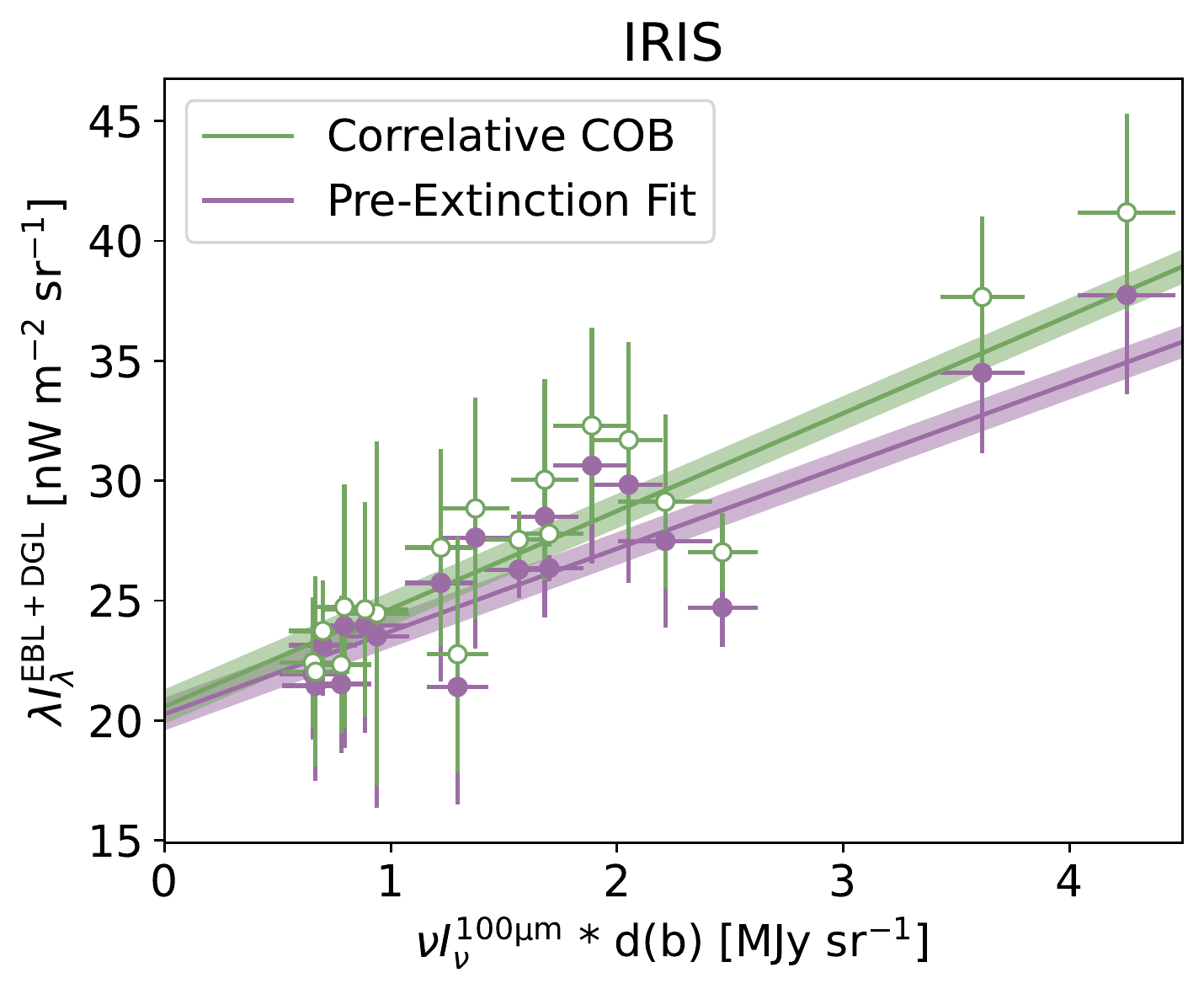} }}%
    \subfloat{{\includegraphics[width=7.5cm]{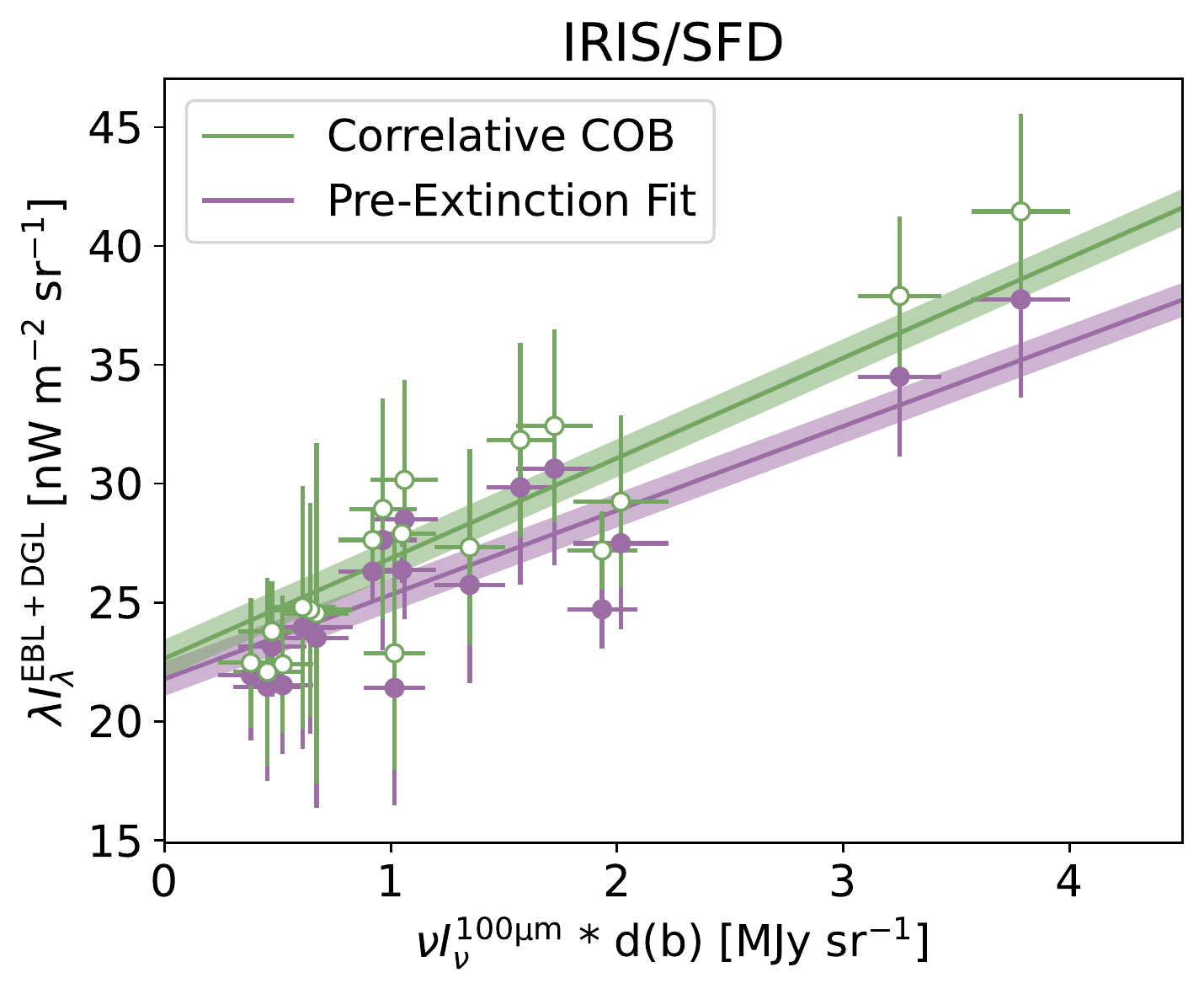} }}%
    \quad
    \subfloat{{\includegraphics[width=7.5cm]{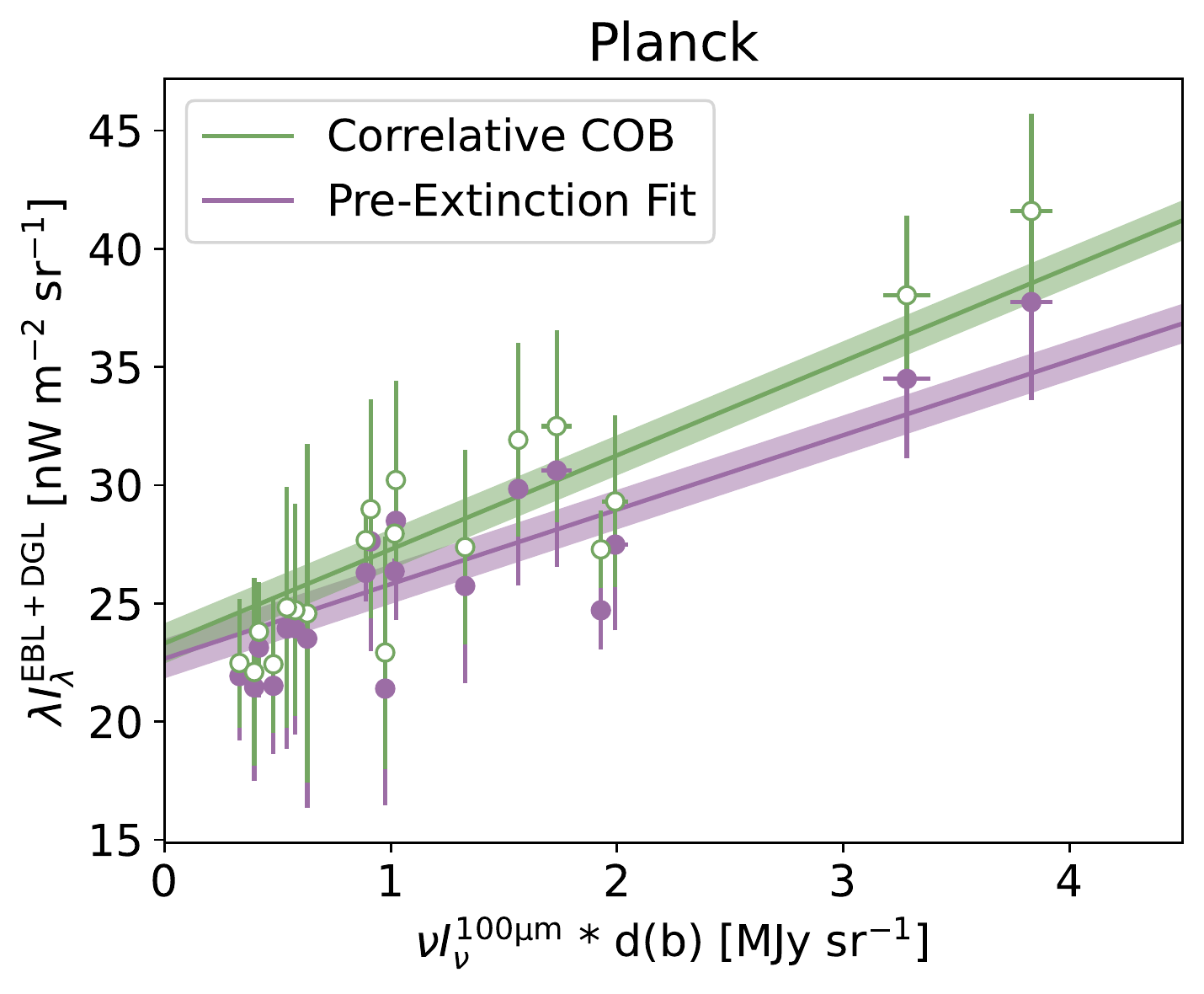} }}%
    \subfloat{{\includegraphics[width=7.5cm]{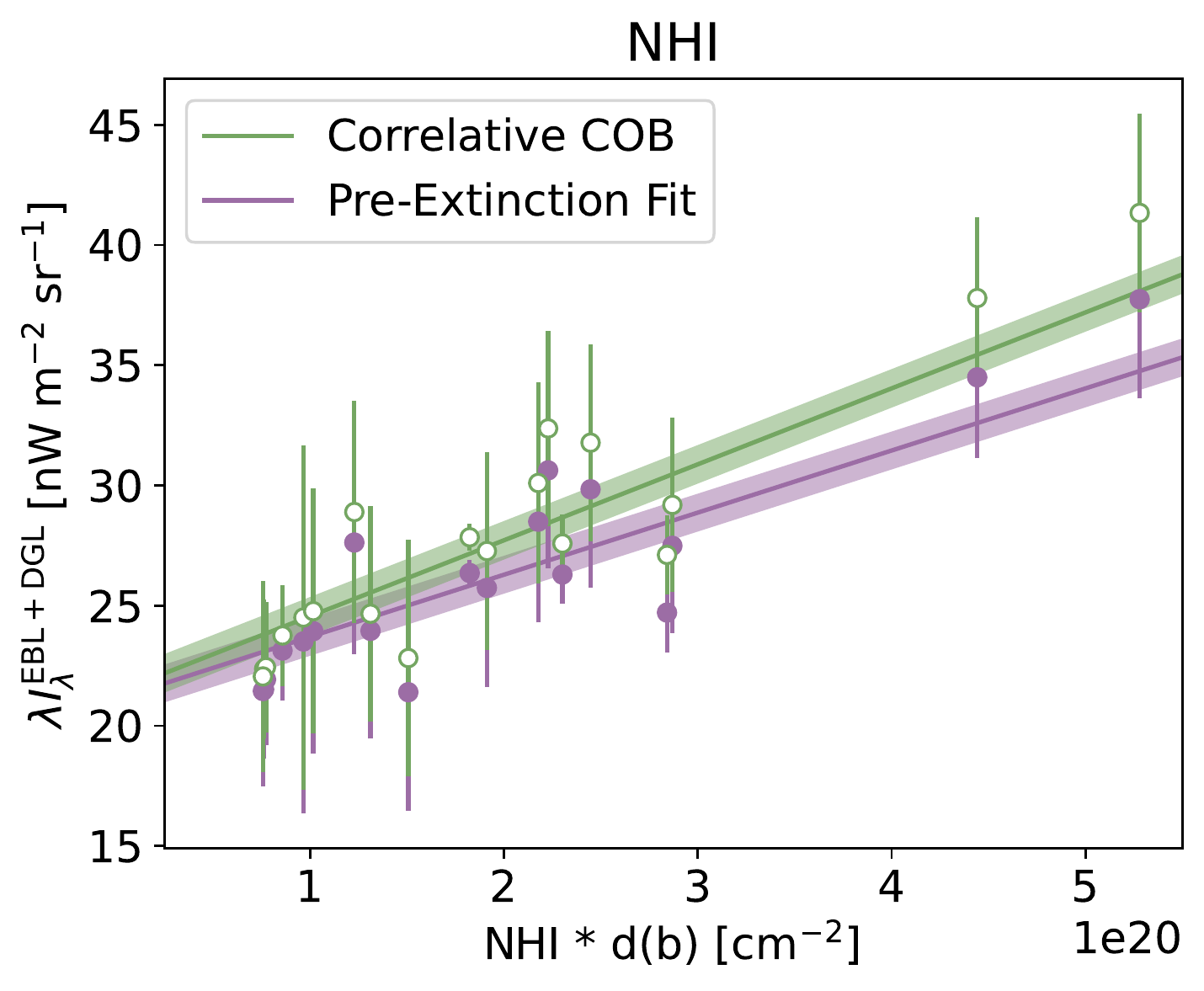} }}%
    \caption{We estimate the COB by fitting a correlation between $\lambda I_{\lambda}^{\mathrm{EBL+DGL}}$ and 100 $\mu$m emission or NHI column density scaled by galactic latitude. Each field is indicated by a filled point, with horizontal and vertical error bars giving $\delta \nu I_{\nu}^{\mathrm{100\mu m}}\cdot d(b)$ (or $\delta\mathrm{NHI}\cdot d(b)$) and $\delta \lambda I_{\lambda}^{\mathrm{EBL+DGL}}$, respectively. The line gives the fit, with the shaded region indicating the RMS error on the fit. While the slope of this fit (purple) is $\nu \beta_{\lambda}$, the intercept is an offset without physical meaning. We then iteratively re-weight this fit to compensate for galactic extinction (green with open points), where the intercept is $\lambda I_{\lambda}^{\mathrm{COB}}$. We perform this procedure four times with our four separate spatial templates (clockwise from top left): IRIS, IRIS/SFD, NHI, and Planck. }%
    \label{fig:cob_correlation}%
\end{figure*}

Using the fitting procedure described in Section \ref{sec:cob_est}, we estimate $\lambda I_{\lambda}^{\mathrm{COB}}$ and $\nu b_{\lambda}$ (as defined by \cite{dirbe6}) as the fit parameters for the four separate spatial templates: IRIS, IRIS/SFD, Planck, and NHI. The fits to the data from all 19 fields are shown in Figure \ref{fig:cob_correlation} both before and after accounting for the expected galactic extinction along the line of sight. The $\lambda I_{\lambda}^{\mathrm{COB}}$ is the extinction-adjusted fit offset and the $\nu \beta_{\lambda}$ is the slope, both of which are listed in Table \ref{tab:fit_results} along with their respective fit errors for each of the four spatial templates. The resulting $\lambda I_{\lambda}^{\mathrm{COB}}$ and $\nu \beta_{\lambda}$ from the four spatial templates are in excellent agreement with each other.

We derive a single best estimate for the COB intensity by computing the mean COB intensity from the four spatial templates, which yields $\lambda I_{\lambda}^{\mathrm{COB}} = 21.98 \pm 1.23\ (\mathrm{stat.}) \pm 1.36\ (\mathrm{cal.})$ \nw. The statistical error is the mean of $\delta \lambda I_{\lambda}^{\mathrm{COB}}$ from the four template fits, while the calibration error is that derived in Table \ref{tab:budget} as a mean for all fields. We simultaneously obtain a $\nu \beta_{\lambda}$ estimate of 5.79 $\pm$ 1.45 \nw/MJy sr$^{-1}$, where the error is the combination of statistical and modeling errors. 

\begin{deluxetable*}{ccccccccc}[htb!]
\tablecaption{For each of our four spatial templates, we calculate $\lambda I_{\lambda}^{\mathrm{COB}}$ [\nw] as the intercept of the extinction-adjusted (green) fit in Figure \ref{fig:cob_correlation} where $\delta \lambda I_{\lambda}^{\mathrm{COB}}$ [\nw] is the statistical error on the intercept. The slope of the non-adjusted (purple) fit is $\nu \beta_\lambda$ [\nw/MJy sr$^{-1}$] with error $\delta \nu \beta_\lambda$ [\nw/MJy sr$^{-1}$]. For the sake of comparison to other measurements, we also calculate $\nu b_\lambda$ and its error $\delta \nu b_\lambda$, which does not contain dependence on $d(b)$ (K. Sano, private communication). For the NHI template only, the slope does not represent $\nu b_\lambda$ as this is the relationship between 100 $\mu$m emission and optical emission and does not apply to NHI column density. Instead, we calculate the relationship between $\lambda I_{\lambda}^{\mathrm{opt}}$ and NHI and its associated error [\nw/cm$^{-2}$].\label{tab:fit_results}}
\tablehead{
\colhead{\textbf{Template}} & \colhead{$\lambda I_\lambda^\mathrm{COB}$} & \colhead{$\delta \lambda I_\lambda^\mathrm{COB}$} & \colhead{$\nu \beta_\lambda$} & \colhead{$\delta \nu \beta_\lambda$} & \colhead{$\nu b_\lambda$} & \colhead{$\delta \nu b_\lambda$} & \colhead{$\lambda I_{\lambda}^{\mathrm{opt}}$/NHI} & \colhead{$\delta \lambda I_{\lambda}^{\mathrm{opt}}$/NHI} }
\startdata
\textbf{IRIS} & 20.58 & 1.46 & 3.45 & 0.86 & 2.73 & 0.77 & ... & ... \\
\textbf{IRIS/SFD} & 22.64 & 1.15 & 3.54 & 0.91 & 3.04 & 0.83 & ... & ... \\
\textbf{Planck} & 23.31 & 1.00 & 3.15 & 0.81 & 2.46 & 0.69 & ... & ... \\
\textbf{NHI} & 21.40 & 1.31 & ... & ... & ... & ... & 2.58 $\times$ 10$^{-20}$ & 0.63 $\times$ 10$^{-20}$ \\
\enddata
\end{deluxetable*}

\vspace{-20pt}
\subsection{Astrophysical and Instrumental Tests}
\label{sec:var_results}
There are several useful checks and validations we can perform with this analysis pipeline. One is to consider what happens if we use a ``standard'' DGL subtraction method, which leads to a different COB estimate. Next, to verify that our COB measurements do not depend on the Milky Way's structure, we search for dependence on the galactic latitude of the fields. Similarly, to constrain the presence of scattered light from IPD in the Edgeworth-Kuiper Belt, we compare our per-field measurements to a model of the IPD \citep{poppe,Poppe_2019b}. We also examine our choice to exclude 150 seconds of data at the beginning of each observing sequence to see how the COB intensity changes with different choices of data cuts. Lastly, we perform a series of jackknife tests based on various physical parameters to detect any effect they may have on the final measurement.

\subsubsection{Direct Subtraction COB Estimate}
\label{sec:sub_results}
\begin{figure*}[htb!]
\centering
\includegraphics[width=5in]{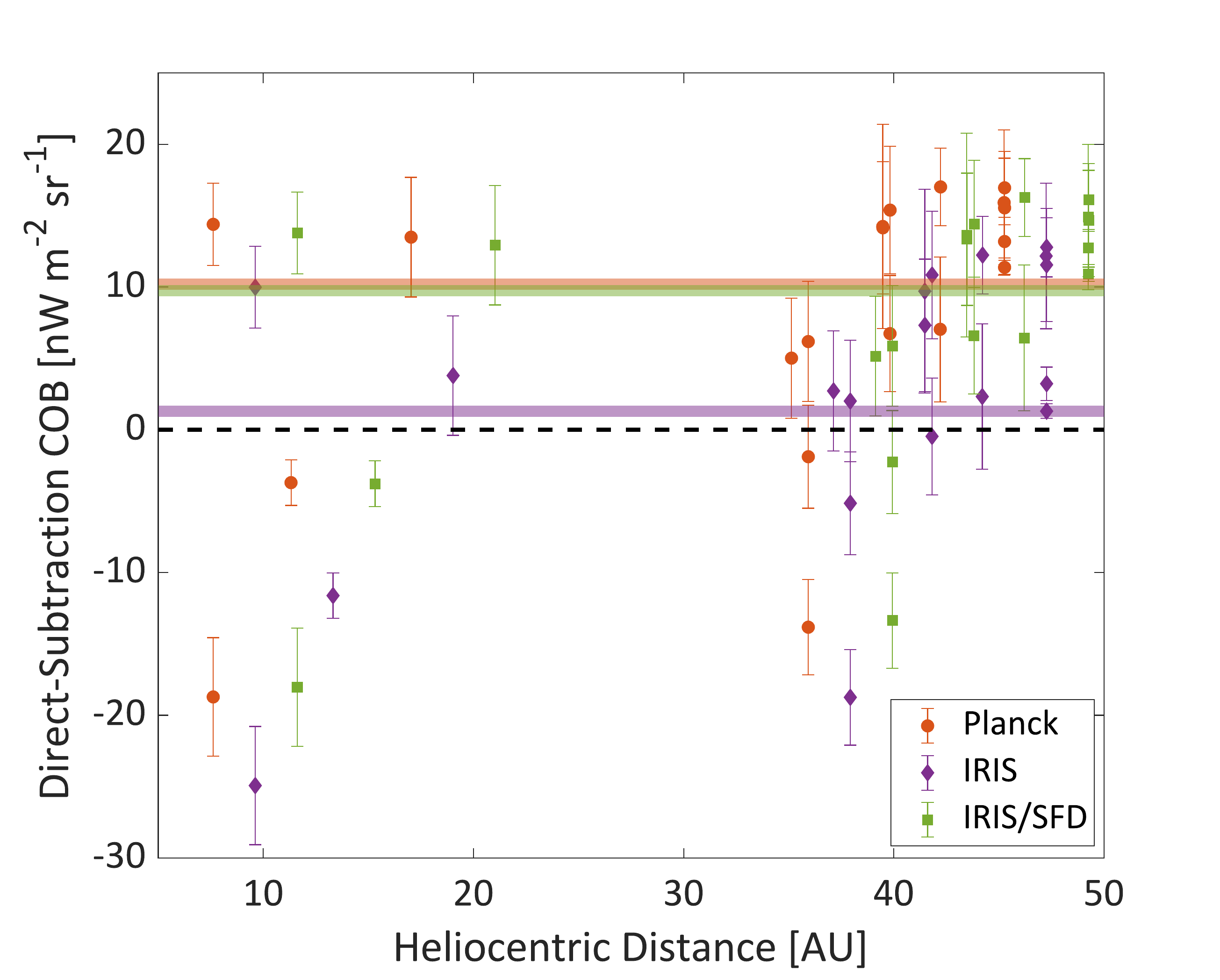}
\caption{Estimate of $\lambda I_{\lambda}^{\mathrm{COB}}$ via direct subtraction of $\lambda I_{\lambda}^{\mathrm{DGL}}$. For the IRIS (purple diamonds), IRIS/SFD (green squares) and Planck (orange circles) spatial templates, we subtract $\lambda I_{\lambda}^{\mathrm{DGL}}$ directly to estimate $\lambda I_{\lambda}^{\mathrm{COB}}$. Points give $\lambda I_{\lambda}^{\mathrm{COB}}$ for each field with statistical error bars. Each template is slightly offset in heliocentric distance for visual clarity. The shaded regions indicate the total combined $\lambda I_{\lambda}^{\mathrm{COB}}$ with combined statistical error from all fields for each template. Significant variation in the DGL between templates makes this method less accurate than correlating directly with FIR emission. \label{fig:cob_sub}}
\end{figure*}
To study the effect of the FIR-optical scaling $\bar{c}_{\lambda}$, we calculate $\lambda I_{\lambda}^{\mathrm{COB}}$ for the IRIS, IRIS/SFD, and Planck templates by directly subtracting the DGL in addition to the other foreground components using the prescription detailed in Section \ref{sec:dgl_direct}. Our per-field measurements are shown in Figure \ref{fig:cob_sub} along with a combined measurement with associated statistical error for each template. 

The direct subtraction estimate of $\lambda I_{\lambda}^{\mathrm{COB}}$ is substantially smaller than our correlative measurement. The primary reason for this is the larger value of $\bar{c}_{\lambda}$, which overproduces the DGL compared with the fit estimate so results in a fainter COB. Figure \ref{fig:b_lam} shows that previous measurements of $\nu b_{\lambda}$ span a range from as low as 5 to as large as 50 \nw / MJy sr$^{-1}$, the choice of which has a significant impact on the resulting COB estimate. Our correlative method is effectively agnostic to this impact as a measurement of $b_{\lambda}$ is a product of our fit, not a contributing parameter. 
The variation in the 100 $\mu$m intensity between the spatial templates also propagates into our estimates in such a way as to produce large scatter in the inferred value of the COB. Additionally, the variation in $\lambda I_{\lambda}^{\mathrm{COB}}$ between the templates is large and some fields produce negative values, which are unphysical. Taken together the direct subtraction COB estimates are more or less consistent with the IGL, which highlights the importance of accurate DGL subtraction to estimates of the COB, even if the other foregrounds are accounted correctly. 

As additional evidence that the correlative method provides a robust estimate of the COB, we note that the NHI correlation is independent of any assumptions about the nature or physics of the scattering dust. The tight agreement between the NHI and thermal dust COB estimates would not occur if $\nu b_{\lambda}$ were very different from our best fitting value.

\subsubsection{Galactic Latitude}
\label{sec:gallat_results}

\begin{figure*}[htb!]
\centering
\includegraphics[width=4in]{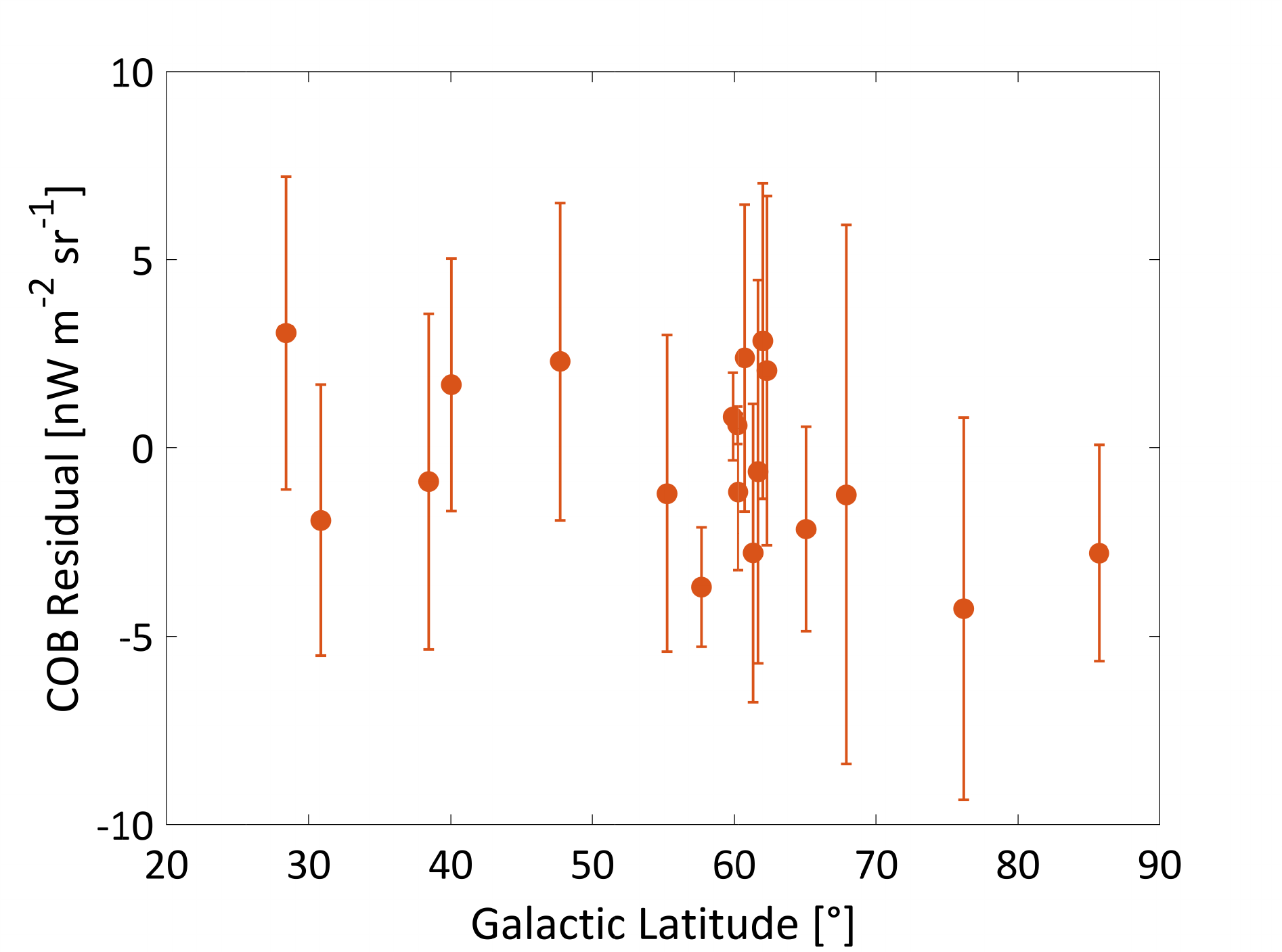}
\caption{A comparison of the residual between our correlative COB measurements with their fit using the Planck template compared to the galactic latitude of each field. Each point gives $\lambda I_{\lambda}^{\mathrm{COB}}$ with the fit subtracted, and the error bars represent $\delta\lambda I_{\lambda}^{\mathrm{COB}}$ based on the set of images for each field. \label{fig:cob_lat}}
\end{figure*}

If we are properly accounting for the variation in galactic structure due to galactic latitude, our COB measurement will not have any dependence on $b$. Figure \ref{fig:cob_lat} gives a comparison of the residual of our correlative COB measurements with their fit for all fields to the galactic latitude of each field. We use the Planck template as an example because the variation between the templates is not large enough to mask any potential trend with field location. 
The Pearson correlation coefficient is -0.41, suggesting at most a weak anti-correlation between the estimated COB and galactic latitude. Having noted that, by construction our central COB value is not directly sensitive to a potential additional variation of the DGL with $b$.

\subsubsection{Interplanetary Dust}
\label{sec:ipd_results}

\begin{figure*}[htb!]
\centering
\includegraphics[width=4in]{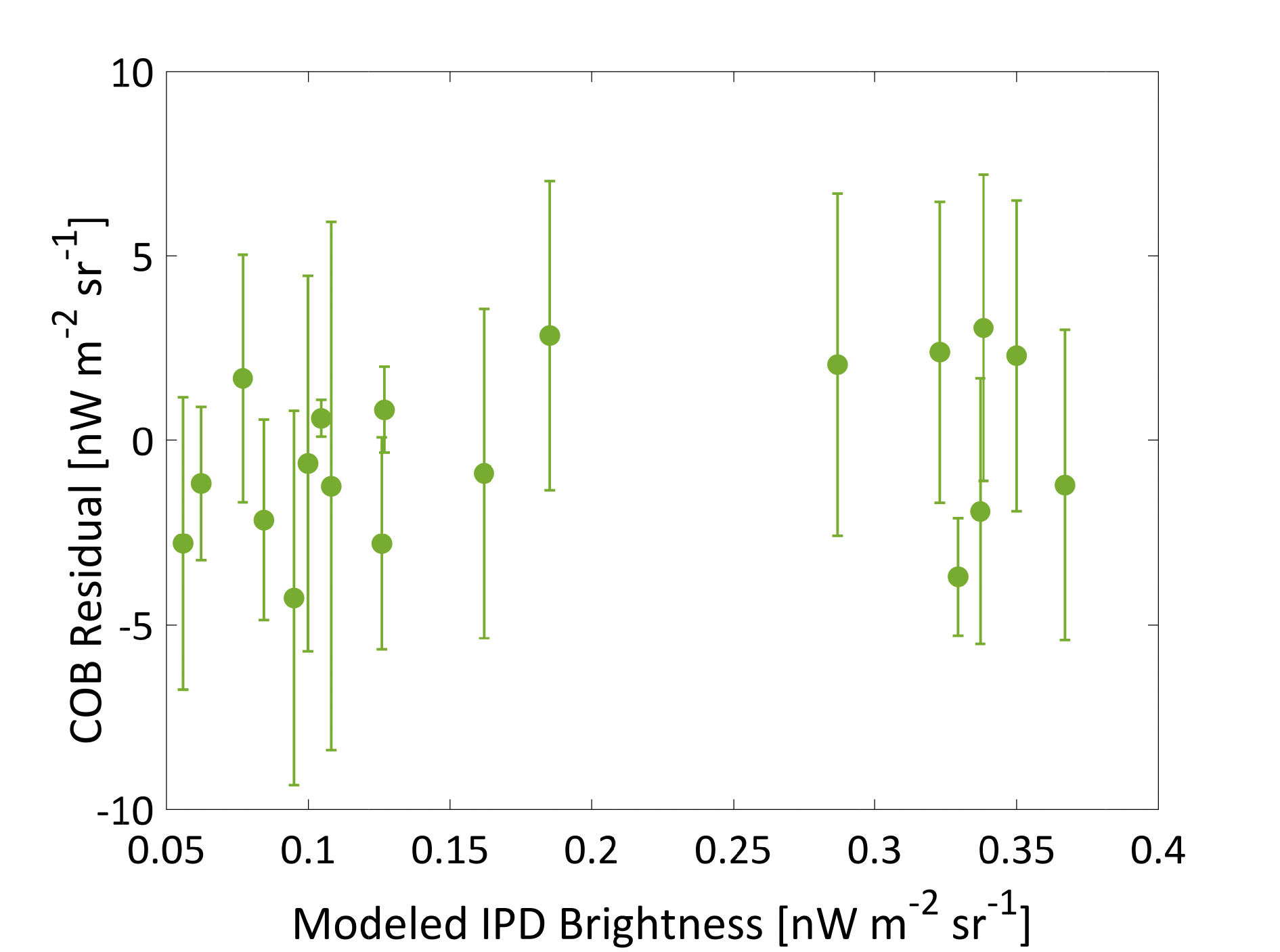}
\caption{A comparison of our correlative COB residual measurements for the Planck template to the interplanetary dust estimated for each field. We do not detect any significant correlation between these quantities. \label{fig:cob_ipd}}
\end{figure*}

Using a model for IPD in the solar system \citep{poppe, Poppe_2019b}, we estimate the surface brightness from sunlight reflected from IPD for each of our fields based on the location of \nh\ at the time the observations were taken (all $>$ 5 AU) and the line of sight to the target. We compare the IPD prediction for each field to the residual of our correlative COB measurements in Figure \ref{fig:cob_ipd}. The Pearson correlation coefficient between these variables is 0.32, suggesting there is no significant relationship between the IPD model and COB residual in these fields. A linear fit between model and residuals gives a slope of 9.37 $\pm$ 4.31, where a unity relation would be expected if the model were correct and zero slope would suggest lack of correlation. The fit slope is consistent with either hypothesis. We conclude that these LORRI data are not sensitive enough to search for light reflected from IPD in the outer solar system, and that at the limit we are able to probe, there is no evidence for such in these data. Since we cannot test the dust model, we do not subtract a surface brightness component associated with IPD from our COB measurement.

\subsubsection{Camera Power-On Data Cut}
\label{sec:camcut_results}
When \cite{Lauer} discovered that first frame power-on effects are important for the LORRI camera, they chose to exclude the first 150 seconds of data from each observation sequence after the camera is first powered on. We have tested this choice against a range of exclusion times from 0 -- 400 seconds to determine the magnitude of any effect this choice may have on the ultimate COB measurement. 

\begin{figure*}[htb!]
\centering
\includegraphics[width=4in]{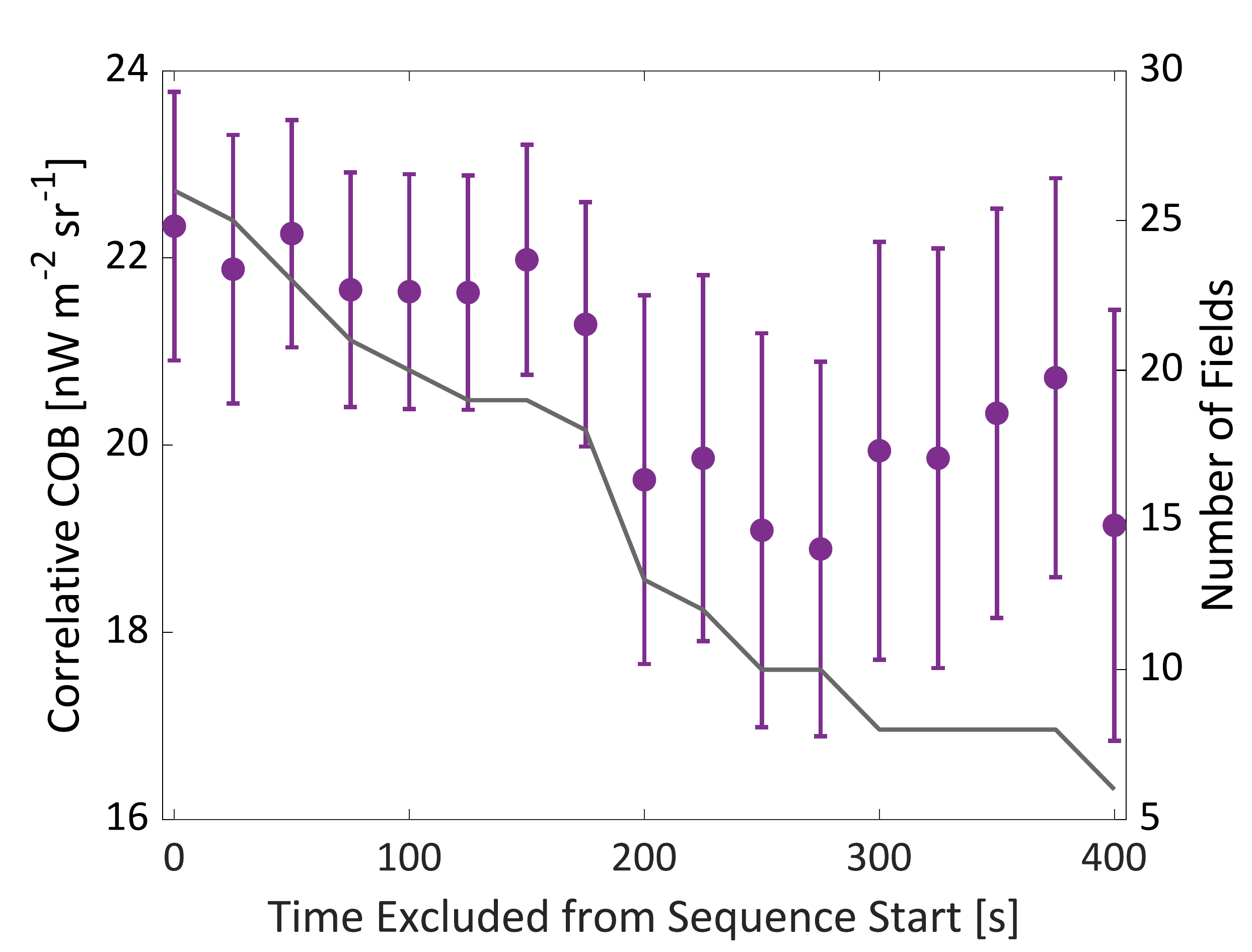}
\caption{We perform the camera power-on data cut for a series of 17 different exclusion times ranging from 0 -- 400 seconds after the start of each observation sequence. We then recalculate the fits shown in Figure \ref{fig:cob_correlation} and derive a new combined $\lambda I_{\lambda}^{\mathrm{COB}}$, shown as the purple points (left axis). The error bars indicate the mean $\delta \lambda I_{\lambda}^{\mathrm{COB}}$ from the four spatial templates. As more data are cut, fewer fields remain from which to fit a measurement as not every field has the total length of observation time required, shown  via the grey line (right axis). This causes increased statistical error. Our choice to exclude 150 seconds of data from each sequence is a stable and robust selection for which statistical error is minimized. \label{fig:cob_time}}
\end{figure*}

After changing the subset of data we are using, we rerun the COB analysis from raw data down through the four spatial template fits and calculate a new combined $\lambda I_{\lambda}^{\mathrm{COB}}$ for each data cut, with the result shown in Figure \ref{fig:cob_time}. 
We find that excluding 0 -- 150 seconds of data has little effect on the resulting COB measurement, although the 150 second exclusion has the smallest statistical error of all trials. Beyond 150 seconds, fewer fields remain from which to draw a measurement and statistical error increases. While the choice of cut does impact the inferred COB, all of the COB measurements from these power-on time cuts agree with each other to 2$\sigma$. We conclude there is an uncertainty of about $1{-}2 \,$ \nw\ associated with the choice of power-on time cut, but it is difficult to assess how this error should be carried since it is within the uncertainty on any given choice.

\vspace{20pt}
\subsubsection{Parameter Jackknife Tests}
\label{sec:jackknife}
We perform a series of jackknife tests in which we split the available science fields approximately in half to test the effect of various physical parameters on our final measurement. For all tests, we repeat the calculation of the correlative $\lambda I_{\lambda}^{\mathrm{COB}}$ and its statistical error $\delta \lambda I_{\lambda}^{\mathrm{COB}}$ for both halves of the data to make a comparison. In Table \ref{tab:jackknife}, we show the results of these tests. 

For the first test, we split our fields into those with heliocentric distance $<$ 37 AU and those with heliocentric distance $>$ 37 AU. This tests dependence of our measurement on the IPD. While fields with lower heliocentric distance produce a slightly higher COB, both sets are indistinguishable within statistical error from each other and from our original measurement. We test dependence on galactic latitude $b$ by dividing our fields into groups with $b$ $<$ 60$\degree$ and $b$ $>$ 60$\degree$. This has the potential to reveal a trend with fainter or brighter DGL. The set with lower $b$ produces a higher COB by $\simmod$1 \nw, but again the results are not significant within their errors. Next, we divide the fields by SEA $<$ 105$\degree$ and SEA $>$ 105$\degree$. This tests our decision to cut data with an SEA $<$ 90$\degree$ as opposed to some other threshold. We see no significant trend as a result of this test. We test for potential dependence on the ISL by dividing our fields based on their masking fraction, which is the percentage of pixels that are masked out of the total number of pixels in a LORRI exposure. We use a threshold of 25\%. Fields with $<$ 25\% of pixels masked produce a slightly lower COB than those with $>$ 25\% of pixels masked, but again with no significant deviation within statistical error. Lastly, we test the fields observed before and after LORRI's software was updated during the period after the Pluto encounter and before the KEM. This tests for any change to the pre-processing pipeline or calibration that could affect our measurement. To search for statistically significant differences in the central values, we compute the $p$-value associated with Welch's $t$-test for each jackknife. As all $p > 0.05$, we conclude there are no significant differences in these tests.

\begin{deluxetable*}{lccc}[htb!]
\tablecaption{For a series of jackknife parameter tests, we re-compute $\lambda I_{\lambda}^{\mathrm{COB}}$, its statistical error, $\delta \lambda I_{\lambda}^{\mathrm{COB}}$, and the $p$-value from Welch's $t$-test. The tests include splitting the available science fields in half by heliocentric distance, galactic latitude, SEA, masking fraction, and before and after the LORRI software was updated post-Pluto encounter. For all tests, $\lambda I_{\lambda}^{\mathrm{COB}}$ demonstrates no significant difference within its statistical error from our original measurement.\label{tab:jackknife}}
\tablehead{
\colhead{\textbf{Jackknife Test}} & \colhead{\textbf{\boldmath$\lambda I_{\lambda}^{\rm COB}$ [nW m$^{-2}$ sr$^{-1}$]}} & \colhead{\textbf{\boldmath$\delta \lambda I_{\lambda}^{\rm COB}$ [nW m$^{-2}$ sr$^{-1}$]}} & \colhead{\textbf{\boldmath$p$-value}} }
\startdata
Heliocentric Distance $<$ 37 AU & 22.40 & 3.58 & \multirow{ 2}{*}{$0.21$}\\ 
Heliocentric Distance $>$ 37 AU & 20.34 & 2.59 & \\\hline
$b$ $<$ 60$\degree$ & 21.39 & 2.28 & \multirow{ 2}{*}{$0.10$}\\ 
$b$ $>$ 60$\degree$ & 19.39 & 2.33 & \\\hline
SEA $<$ 105$\degree$ & 20.14 & 1.81 & \multirow{ 2}{*}{$0.19$}\\ 
SEA $>$ 105$\degree$ & 21.81 & 2.86 & \\\hline
Mask Fraction $<$ 25\% & 19.98 & 2.61 & \multirow{ 2}{*}{$0.70$}\\ 
Mask Fraction $>$ 25\% & 20.46 & 2.31 & \\\hline
Before Software Update & 19.19 & 2.76 & \multirow{ 2}{*}{$0.39$} \\ 
After Software Update & 20.33 & 2.40 & \\
\enddata
\end{deluxetable*}

\vspace{-20pt}
\subsection{Comparison to Previous Measurements}
\label{sec:comp_results}
\begin{figure*}[htb!]
\centering
\noindent\includegraphics[width=0.49\textwidth]{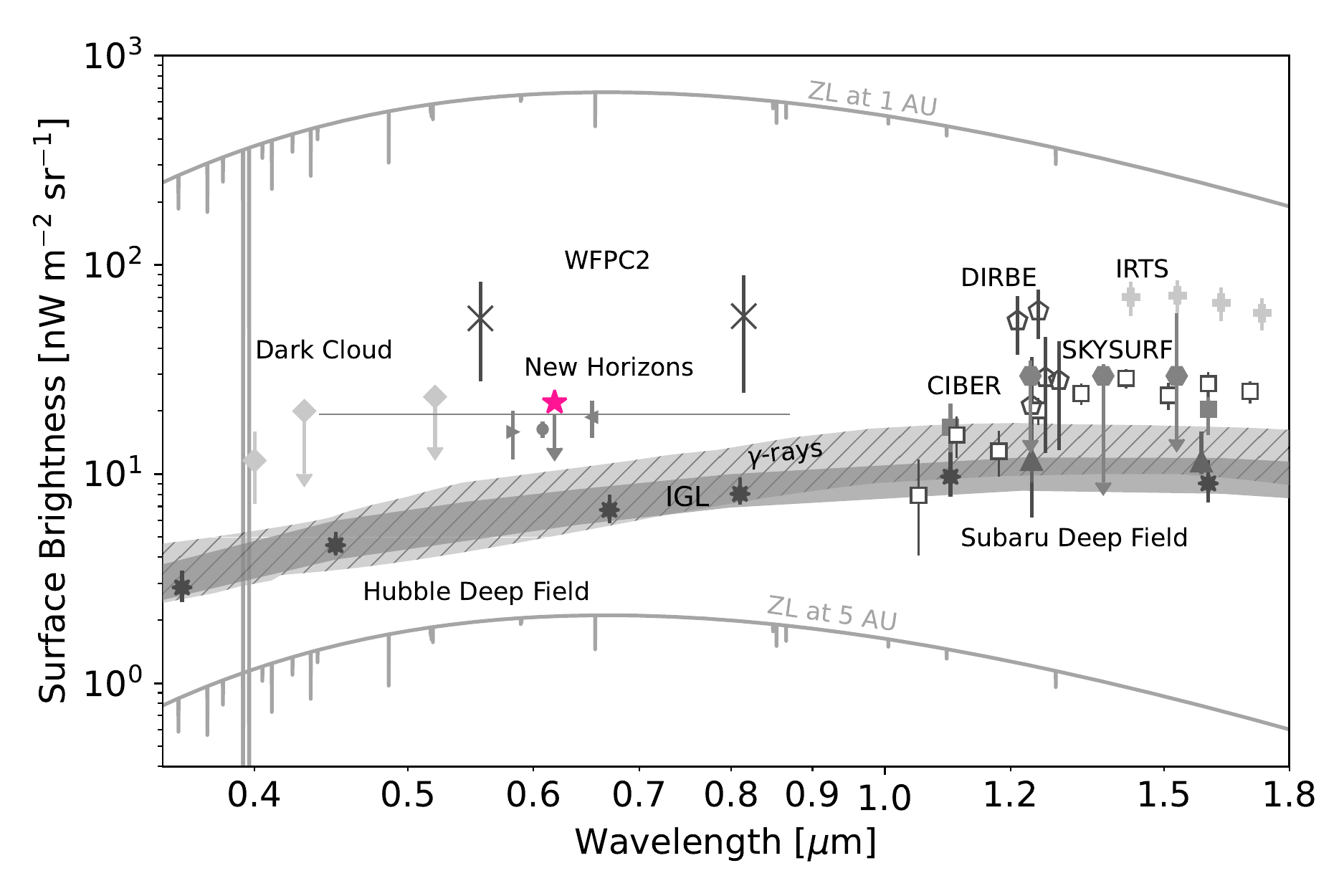}
\noindent\includegraphics[width=0.49\textwidth]{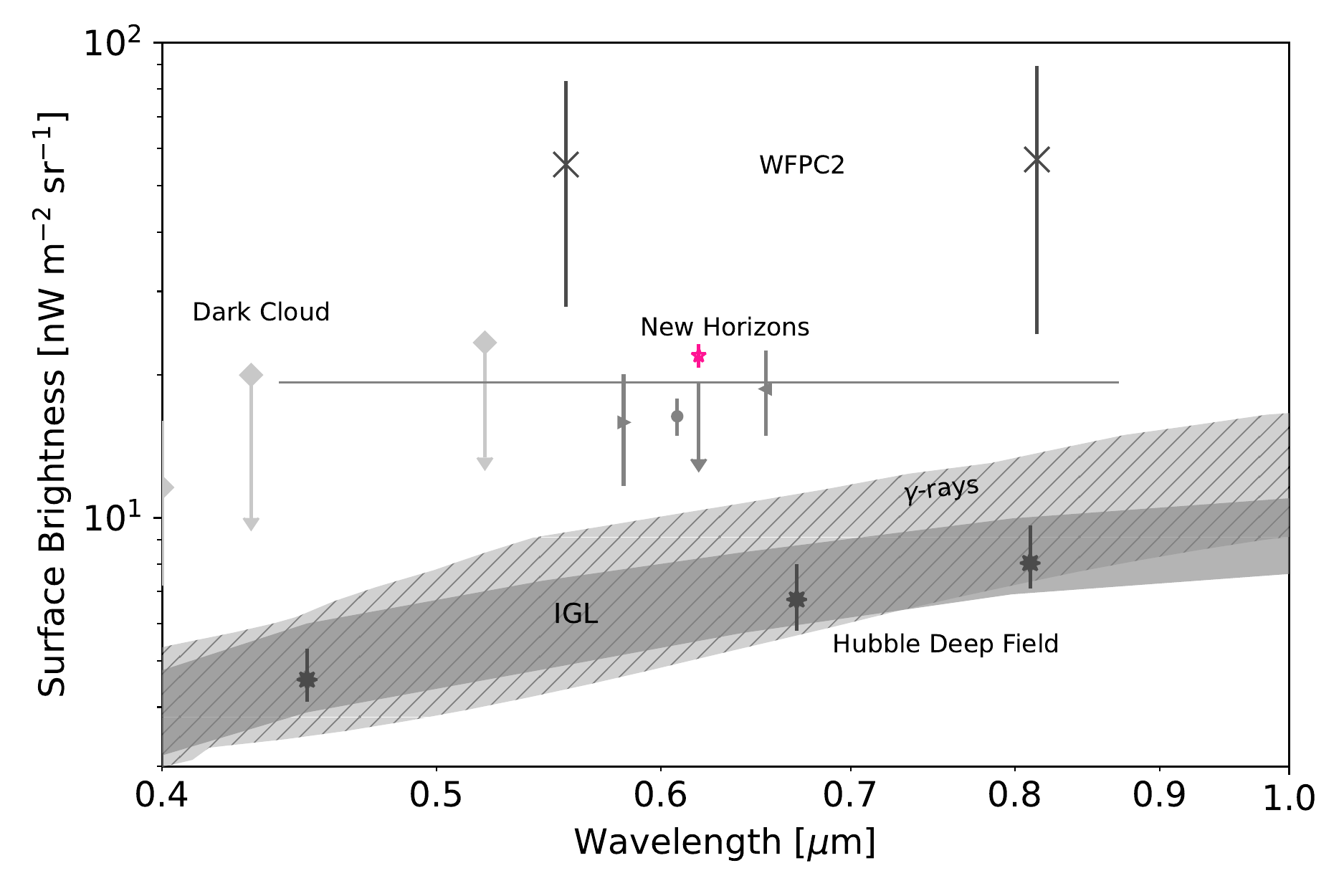}
\caption{Left: Comparison of previous COB and IGL measurements to the measurement we present here. The COB has been constrained or measured using the dark cloud method (light grey diamonds) \citep{dark_cloud}, WFPC2 on HST (dark grey crosses) \citep{Bernstein}, LORRI on \nh\ (upper limit \citep{nature}, left and right triangles showing measurements made using two different models of the DGL \citep{Lauer}, and circle \citep{lauer_2022} -- horizontal error bar indicates wavelength range of LORRI), CIBER (filled and open squares) \citep{Zemcov2014,ciber}, a combination of DIRBE and 2MASS data (open pentagons) \citep{dirbe1,dirbe2,dirbe3,dirbe4,dirbe5,dirbe6}, IRTS (light grey pluses) \citep{irts}, and SKYSURF, a panchromatic archival HST measurement (dark grey hexagons) \citep{skysurf_1,skysurf_2}. The hashed region gives constraints on COB values from a combination of HESS \citep{hess}, Fermi-LAT \citep{fermilat_gamma}, MAGIC \citep{magic_gamma}, and GeV-TeV \citep{gev_tev_gamma} $\gamma$-ray observations. The filled region gives the upper limit on the IGL from galaxy counts based on observations from the Hubble Deep Field (dark grey asterisks) \citep{madau_hst_galcounts,fazio_hst_galcounts} and Subaru Deep Field (triangles) \citep{keenan_subaru_galcounts,totani_subaru_galcounts}. For comparison, the intensity of the ZL at 1 and 5 AU is also shown to highlight how challenging it is to accurately measure the COB from 1 AU \citep{kurucz,nature}. We show our new measurement as a pink star, with statistical error bars (1.23 \nw) too small to be seen on this plot. While slightly higher than the previous upper limit and measurements made using \nh, this measurement is consistent with other direct measurements that show a significant excess in brightness over the expected IGL. Right: Restricted axes version of the COB compilation plot focusing on optical wavelengths to provide a clear comparison to previous measurements. }\label{fig:cob_meas_new}
\end{figure*}

We put our COB measurement in the context of previous EBL measurements in Figure \ref{fig:cob_meas_new}. Our measurement is compatible with the previous measurements made using LORRI, and is in general agreement with other photometric COB measurements including those made using the dark cloud method \citep{dark_cloud}. However, like other recent independent determinations with LORRI \citep{Lauer, lauer_2022}, it is in strong tension with the $\gamma$-ray constraints and the IGL. 

\begin{figure*}[htb!]
\centering
\includegraphics[width=6in]{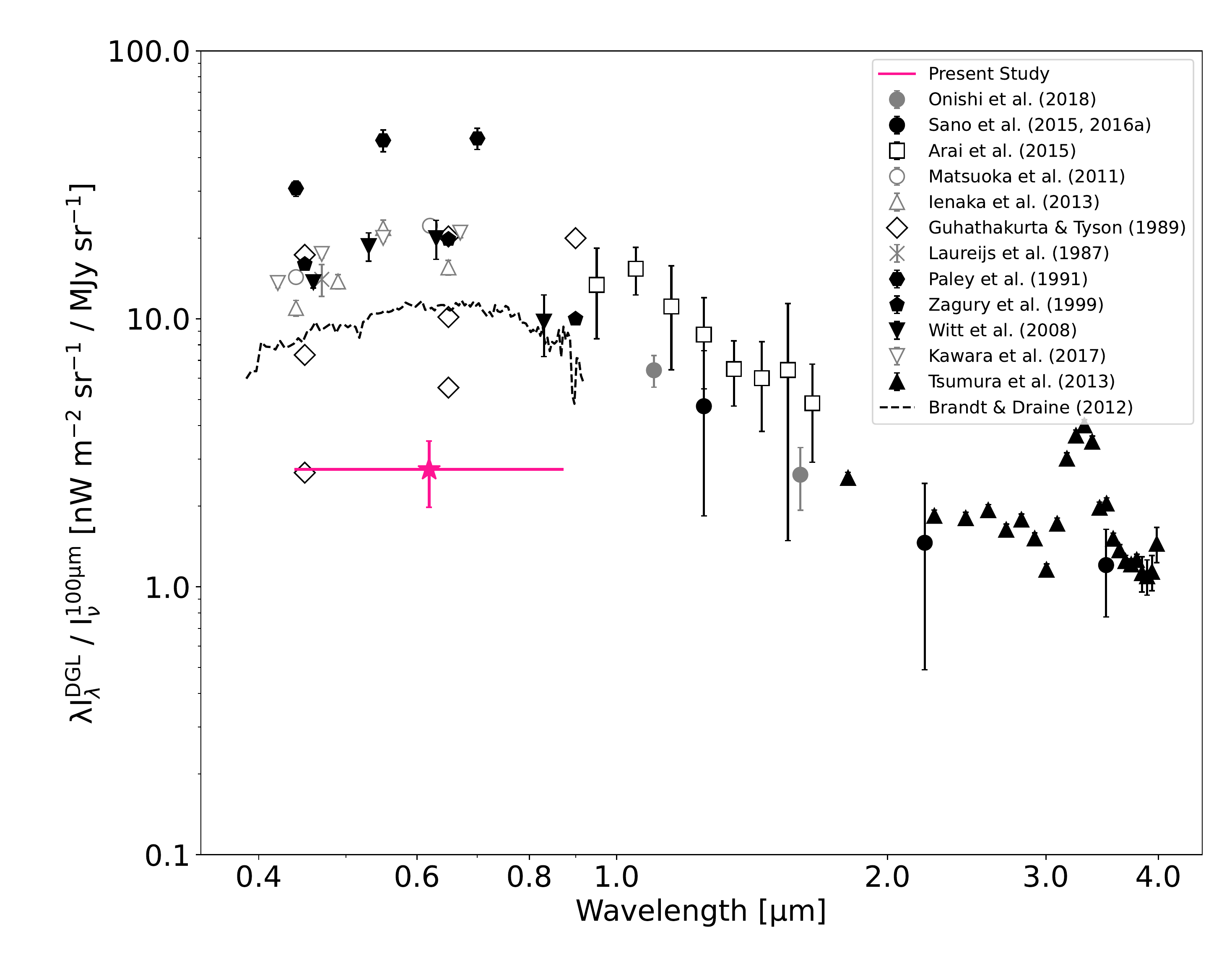}
\caption{Comparison of $\nu b_{\lambda}$ with previous measurements. The pink star gives our estimate of $\nu b_{\lambda}$ combined from the estimates made using the IRIS, IRIS/SFD, and Planck templates. Again, the horizontal bar gives LORRI's wavelength range. The error bars indicate the combined $\delta \nu b_{\lambda}$. Previous studies include \cite{onishi, dirbe5, dirbe6, arai, pioneer_2, ienaka, guhathakurta, laureijs, paley, zagury, witt_2008, kawara, tsumura_2013, brandt_draine_2012}. \label{fig:b_lam}}
\end{figure*}

In Figure \ref{fig:b_lam}, in order to facilitate a comparison to previous measurements, we calculate $\nu b_\lambda$ as a version of $\nu \beta_\lambda$ that does not have any dependence on $d(b)$ (K. Sano, private communication). We estimate $\nu b_\lambda$ = 2.74 $\pm$ 0.76 \nw/MJy sr$^{-1}$, with the individual templates' measurements listed in Table \ref{tab:fit_results}. Our estimate is significantly lower than several independent determinations in the literature. We have investigated possible causes for this, including: 
\begin{itemize}
\item An instrumental component, such as an misestimation of the size of the extended response function. Such an effect would change the relative response between point sources and extended sources, and so would cause a miscalibration of extended emission \citep{Griffin2013}. However, this would increase $b_{\lambda}$, causing a larger observed signal towards fields with larger surface brightness. We conclude misestimation of the beam cannot explain the low $b_{\lambda}$. Instrumental effects unrelated to the coupling of the detector to astrophysical signal would not be correlated with observed surface brightness.

\item Interplanetary dust, if it were unexpectedly bright and by chance anti-correlated with galactic latitude in our specific fields, could cause a low value of $b_{\lambda}$. However, Figure \ref{fig:cob_ipd} demonstrates the lack of IPD signal at an amplitude sufficient to explain the $b_{\lambda}$ discrepancy in these data.

\item Differences in the estimation of residual starlight below the detection threshold between different analyses could cause systematic overestimates of $b_{\lambda}$ compared with our analysis. Measurements of $b_{\lambda}$ require an estimate of residual ISL specific to that measurement to be subtracted. The ISL amplitude in a field is correlated with the DGL amplitude, since both depend on galactic latitude. As a result, errors in the estimation of ISL below a measurement's detection limit could cause artificial boosting of $b_{\lambda}$ for that measurement. We have performed a calculation where we apply progressively brighter star masking thresholds in our analysis, and find that $b_{\lambda}$ does increase with the cut magnitude, as expected. As a point of comparison, we find that $\nu b_{\lambda} = 10$ \nw/MJy sr$^{-1}$ when stars brighter than $m_{G}=11$ are masked. However, at this masking threshold the excess ISL in our fields from $m_{G} > 11$ stars would be $\sim 100$ \nw, a level so large that it would be noticed in the previous measurements. We conclude that residual ISL in existing measurements of $b_{\lambda}$ is an unlikely explanation for the discrepancy.

\item The effect of the CIB zero-point on the $100 \, \mu$m template used in the $b_{\lambda}$ scaling. To examine this, we test an alternative CIB intensity subtraction in our DGL estimation for the IRIS template of 0.24 MJy sr$^{-1}$ \citep{penin}. The IRIS-derived $\lambda I_{\lambda}^{\mathrm{COB}}$ decreases by $<$ 1 
\nw, and the total $\lambda I_{\lambda}^{\mathrm{COB}}$ decreases by $<$ 0.2 \nw. Additionally, there is no change to $b_{\lambda}$. This demonstrates that neither of these quantities are particularly sensitive to the CIB zero-point.

\item Changing properties of the scattering properties of galactic dust with height above the galactic plane. Most previous determinations of $b_{\lambda}$ have been towards relatively bright cirrus regions with higher optical depths than our fields \citep{leinert}. The few determinations of the DGL scaling towards very faint fields have found generally lower scaling values \citep[e.g.][]{Zemcov2014}, suggesting there may be some additional dependence that increases the dispersion in $b_{\lambda}$ along different sight lines. As a check of our value of $b_{\lambda}$, we test for the value of $b_{\lambda}$ that would fully decorrelate the residual intensity as a function of galactic latitude, i.e. cause the slope in Figure \ref{fig:cob_lat} to be 0. This test is not ideal since galactic latitude is an inferior proxy to the scattering dust, but provides a point of comparison with a roughly independent abscissa. We find that $b_{\lambda}$ \simmod\ 5 \nw/MJy sr$^{-1}$ decorrelates the points in Figure \ref{fig:cob_lat}. While this is closer to previously measured values of $b_{\lambda}$, it cannot fully account for the discrepancy and indicates that any additional dependence on galactic latitude cannot resolve these measurements.
\end{itemize}

Due to our lack of knowledge of the input fiducial values, we do not formally carry uncertainties from these effects on our quoted COB result, but we estimate that a reduction of as much as several \nw\ in the final COB brightness could result from combinations of these kinds of effects. Further, isotropic offsets in the DGL brightness are difficult to constrain at the level of the CIB brightness, and will impose uncertainties at the \nw\ level in the COB at our current level of understanding of the CIB absolute intensity. We conclude that the DGL scaling likely dominates the COB measurement error budget, and that more work should be done to constrain the $b_{\lambda}$ relation at optical wavelengths in the future.

Measurements of the relation between optical brightness and NHI column density are uncommon in the literature \citep{leinert}, but \citet{Toller1981} finds a relationship of:
\begin{equation}
\lambda I_{\lambda}^{\rm DGL} = (2.9 \, {\rm nW} \, {\rm m}^{-2} \, {\rm sr}^{-1}) \cdot \left( \frac{N_{\rm HI}}{10^{20} \, {\rm atoms} \, {\rm cm}^{-2}} \right)
\end{equation}
with an unassessed uncertainty from Pioneer 10 data \citep{leinert}. To provide a point of comparison, we recompute the scaling excluding the effect of $d(b)$ and find a value of $1.92\pm0.52$ \nw/$10^{20} \, {\rm atoms} \, {\rm cm}^{-2}$. The excellent match between the COB brightness inferred from the NHI and thermal IR templates is strong evidence that the COB is unexpectedly high compared with models and galaxy counts.

In addition to astrophysical explanations, it is possible something about the LORRI instrument and detector are causing a systematic misestimate of the COB. Though both we and others have attempted to bound or rule out these effects, in some cases the data are not sufficient in themselves to fully assess the possible systematic uncertainty. One such effect is dark current. It is well known that CCDs exposed to cosmic rays will exhibit increased dark current over time \citep{ccds}. Though the LORRI reference pixels do not seem to have changed their characteristics over the course of the mission \citep{Symons2022}, it could be that radiation damage has differentially impacted the dark current in the light and reference pixels at a level that is difficult to observe. However, \cite{Lauer} performed a thorough test of LORRI's dark current and detected no significant change. A possibly related issue is the observed relationship between the reference pixels and the light pixels discussed in Section \ref{sec:refcorr}. Though we derive and correct for an empirical relationship between these quantities, it is puzzling that there is a systematic offset between them in the first place. Applying the nominal surface brightness gain to the offset between the pixel populations, we find this offset corresponds to $16.9$ \nw, which in amplitude could explain the discrepancy between the measured COB and the expected IGL. Finally, the relaxation time of the detector following power-on appears to have a time constant of about 100 seconds, but we are not able to track behavior over very long time scales. It is possible that the detector response following power-on has multiple time constants that would only become apparent when the instrument is powered for long time scales that would source unaccounted systematics in this measurement. The data that are available in the archive are not sufficient to constrain these kinds of effects beyond what we have done. Though we have no evidence that any of the corrections we apply are incorrect, these issues do highlight the difficulty associated with systematic instrumental effects as well as the need for a dark shutter to help reliably track subtle changes in the instrument over years.

Perhaps the most straightforward explanation for an excess of diffuse emission is that our IGL expectation is incorrect and the galaxy counts have a deficit \citep{conselice}. If this is true, upcoming JWST results will likely provide at least a partial resolution due to the telescope's unprecedented ability to detect faint, previously unseen galaxy populations \citep{jwst_og}. Also of concern is if known galaxy populations have significant extended diffuse emission that has not been properly measured. 
IHL has also been extremely difficult to measure because it is both very faint and intrinsically diffuse. However, IHL from low-redshift sources has the potential to explain excess emission \citep{cooray_2012,Zemcov2014, yun_ting}. 
Another proposed source of diffuse emission is faint compact objects (FCO), which could take the form of mini-quasars. These low-redshift objects are proposed to source a large amount of baryonic mass despite being difficult to detect \citep{fco}. JWST should be able to detect FCOs directly if they are the correct explanation. High-redshift primordial black holes are another proposed source. Also referred to as direct collapse black holes (DCBH), these objects may also provide an explanation for high-mass, high-redshift quasars \citep{dchb}. Even though current $\gamma$-ray measurements tend to align more closely with the expected IGL, a dense population of dark matter particles could have the potential to prevent pair-production as $\gamma$-rays travel long distances, which could result in an under-estimate of $\gamma$-ray attenuation. This could result in $\gamma$-ray measurements more aligned with photometric EBL measurements than the IGL expectation \citep{gamma_cosmo}. 
An alternative explanation for the LORRI COB excess is an origin
related to particle decays, especially axion-like particles (ALPs)
with a mass in the range of 0.5 eV to 10 eV
\citep{gong,kohri,bernala}. In addition
to the mean intensity, such decays are expected to leave a large
anisotropy signal in the COB and can
be measured with anisotropy power spectra (e.g., CIBER: \citealt{Zemcov2014}; HST: \citealt{mitchell}). A recent analysis of COB
intensity and fluctuations
power spectra find evidence for ALP decays of $\simmod$9.1 eV particles at the
2$\sigma$ level \citep{bernalb}. It is expected that the shorter
wavelength optical and UV COB and anisotropy
measurements can further constrain dark matter decays as a source of
the intensity excess and will likely be targets for upcoming
sub-orbital and space-based measurements using the small
satellite architecture. 
Any of these proposed sources or some combination of all of them could together make up the observed excess. 

Improved targeted measurements from more capable instruments will be necessary to resolve the current discrepancies between IGL and the photometrically determined COB. As of this writing, JWST has recently returned its first data, including a deep field observation that will likely revolutionize our understanding of galaxies in the universe \citep{jwst,jwst_og}. Upcoming missions such as SPHEREx \citep{spherex}, the first NIR all-sky spectral survey, and Euclid \citep{euclid}, which will also measure galaxy redshifts, will provide unique opportunities for next-generation measurements of the EBL. However, these missions will still be located at 1 AU and will suffer from the same foregrounds that have dogged COB measurements for decades. Even if ZL can be handled, both this COB measurement and previous measurements are critically dependent on the characterization and subtraction of the DGL, and under (or over) subtraction of DGL has an outsized impact on the scientific interpretation. Further study of the scaling between optical and FIR emission along with the scaling's dependence on sky position is necessary to resolve these disparate measurements. A dedicated small probe to the outer solar system, or a piggy-back instrument on a similar planetary or heliophysics mission, would be able to provide the best possible measurement \citep{Zemcov_PASP, Cooray_2009}. One particularly intriguing mission concept is the Interstellar Probe \citep{isp}. The proposed 50-year mission into interstellar space would provide an unparalleled opportunity to shed light on the EBL 
in the darkness between stars. 

\section*{Acknowledgments} 
The work of T.S., M.Z., A.C., and A.R.P.~was supported by the New Frontiers Data Analysis Program (NFDAP) under NASA grant 80NSSC18K1557. 
We would like to thank the LORRI team for advice, and the PDS archive team for help with both queries about the served LORRI data and assistance ingesting our final data products. 
The authors thank E. R. Imata for his assistance in archiving study results to the NASA Planetary Data System.
Thanks to undergraduate research assistants S. Venuto, S. Thayer, A. Dignan, D. Houlihan, and A. Bush for their work on this project and V.~Gorjian for comments that helped improve this work.
This work has made use of data from the European Space Agency (ESA) mission
{\it Gaia} (\url{https://www.cosmos.esa.int/gaia}), processed by the {\it Gaia}
Data Processing and Analysis Consortium (DPAC,
\url{https://www.cosmos.esa.int/web/gaia/dpac/consortium}). Funding for the DPAC
has been provided by national institutions, in particular the institutions
participating in the {\it Gaia} Multilateral Agreement.
EBHIS is based on observations with the 100-m telescope of the MPIfR (Max-Planck-Institut für Radioastronomie) at Effelsberg. The Parkes Radio Telescope is part of the Australia Telescope which is funded by the Commonwealth of Australia for operation as a National Facility managed by CSIRO.

\software{Astrobase \citep{astrobase}, Astropy \citep{astropy,astropy_2}, gdpyc \citep{gdpyc}, Matplotlib \citep{matplotlib}, NumPy \citep{numpy}, pandas \citep{pandas}, and SciPy \citep{scipy}.}

\newpage
\bibliography{ref}

\end{document}